\documentclass[12pt]{article}

\PassOptionsToPackage{unicode}{hyperref}
\PassOptionsToPackage{hyphens}{url}
\PassOptionsToPackage{dvipsnames,svgnames,x11names}{xcolor}
\usepackage{amsmath,amssymb}
\usepackage{iftex}
\ifPDFTeX
  \usepackage[T1]{fontenc}
  \usepackage[utf8]{inputenc}
  \usepackage{textcomp} 
\else 
  \defaultfontfeatures{Scale=MatchLowercase}
  \defaultfontfeatures[\rmfamily]{Ligatures=TeX,Scale=1}
\fi
\ifPDFTeX\else
\fi
\IfFileExists{upquote.sty}{\usepackage{upquote}}{}
\IfFileExists{microtype.sty}{
  \usepackage[]{microtype}
  \UseMicrotypeSet[protrusion]{basicmath} 
}{}
\usepackage{xcolor}
\usepackage[margin=1in]{geometry}
\usepackage{graphicx}
\makeatletter
\def\maxwidth{\ifdim\Gin@nat@width>\linewidth\linewidth\else\Gin@nat@width\fi}
\def\maxheight{\ifdim\Gin@nat@height>\textheight\textheight\else\Gin@nat@height\fi}
\makeatother
\setkeys{Gin}{width=\maxwidth,height=\maxheight,keepaspectratio}
\makeatletter
\def\fps@figure{htbp}
\makeatother
\usepackage{setspace}
\usepackage{graphicx}
\usepackage{subcaption}
\usepackage{caption}
\usepackage{epsfig}
\usepackage{psfrag}
\usepackage{wrapfig}
\usepackage[all]{xy}
\usepackage{bbm}
\usepackage{mathtools}
\usepackage{float}
\usepackage{longtable,booktabs}
\usepackage{diagbox}
\usepackage{tikz}
\usetikzlibrary{positioning,shapes.geometric}
\usetikzlibrary{shapes,chains}
\usetikzlibrary[calc]
\usepackage{subcaption}
\usepackage{array}
\usepackage{pdfpages}
\usepackage[linesnumbered,ruled,vlined]{algorithm2e}

\def\Var{{\rm Var}\,}
\def\E{{\rm E}\,}
\def\arg{{\rm arg}\,}
\def\Cov{{\rm Cov}\,}

\ifLuaTeX
  \usepackage{selnolig}  
\fi
\usepackage[]{natbib}
\IfFileExists{bookmark.sty}{\usepackage{bookmark}}{\usepackage{hyperref}}
\IfFileExists{xurl.sty}{\usepackage{xurl}}{} 
\hypersetup{
  pdftitle={Causal Inference in Longitudinal Data under Unknown
Interference},
  pdfauthor={Ye Wang; Michael Jetsupphasuk},
  pdfkeywords={Causal inference, Design-based, Interference, Spillover
effect, Longitudinal data, IPTW estimator, Marginal structural models},
  colorlinks=true,
  linkcolor={Maroon},
  filecolor={Maroon},
  citecolor={Blue},
  urlcolor={blue},
  pdfcreator={LaTeX via pandoc}}

\title{Causal Inference in Longitudinal Data under Unknown
Interference}
\author{Ye Wang\footnote{Assistant Professor, Department of Political
  Science, University of North Carolina at Chapel Hill, \href{yewang@unc.edu}{yewang@unc.edu}.} \and Michael Jetsupphasuk\footnote{PhD Candidate,
  Department of Biostatistics, University of North Carolina at Chapel
  Hill, \href{jetsupphasuk@unc.edu}{jetsupphasuk@unc.edu}.}}
\date{}

\begin{document}
\maketitle
\begin{abstract}
In longitudinal studies where units are embedded in space or networks, interference may arise, meaning that a unit's outcome can depend on treatment histories of others. The presence of interference poses significant challenges for causal inference, particularly when the interference structure---how a unit's outcome responds to others' influences---is complex, heterogeneous, and unknown. This paper develops a general framework for identifying and estimating both direct and spillover effects of treatment histories under minimal assumptions about the interference structure. We introduce a class of causal estimands that capture the effects of treatment histories at any specified proximity level and show that they can be represented by a modified marginal structural model. Under sequential exchangeability, these estimands are identifiable and can be estimated using inverse probability weighting. We derive conditions for consistency and asymptotic normality of the estimators and provide procedures for constructing asymptotically conservative confidence intervals. The method's utility is demonstrated through applications in both social science and biomedical settings.
\end{abstract}

\newtheorem{assumption}{Assumption}
\newtheorem{theorem}{Theorem}
\newtheorem{proposition}{Proposition}
\newtheorem{remark}{Remark}
\newtheorem{lemma}{Lemma}
\newpage

\section{Introduction}\label{introduction}
Empirical researchers are often interested in estimating the causal effects of treatment histories in longitudinal data. For example, public health researchers may seek to determine whether a history of smoking increases the likelihood of developing cardiovascular diseases, while social scientists want to understand how past experiences with government programs shape voting behavior. Under the standard assumption of sequential exchangeability---in either experimental or observational studies---these effects can be represented by a marginal structural model (MSM) and consistently estimated through inverse probability weighting (IPW) \citep{robins2000marginal, hernan2010causal}.

However, this classic approach relies on the stable unit treatment value assumption (SUTVA), which rules out the possibility that the outcome of a unit is affected by treatment histories of others, a phenomenon known as spillover or interference \citep{cox1958planning}. In contexts where units are embedded in geographic space or social networks, interference is likely present, and researchers may lack knowledge of the interference structure---the relationship between a unit's outcome and the treatments of others. In reality, this relationship can be complex and heterogeneous across units. For example, some individuals' health may be influenced only by secondhand smoke from immediate contacts in the social network, whereas others who play a more active role in the community may also be affected by the smoking behavior of individuals to whom they are not directly connected. In such settings, which we refer to as ``unknown interference,'' two key questions arise: First, what can still be learned using the classic approach? Second, how can we identify and estimate spillover effects generated by one unit's treatment on the outcome of others?

In this paper, we present a general framework to address these questions. We introduce a novel estimand, the average marginalized response (AMR), which is defined as the average, across all units in the sample, of the expected outcomes of their neighbors at a proximity level of $d \geq 0$, under a fixed treatment history for the unit. The AMR remains well-defined for any user-specified proximity metric, such as spatial distance, network path length, or shared group membership, even when the interference structure is unknown.\footnote{The proximity metric is used solely to index spillover effects and does not impose any restrictions on the form or complexity of the interference structure.} When $d = 0$, the AMR reduces to the expected outcome of each unit under a fixed treatment history of its own, averaged over the sample. For any $d$, the AMR is a function of treatment history alone, thus can be represented by a MSM, where linear combinations of the parameters describe the causal effect of specific treatment histories on the expected outcome of the unit itself (when $d = 0$) or its neighbors at the proximity level of $d$ (when $d > 0$) under the prevailing treatment assignment policy.

We then show that both the AMR and the associated MSM parameters can be identified under the standard assumption of sequential exchangeability. This identification result enables estimation of MSM parameters using IPW estimators, such as weighted least squares (WLS), where we regress the average outcome of a unit's neighbors at the proximity level of $d$ on indicators of its own treatment history, weighted by the inverse probability of that history being observed. When $d = 0$, the estimator coincides with the classic approach, so its results can be interpreted as estimated effects for the AMR at $0$. Estimates for $d > 0$ allow researchers to examine how the effects of different treatment histories extend to neighbors at varying levels of proximity. 

We prove that this estimator exhibits desirable large-sample properties---specifically, consistency and asymptotic normality---when either the influence of units' treatments decays sufficiently quickly over proximity or neighborhood sizes grow slowly with the sample size, using recently developed central limit theorems for dependent data \citep{kojevnikov2021limit, leung2022causal}. In addition, we provide a heteroskedasticity and autocorrelation-consistent (HAC) variance estimator that is guaranteed to be conservative, enabling researchers to construct Wald-style confidence intervals with asymptotically valid coverage. 

We explore several extensions of the method, including evaluating the identification assumptions using placebo tests and allowing for diffusion or dependence in treatment assignment. We also discuss the implications of misspecifying the MSM and having measurement error in the proximity metric. We test the method's performance via Monte Carlo simulations and demonstrate its application through two empirical studies. The first study examines whether wind turbine proposals generate spillover effects on political behavior among Canadian voters. The second study investigates whether exposure to second-hand smoke through social connections can impact health outcomes in the Framingham Heart Study.

This paper contributes to several strands of the causal inference literature. Similar to the result in \citet{savje-etal2018-unknown-interference} for cross-sectional settings, we show that the classic approach in longitudinal analysis \citep{robins2000marginal, hernan2010causal} yields estimates with a causal interpretation even under interference. We further extend this approach to estimate spillover effects at various proximity levels. The definition of the AMR generalizes the concept of marginalized causal effects \citep{hudgens_halloran08, hu2022average, papadogeorgou2020causal, wang2020design} to longitudinal settings, and we clarify its connection to MSMs.

Compared to methods based on ``exposure mappings'' \citep{ogburn2017causal, jiang2023dynamic, jetsupphasuk2025estimating}, which assume that the influence of other units follows a known functional form, our approach requires only minimal knowledge of the interference structure and remains valid when this structure differs across units. It is also robust to dependence in confounders and contagion in outcomes and avoids the additional step of computing exposure probabilities \citep{aronow2017estimating, forastiere2021identification}. We demonstrate how the existing inference procedure developed for exposure mappings \citep{kojevnikov2021limit, leung2022causal, gao2023causal} can be applied to estimates of spillover effects in our framework. By incorporating more information from treatment assignment, our approach complements methods built upon outcome models, which are commonly employed in both network and spatio-temporal analyses \citep{ogburn2020causal, reich2021review}.

The rest of the paper is organized as follows: Section \ref{the-framework} describes the basic framework and defines the estimands. Section \ref{identification-and-estimation} discusses identification and introduces the estimators. Section \ref{large-sample-theory} establishes large-sample properties of the proposed estimators and develops methods for conducting statistical inference. Section \ref{extensions} explores possible extensions. Section \ref{application} presents results from two empirical analyses. Section \ref{conclusion} concludes.

\section{The Framework}\label{the-framework}
\subsection{Set Up}\label{set-up}
We focus on complete longitudinal data from either experiments or observational studies with $N$ units spanning over $T + 1$ periods, where $N \gg T$.\footnote{We discuss scenarios with a large $T$ in Section~\ref{extensions}.} Throughout the paper, uppercase letters denote random variables, lowercase letters represent their realizations, and boldface indicates vectors or matrices. Subscripts refer to specific units, while superscripts in parentheses indicate specific time periods. For each unit $i \in \mathcal{N} = \{1,2,\dots,N\}$ in period $t \in \{0, 1, 2, \dots ,T\}$, we observe the treatment status $A_{i}^{(t)} \in \{0,1\}$ and a vector of time-varying confounders $\mathbf{L}_{i}^{(t)}$, which can also include time-invariant variables. For simplicity, we assume that $A_{i}^{(0)} = 0$ for all $i$. The outcome $Y_{i}$ is measured in the final period $T$. We assume that units in the data are embedded in a fixed social network or geographic space, $\mathcal{G}_{N}$, with a time-invariant proximity metric $d_{ij}$ defined for every pair of units $i$ and $j$. 

\begin{figure}[htp]
 \begin{center}
 \caption{A DAG illustration}
 \label{fig:dag}
\begin{tikzpicture}[
scale=0.85,dot/.style={fill,draw,circle,minimum width=1pt},
arrow style/.style={->,line width=1pt, shorten <=2pt,shorten >=2pt, lightgray},
arrow1 style/.style={->,line width=1.2pt, shorten <=2pt,shorten >=2pt},
arrow2 style/.style={->,line width=1.2pt, shorten <=2pt,shorten >=2pt, red},
arrow3 style/.style={->,line width=1.2pt, shorten <=2pt,shorten >=2pt, blue},
arrow4 style/.style={->,line width=1.2pt, shorten <=2pt,shorten >=2pt, purple},
arrow5 style/.style={<->,line width=1pt, dashed, shorten <=2pt,shorten >=2pt },]
\node [fill=black, dot, label=below right: $Y_{i}$] (y2) at (8,5) {};
\node [fill=black, dot, label=above right: $Y_{j}$] (y4) at (8,1) {};
\node [fill=gray, dot, label=below right: $\mathbf{L}_{i}^{(T)}$] (y1) at (2,5) {} edge [arrow style, bend left=20] (y2) edge [arrow style, bend left=20] (y4);
\node [fill=gray, dot, label=above right: $\mathbf{L}_{j}^{(T)}$] (y3) at (2,1) {} edge [arrow style, bend right=20] (y4) edge [arrow style, bend right=20] (y2);

\node [fill=white, dot, label=below left: $A_{i}^{(T)}$] (a2) at (6,5) {} edge [arrow2 style] (y2) edge [arrow3 style] (y4);
\node [fill=white, dot, label=below left: $A_{i}^{(T-1)}$] (a1) at (0,5) {} edge [arrow style] (y1) edge [arrow2 style, bend left=30] (y2) edge [arrow1 style, bend left=20] (a2) edge [arrow style] (y3) edge [arrow3 style] (y4);
\node [fill=white, dot, label=above left: $A_{j}^{(T)}$] (a4) at (6,1) {} edge [arrow2 style] (y4) edge [arrow3 style] (y2);
\node [fill=white, dot, label=above left: $A_{j}^{(T-1)}$] (a3) at (0,1) {} edge [arrow style] (y1) edge [arrow style] (y3) edge [arrow1 style, bend right=20] (a4) edge [arrow2 style, bend right=30] (y4) edge [arrow3 style] (y2);

\draw[arrow1 style] (y1) -- (a2);
\draw[arrow1 style] (y3) -- (a4);
\end{tikzpicture}
 \end{center}
 \textit{Notes:} The DAG depicts two units, $\{i, j\}$, observed over two periods, $\{T-1, T\}$. Variables are represented by circles and causal paths by arrows. White circles denote treatment, gray circles represent time-varying confounders, and black circles correspond to outcomes. Red arrows illustrate the effect of a unit's own treatment history, while blue arrows capture spillover effects due to interference. Black arrows indicate relationships influencing treatment assignment, and gray arrows represent other potential dependencies between variables under the assumptions outlined in Section~\ref{identification-and-estimation}.
\end{figure}

We adopt the potential outcomes framework and allow for an unknown interference structure. We represent unit $i$'s treatment history between period $1$ and period $T$ by $\mathbf{A}_i^{(1:T)} = \left(A_{i}^{(1)},A_{i}^{(2)},\dots,A_{i}^{(T)}\right)$. For any subset $\mathcal{S} \subseteq \mathcal{N}$, $\mathbf{A}_{\mathcal{S}}^{(1:T)}$ denotes the collection of $\mathbf{A}_i^{(1:T)}$ for $i \in \mathcal{S}$, an $(|\mathcal{S}|\times T)$-dimensional vector. Both the outcome variable $Y_{i}$ and time-varying confounders $\mathbf{L}_{i}^{(t)}$ are functions of the full treatment histories of all the $N$ units, $\mathbf{A}_{\mathcal{N}}^{(1:T)}$: 
\vspace{-1em}
\begin{equation}
    Y_{i} = Y_{i}\left(\mathbf{A}_{\mathcal{N}}^{(1:T)}\right), \mathbf{L}_{i}^{(t)} = \mathbf{L}_{i}^{(t)}\left(\mathbf{A}_{\mathcal{N}}^{(1:T)}\right),\vspace{-0.5em}
\end{equation}
where the functional forms are unknown to the researcher and may vary across units. Since $\mathbf{A}_{\mathcal{N}}^{(1:T)}$ is an $(N\times T)$-dimensional vector, there could be $2^{N\times T}$ different potential values for each $Y_{i}$ or $\mathbf{L}_{i}^{(t)}$, in contrast to $2^{T}$ values in the classic setting without interference. Figure \ref{fig:dag} depicts the relationships across variables in our framework using a directed acyclic graph (DAG) with two units and two time periods. The arrows from $A_{i}^{(T-1)}$ and $A_{i}^{(T)}$ to $Y_{i}$ represent the causal effect of unit $i$'s own treatment history, while the arrows from $A_{j}^{(T-1)}$ and $A_{j}^{(T)}$ to $Y_{i}$ represent spillover effects due to interference.

\begin{figure}
 \begin{center}
 \caption{Simulated Data and Treatment Assignment}
 \label{fig:space}
\includegraphics[width=.45\linewidth, height=.5\linewidth]{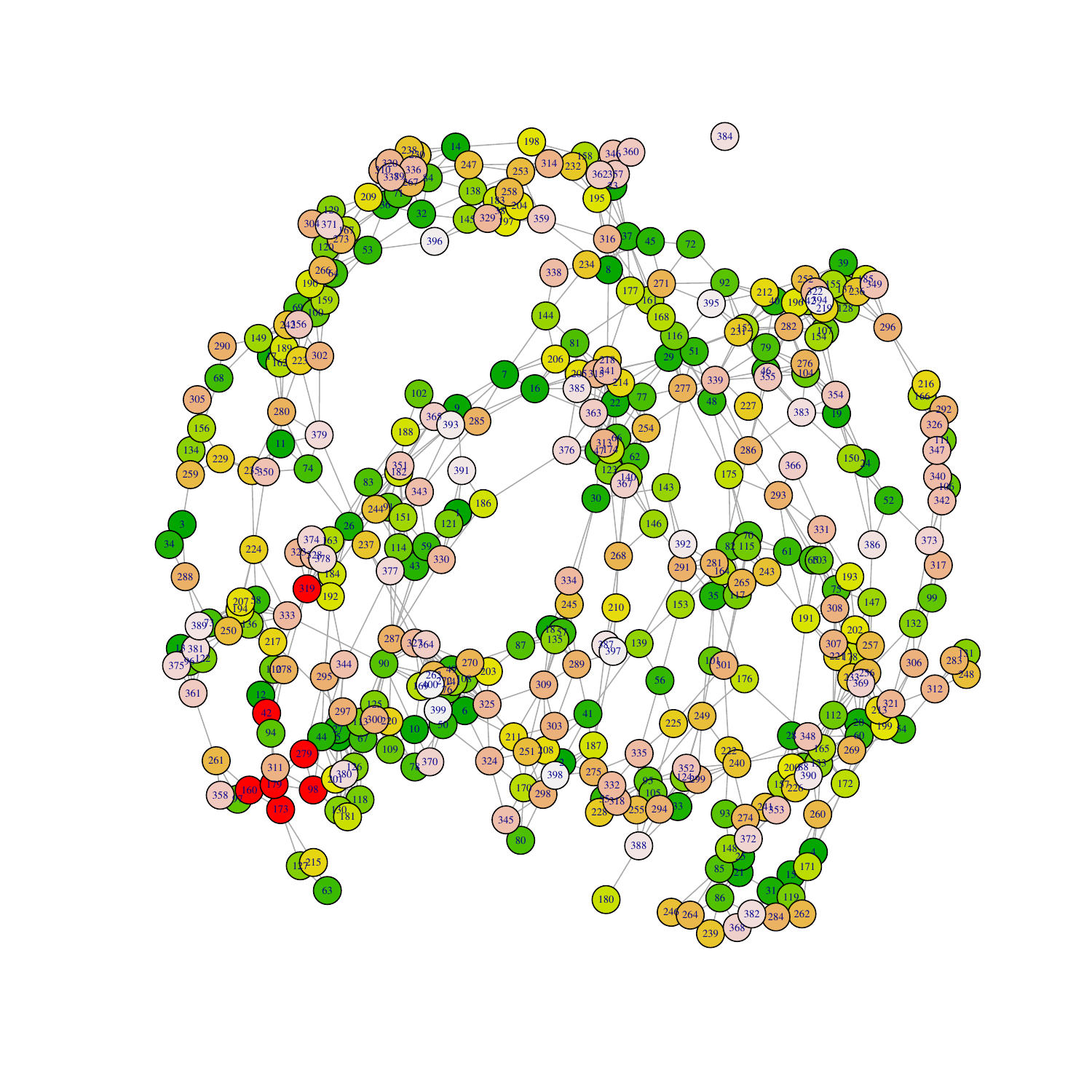}
\includegraphics[width=.45\linewidth, height=.5\linewidth]{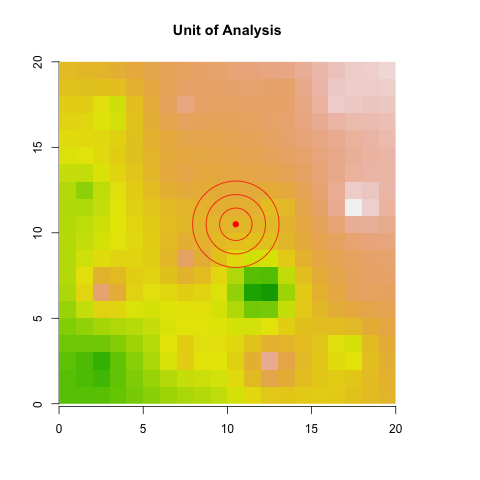} \\
\includegraphics[width=.45\linewidth, height=.5\linewidth]{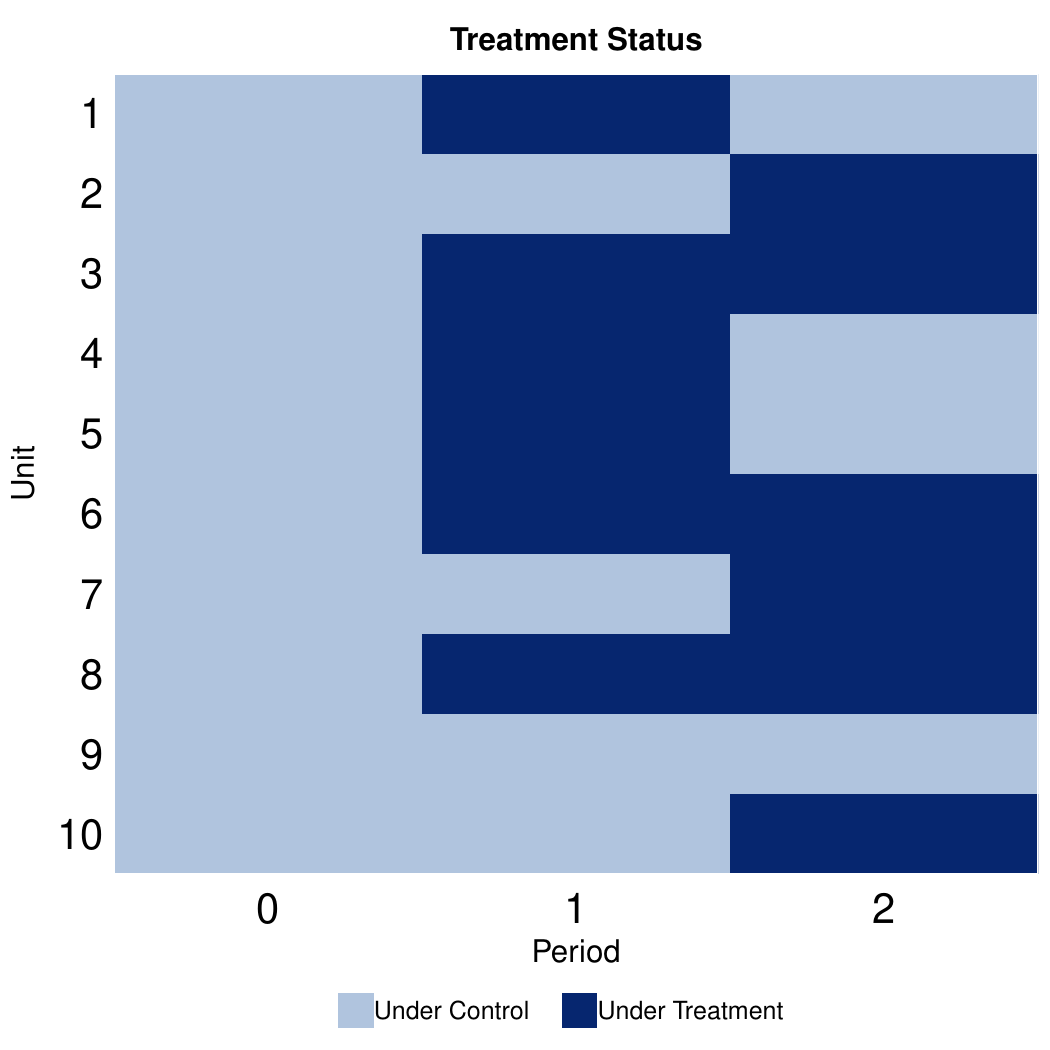} 
\includegraphics[width=.45\linewidth, height=.5\linewidth]{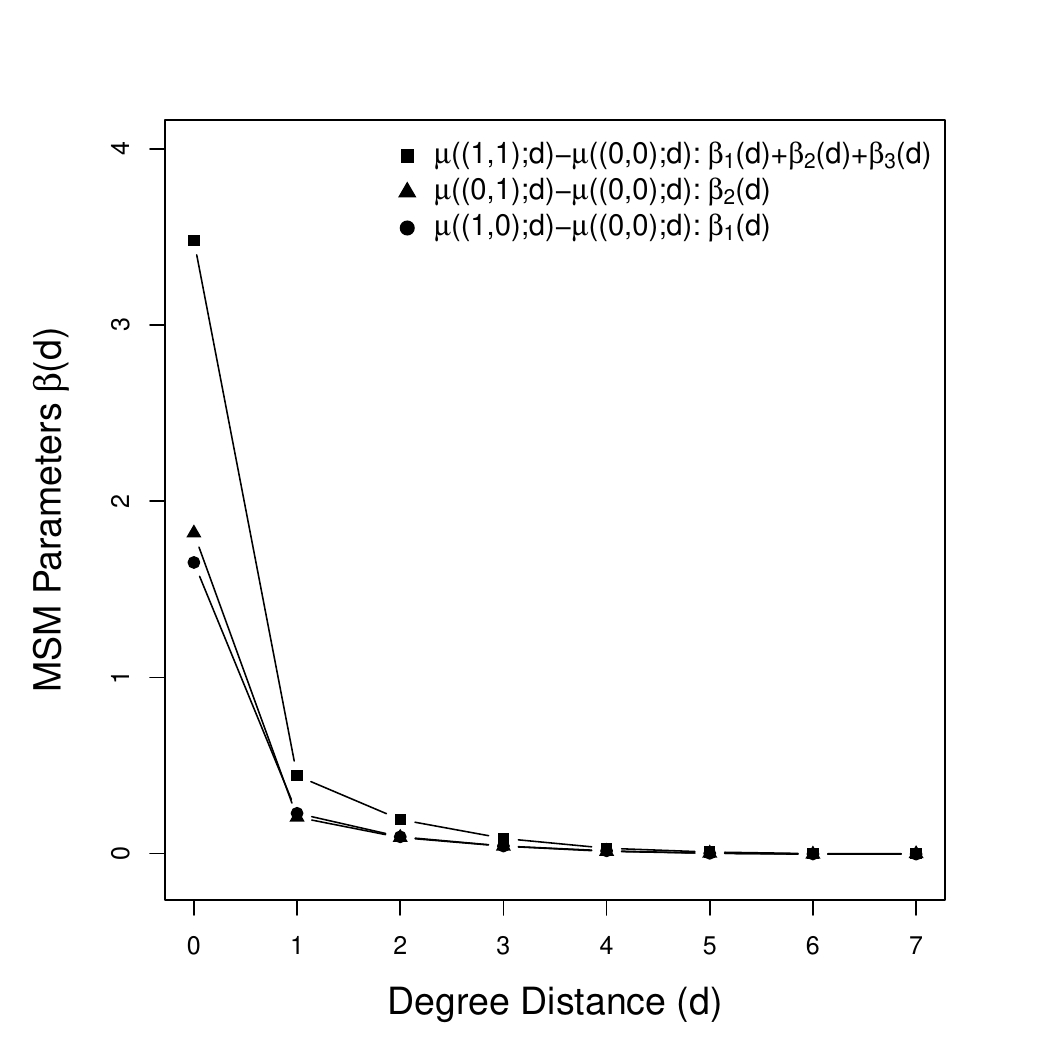}
 \end{center}
\textit{Notes:} The top plots show the locations of the $400$ units from the simulation in either a social network (left) or a $20\times 20$ spatial raster object (right), where the colors indicate the outcome's variation. Red colors on the left mark a node ($42$) and its second-degree neighbors in the network, while red circles on the right show the construction of $\Omega_i(d)$ around the spotted unit. The bottom-left plot depicts the treatment assignment history for the first $10$ units in the sample, while the bottom-right plot presents combinations of MSM parameters that capture various types of causal effects.
\end{figure}

\paragraph*{An illustration} We provide a concrete example for concepts defined above using simulated data with $N = 400$ and $T = 2$. The units can be understood as either nodes in a social network (top-left plot of Figure \ref{fig:space}) or tiles in a spatial raster (top-right plot of Figure \ref{fig:space}). As shown in the bottom-left plot of Figure \ref{fig:space}, there are four possible treatment histories: $\mathbf{A}_i^{(1:2)} \in \{(0, 0), (0, 1), (1, 0), (1, 1)\}$. Consider the outcome of an arbitrary unit $i$ (e.g., unit $42$). Without further restrictions, its value is jointly determined by the treatment histories of all the $400$ units: $Y_{42} = Y_{42}\left(\mathbf{A}_{\{1,2,\dots,400\}}^{(1:2)}\right) = Y_{42}\left(\mathbf{A}_1^{(1:2)}, \mathbf{A}_2^{(1:2)}, \dots, \mathbf{A}_{400}^{(1:2)}\right)$.\footnote{Details of the data generating process are described in the Supplementary Material.} 

\subsection{Estimands}\label{estimands}
Our goal is to understand how the treatment history of a unit influences its own outcome or the outcome of its neighbors in $\mathcal{G}_{N}$ under the prevailing treatment assignment policy. Let $\mathbf{A}_{\mathcal{N}\setminus \{j\}}^{(1:T)}$ represent the remaining part in $\mathbf{A}_{\mathcal{N}}^{(1:T)}$ after removing unit $j$'s treatment history between periods $1$ and $T$. Then, we use $Y_i\left(\mathbf{a}^{(1:T)}, \mathbf{A}_{\mathcal{N}\setminus \{j\}}^{(1:T)}\right) = Y_i\left(\mathbf{A}_j^{(1:T)}=\mathbf{a}^{(1:T)}, \mathbf{A}_{\mathcal{N}\setminus \{j\}}^{(1:T)}\right)$ to denote the potential outcome of unit $i$ when the treatment history of unit $j$ is fixed at $\mathbf{a}^{(1:T)}$ but all other units' treatment histories are generated independently from the policy. 

To construct meaningful estimands, we take the expectation of $Y_i\left(\mathbf{a}^{(1:T)}, \mathbf{A}_{\mathcal{N}\setminus \{j\}}^{(1:T)}\right)$ over $\mathbf{A}_{\mathcal{N}\setminus \{j\}}^{(1:T)}$. This yields the \textbf{marginalized potential outcome} for unit $i$ given $j$'s treatment history:
\vspace{-1em}
\begin{equation}
    \mu_{i;j}\left(\mathbf{a}^{(1:T)}\right) \coloneqq \E\left[Y_{i}\left(\mathbf{a}^{(1:T)}, \mathbf{A}_{\mathcal{N}\setminus \{j\}}^{(1:T)}\right)\right].\vspace{-0.5em}
\end{equation}
This quantity can be defined for every pair of $i$ and $j$, including cases where $i = j$. In what follows, $\E[\cdot]$ always denotes the expectation over treatment histories conditional on other variables that may affect the outcome, such as pre-treatment covariates and measurement error. In Section~\ref{app:inter} of the Supplementary Material, we show that this definition extends to expectations taken over all variables affecting the outcome, with our method still applying.

Next, for each unit $j$, we aggregate $\mu_{i;j}\left(\mathbf{a}^{(1:T)}\right)$ over a pre-specified set of $j$'s neighbors denoted as $\Omega_j(d)$, where $d$ is a specific value of the proximity metric. This leads to a linear mapping $\mu \left(\cdot; \Omega_j(d)\right)$ from the non-random vector $\big\{\mu_{i;j}\left(\mathbf{a}^{(1:T)}\right)\big\}_{i \in \mathcal{N}}$ to a real number: 
\vspace{-1em}
\begin{align}\label{eq:AMRj}
\mu \left(\big\{\mu_{i;j}\left(\mathbf{a}^{(1:T)}\right)\big\}_{i \in \mathcal{N}}; \Omega_j(d)\right) & \coloneqq \frac{\sum_{i=1}^N \mathbf{1}\{i \in \Omega_j(d)\}\mu_{i;j}\left(\mathbf{a}^{(1:T)}\right)}{|\Omega_j(d)|},\vspace{-0.5em}
\end{align}
where $\mathbf{1}\{\cdot\}$ is the indicator function and $|\Omega_j(d)| = \sum_{i=1}^N \mathbf{1}\{i \in \Omega_j(d)\}$ represents the number of units in $\Omega_j(d)$. The form of $\Omega_j(d)$ can be chosen by the researcher based on the type of spillover effects under study. In social networks, a natural choice is the $d$th-degree neighbors of each $j$: $\Omega_j(d) \coloneqq \{i \in \mathcal{N}: d_{ij} = d\}$. In spatial settings, one option is the ``circle average'' introduced by \citet{wang2020design}, where $d$ stands for any distance value and $\Omega_i(d)$ represents a ``doughnut'': $\{i \in \mathcal{N}: d - \kappa < d_{ij} \leq d\}$, with $\kappa$ being a fixed constant. In either case, the quantity defined in (\ref{eq:AMRj}) captures the expected response of unit $j$'s neighbors whose proximity to $i$ is (approximately) $d$ to $i$'s treatment history. With a slight abuse of terminology, we will refer to units belonging to $\Omega_j(d)$ as $j$'s $d$th-degree neighbors henceforth. When $d = 0$, $\Omega_j(d)$ only includes unit $j$ itself, and the quantity reduces to $Y_{j;j}\left(\mathbf{a}^{(1:T)}\right)$, unit $j$'s expected outcome given its own treatment history.

In the top-left plot of Figure \ref{fig:space}, all the second-degree neighbors ($d=2$) of unit $42$ are marked in red. In the top-right plot, a specific unit $j$ is highlighted in red, while red circles indicate the range of $\Omega_j(d)$ (in the form of doughnuts with $\kappa = 1$) for $d \in \{0, 1, 2, 3\}$. Both $d$ and $\Omega_j(d)$ can be adapted to specific contexts. For example, $d$ may take discrete values indicating whether two units are from the same street or district, or represent traffic accessibility or even cultural proximity between units. $\Omega_j(d)$ may include all units within the proximity range of $d$, or be further restricted to units with specific covariate profiles to capture conditional effects.\footnote{These choices do impact statistical inference. We discuss this issue in Section \ref{large-sample-theory}.}  We discuss potential impacts of measurement error of $d$ in Section~\ref{extensions} and summarize choices of $d$ and $\Omega_j(d)$ in Section~\ref{app:exp} of the Supplementary Material. When the context is clear, we represent the value of $\mu(\cdot; \Omega_j(d))$ with $\mu_j(\cdot; d)$.

It is worth noting that the definition of $\Omega_j(d)$ does not require any knowledge of the interference structure---specifically, the functional form of $Y_{i}\left(\mathbf{A}_{\mathcal{N}}^{(1:T)}\right)$. It remains well-defined for any proximity metric chosen by the researcher. Instead of explicitly studying how a unit's outcome is influenced by the others---typically formalized through an exposure mapping---our framework examines the effects of a unit's treatment history on a predefined set of neighbors. We do not assume that any $\Omega_j(d)$ or their collection captures all the influences generated by a unit. Overlap of $\Omega_j(d)$ across the units is permitted, hence each unit may contribute to multiple $\mu_j \left(\Big\{\mu_{i;j}\left(\mathbf{a}^{(1:T)}\right)\Big\}_{i \in \mathcal{N}}; d\right)$ values if it is influenced by more than one neighbor.

Finally, by taking the average of $\mu_j \left(\big\{\mu_{i;j}\left(\mathbf{a}^{(1:T)}\right)\big\}_{i \in \mathcal{N}}; d\right)$ across all units in the sample, we obtain the \textbf{average marginalized response (AMR)}:
\vspace{-1em}
\begin{align}\label{eq:AMR}
\mu\left(\mathbf{a}^{(1:T)};d\right) \coloneqq \frac{1}{N}\sum_{j=1}^N \mu_j \left(\Big\{\mu_{i;j}\left(\mathbf{a}^{(1:T)}\right)\Big\}_{i \in \mathcal{N}}; d\right).\vspace{-0.5em}
\end{align}
Given a pre-specified range $\mathcal{D}$, the AMR can be defined for each $d \in \mathcal{D}$ and enables researchers to assess how spillover effects vary with proximity without making assumptions about their magnitude. For instance, we could set $\mathcal{D} = \{0, 1, 2, 3, 4\}$ if the goal is to test whether spillover effects only extend to one's third-degree neighbors. Moreover, the AMR at any $d$ depends only on the treatment history $\mathbf{a}^{(1:T)}$ and can therefore be represented by a saturated marginal structural model (MSM) when $T$ is not too large:
\vspace{-1em}
\begin{align}\label{eq3}
\mu\left(\mathbf{a}^{(1:T)};d\right) = \mathbf{m}\left(\mathbf{a}^{(1:T)}\right)'\boldsymbol{\beta}(d).\vspace{-1em}
\end{align}
Here, $\mathbf{m}\left(\mathbf{a}^{(1:T)}\right)$ consists of indicators for the treatment status in each period and their interactions, and $\boldsymbol{\beta}(d)$ is a vector of parameters specific to $d$. When $d = 0$, Equation~(\ref{eq3}) resembles a classic MSM, where the parameters describe the direct effects of a unit's treatment histories on its own expected outcome. When $d > 0$, $\boldsymbol{\beta}(d)$ captures spillover effects generated by treatment histories on the expected outcome of the unit's $d$th-degree neighbors.\footnote{As in a conventional MSM, these quantities include the effects mediated through the time-varying confounders, represented by gray arrows in Figure \ref{fig:dag}.} $\{\boldsymbol{\beta}(d)\}_{d \in \mathcal{D}}$ thus provides a comprehensive view of both direct and spillover effects in the sample. 

Similar to other marginalized estimands in the literature \citep{hudgens_halloran08, savje-etal2018-unknown-interference, hu2022average, wang2020design}, these causal quantities reflect the effect of manipulating a single unit's treatment history under the current treatment assignment policy, rather than contrasting hypothetical scenarios where all units receive specific treatments. They are inherently sample-specific (though we omit $\mathcal{N}$ in the subscript for simplicity) and ``descriptive'' of the status quo, not ``prescriptive'' for identifying optimal policies \citep{kennedy2019nonparametric}. As such, $\boldsymbol{\beta}(d)$ does not represent parameters that are invariant to policy changes. Nevertheless, it remains meaningful under the prevailing policy, capturing all causal pathways through which treatment influences outcomes, including direct effects, spillover effects, changes in time-varying confounders, and shifts in treatment distributions over time. In Section~\ref{app:welfare} of the Supplementary Material, we illustrate how the AMRs can inform the direction of policy changes that improve the average expected outcome, also known as welfare in the literature \citep{hu2022average, viviano2020policy}.

\begin{table}[!t]
\caption{Examples of AMR} \label{tab_ame}
  \begin{longtable}[]{@{}lcccc@{}}
  \toprule
  $\mathbf{a}^{1:2}$ & $(0, 0)$ & $(0, 1)$ & $(1, 0)$ & $(1, 1)$ \tabularnewline
  \midrule
  \endhead
  $AMR$ & $\mu\left((0, 0); d\right)$ & $\mu\left((0, 1); d\right)$ & $\mu\left((1, 0); d\right)$ & $\mu\left((1, 1); d\right)$ \tabularnewline
  Associated $\boldsymbol{\beta}(d)$ & $\beta_{0}(d)$ & $\beta_{0}(d) + \beta_{2}(d)$ & $\beta_{0}(d) + \beta_{1}(d)$ & $\beta_{0}(d) + \beta_{1}(d) + \beta_{2}(d) + \beta_{3}(d)$ \tabularnewline
  \bottomrule
  \end{longtable}
\end{table}

\paragraph*{An illustration (continued)} In the example introduced previously, we can define a series of AMRs for each value of $\mathbf{a}^{(1:T)}$ and $d \in \mathcal{D} = \{0, 1, \dots, 7\}$, where $d$ is measured by the length of the shortest path connecting two units in the social network and the Euclidean distance between two units in the spatial raster. $\Omega_j(d)$ is defined as $j$'s $d$th-degree neighbors in the former case and a doughnut surrounding $j$ with $\kappa = 1$ in the latter, as depicted in the top plots of Figure \ref{fig:space}. The top row of Table \ref{tab_ame} presents AMRs defined under each of the four possible treatment histories, which can be summarized by the following MSM at any $d$:
\vspace{-1em}
$$
\mu\left(\mathbf{a}^{(1:T)};d\right) = \beta_{0}(d) + \beta_{1}(d)a^{(1)} + \beta_{2}(d)a^{(2)} + \beta_{3}(d)a^{(1)}a^{(2)},\vspace{-1em}
$$
with $\mathbf{m}\left(\mathbf{a}^{(1:T)}\right) = \left(1, a^{(1)}, a^{(2)}, a^{(1)}a^{(2)}\right)'$. Since this model is saturated, the combination of its parameters can recover each of the AMRs, as presented in the bottom row of Table \ref{tab_ame}. These parameters can also be used to construct the causal contrast between two treatment histories. For instance, the effect of $(1, 1)$ relative to $(0, 0)$ equals $\beta_{1}(d) + \beta_{2}(d) + \beta_{3}(d)$ in the model.

We specify an effect function that emanates from each unit and declines in $d$ exponentially. Effects received by each unit are additive across other units in the sample and amplified by an idiosyncratic constant that represents treatment effect heterogeneity. The effects in each period carry over into the next, declining by $50\%$ in magnitude. In period $t$, the probability for unit $i$ to receive the treatment depends on its treatment status and the value of a time-varying confounder in the previous period. We repeat the simulation $1,000$ times and approximate each $\mu_{i;j}\left(\mathbf{a}^{(1:T)}\right)$ by the average of $Y_{i}\left(\mathbf{a}^{(1:T)}; \mathbf{A}_{\mathcal{N}\setminus \{j\}}^{(1:T)}\right)$ over the simulations. The AMRs are then constructed in line with their definition in Equation~(\ref{eq:AMR}). We plot several linear combinations of the MSM parameters against proximity levels in the bottom-right plot of Figure \ref{fig:space}. As shown, each curve captures how the effects generated by a specific treatment history, relative to the benchmark $(0, 0)$, vary with proximity.\footnote{The two histories, $(1, 0)$ and $(0, 1)$, produce spillover effects through different mechanisms. For $(0, 1)$, the effect arises from the spillover of contemporaneous treatments. For $(1, 0)$, it reflects both the spillover of past treatments and their influence on the distribution of contemporaneous treatments.}

\section{Identification and Estimation}\label{identification-and-estimation}
\subsection{Identification Assumptions}\label{identification-assumptions}
To identify the AMR and parameters in the associated MSM, we impose the assumption of sequential exchangeability:

\begin{assumption}[Sequential exchangeability]\label{assum:si}
\vspace{-1em}
\[
  \Big\{Y_{i}\left(\mathbf{a}^{(1:T)}, \mathbf{A}_{\mathcal{N}\setminus \{j\}}^{(1:T)}\right), \mathbf{L}_{i}^{(t+1)}\left(\mathbf{a}^{(1:T)}, \mathbf{A}_{\mathcal{N}\setminus \{j\}}^{(1:T)}\right)\Big\} \perp A_{j}^{(t)} \mid \left(\mathbf{A}_{j}^{(1:(t-1))}, \mathbf{L}_{j}^{(1:t)}\right).\vspace{-1em}
\]
for any $i$, $j$, $t$, and $\mathbf{a}^{(1:T)}$.
\end{assumption}

Assumption~\ref{assum:si} states that for any unit $j$, its treatment status in period $t$, $A_{j}^{(t)}$, is independent to the distribution of any unit's potential outcomes or the potential values of time-varying confounders in period $t+1$---rather than just its own---conditional on its history of treatment assignment and time-varying confounders.\footnote{Our identification result only requires Assumption~\ref{assum:si} to hold for the potential outcomes. However, the broader formulation allows for the possibility of evaluating the assumption's validity through placebo tests, as discussed in Section~\ref{extensions}.} The assumption reduces to the familiar version in the absence of interference, where $Y_{i}\left(\mathbf{a}^{(1:T)}, \mathbf{A}_{\mathcal{N}\setminus \{j\}}^{(1:T)}\right) = Y_{i}\left(\mathbf{a}^{(1:T)}\right)$, $\mathbf{L}_{i}^{(t+1)}\left(\mathbf{a}^{(1:T)}, \mathbf{A}_{\mathcal{N}\setminus \{j\}}^{(1:T)}\right) = \mathbf{L}_{i}^{(t+1)}\left(\mathbf{a}^{(1:T)}\right)$, and units are independent from each other. Assumption~\ref{assum:si} is violated if the treatment assignment of unit $j$ depends on the history of another unit, thereby excluding any form of diffusion or dependence in treatment across units \citep{vanderweele2013social, savje-etal2018-unknown-interference}. This requirement is naturally satisfied in experiments and observational studies where a unit's treatment status is determined independently and commonly adopted in the literature \citep[e.g.,][]{liu2016inverse}. In such cases, Assumption~\ref{assum:si} holds even in the presence of ``contextual factors,'' where the confounders are correlated across units \citep{vanderweele2013social, egami2024identification},\footnote{A common example is sources of ``spatial confounding'' discussed in \citet{papadogeorgou2019adjusting} and \citet{reich2021review}.} or when contagion exists and the confounders affect another unit's outcome \citep{ogburn2017vaccines}. In Section~\ref{extensions}, we explore ways to extend the method to settings where treatment diffusion exists or treatment assignment is correlated across units.

We assume that positivity always holds. Denoting $\left(\mathbf{A}_{j}^{(1:(t-1))}, \mathbf{L}_{j}^{(1:t)}\right)$ as $\mathbf{V}_{j}^{(t)}$, the assumption can be stated as:
\begin{assumption}[Positivity]\label{assum:po}  
For any $j$ and $t$, there exists some $\eta > 0$ such that 
\vspace{-0.5em}
$$
1 - \eta < P\left(A_{j}^{(t)} = 1 \mid \mathbf{V}_{j}^{(t)}\right) < \eta.\vspace{-1em}
$$ 
\end{assumption}
With Assumptions~\ref{assum:si} and~\ref{assum:po} satisfied, the propensity score for unit $j$'s treatment history $\mathbf{a}^{(1:T)}$ can be expressed as $e\left(\mathbf{A}_j^{(1:T)} = \mathbf{a}^{(1:T)}; \mathbf{V}_{j}^{(1:T)}\right) = \prod_{s=1}^T P\left(A_{j}^{(s)} = a^{(s)} \mid \mathbf{V}_{j}^{(s)}\right)$.
These probabilities are known in experimental settings and need to be estimated from data in observational studies. 

Next, we impose the restriction that any time-varying confounder is influenced only by treatment histories up to the present and not by future treatment statuses. This assumption, commonly referred to as no anticipation, is formally stated as follows:
\begin{assumption}[No anticipation]\label{assum:na}  
For any treatment history $\mathbf{a}^{(1:T)}$ and $\tilde{\mathbf{a}}^{(1:T)}$ and any period $t$,
\vspace{-1em}
$$
\mathbf{L}_{i}^{(t+1)}\left(\mathbf{a}^{(1:T)}\right) = \mathbf{L}_{i}^{(t+1)}\left(\tilde{\mathbf{a}}^{(1:T)}\right) \text{ if } \mathbf{a}^{(1:t)} = \tilde{\mathbf{a}}^{(1:t)}.
$$ 
\end{assumption}
Although Assumption~\ref{assum:na} is not essential for causal identification, it simplifies estimation and facilitates placebo tests (see Section~\ref{extensions} for details). In Section~\ref{extensions}, we discuss possible relaxations of this assumption. We present structural equations that describe the relationships among variables under Assumptions~\ref{assum:si}–\ref{assum:na} and discuss their connection to the framework in \citet{ogburn2017causal} in Section~\ref{app:inter} of the Supplementary Material.

\subsection{Estimator}\label{algorithm}
The key insight of the paper is that assumptions introduced in the previous subsection are sufficient for identifying the AMR at any proximity level $d$ and the parameters in the associated MSM. Define the \textbf{transformed outcome} of unit $j$ at $d$ as
\vspace{-1em}
\begin{align}\label{eq:to}
    \mu_j(d) = \mu\left(\big\{Y_{i}\big\}_{i \in \mathcal{N}};\Omega_j(d)\right) = \frac{\sum_{i=1}^N \mathbf{1}\{i \in \Omega_j(d)\}Y_{i}}{|\Omega_j(d)|},\vspace{-1em}
\end{align}
which represents the output of the same mapping $\mu(\cdot; \Omega_j(d))$, but applied to the vector of observed outcomes rather than marginalized potential outcomes. We then establish the following identification result:
\begin{proposition}\label{thm:id}
Under Assumptions~\ref{assum:si}-\ref{assum:na},
\vspace{-1em}
\begin{align}\label{eq:id}
    \mu\left(\mathbf{a}^{(1:T)};d\right) = \frac{1}{N} \sum_{j=1}^N \E\left[\frac{\mathbf{1}\{\mathbf{A}_j^{(1:T)} = \mathbf{a}^{(1:T)}\}\mu_j(d)}{e\left(\mathbf{A}_j^{(1:T)} = \mathbf{a}^{(1:T)}; \mathbf{V}_{j}^{(1:T)}\right)}\right]. \vspace{-1em}
\end{align}
\end{proposition}
Since each parameter in the MSM is a linear combination of the AMRs, they can be similarly identified from the data. The proofs are in Section~\ref{app:id} of the Supplementary Material.

As in the classic case, these identification results suggest that $\boldsymbol{\beta}(d)$ can be estimated using IPW. Specifically, we rely on the weighted least squares (WLS) algorithm to obtain:
\vspace{-1em}
\begin{align}\label{eq:estr}
    \hat{\boldsymbol{\beta}}(d) = \arg \min_{\boldsymbol{\beta}(d)} \sum_{j = 1}^N w_j \left(\mu_j(d) - \mathbf{m}\left(\mathbf{A}_j^{(1:T)}\right)'\boldsymbol{\beta}(d)\right)^2,\vspace{-1em}
\end{align}
where $w_j = \frac{\prod_{s=1}^T P\left(A_{j}^{(s)} \mid \mathbf{A}_{j}^{(1:(s-1))}\right)}{e\left(\mathbf{A}_{j}^{(1:T)}; \mathbf{V}_{j}^{(1:T)}\right)}$ is the stabilized weight \citep{cole2008constructing}. If $e\left(\mathbf{A}_{j}^{(1:T)}; \mathbf{V}_{j}^{(1:T)}\right)$ is unknown, as in observational studies, then we replace it with an estimate in $w_j$. In the following section, we show that this estimator possesses desirable large-sample properties, such as consistency and asymptotic normality, under mild restrictions on the dependence across units caused by interference. When $d = 0$, $\mu_j(d) = Y_j$, and Equation~(\ref{eq:estr}) becomes the classic estimator in the absence of interference. Therefore, results from the classic estimator can be seen as estimated effects for the AMR defined at $0$.

Beyond its validity under unknown interference, the proposed method offers two other advantages over approaches based on exposure mappings. First, it only requires researchers to estimate the probability of receiving treatment, rather than the probability of attaining a specific level of exposure, which often involves computing a convolution.\footnote{For instance, in a common exposure mapping where exposure is defined as the average of treatment statuses among a unit's first-degree neighbors, computing this probability is equivalent to estimating the distribution of the sum of multiple random variables.} Moreover, the exposure probabilities can be close to 0 or 1 even when Assumption~\ref{assum:po} is satisfied. Second, because identification relies on Assumption~\ref{assum:si}, $e\left(\mathbf{A}_{j}^{(1:T)}; \mathbf{V}_{j}^{(1:T)}\right)$ does not need to account for variables influencing the treatment statuses of a unit's neighbors \citep{forastiere2021identification}. Steps 1-4 in Algorithm~\ref{alg} summarize the estimation procedure for the proposed method. We provide justifications for Steps 4 and 5 in the following section. 

\begin{figure}
\begin{algorithm}[H]\label{alg}
\For {$d \in \mathcal{D}$}{ 
\For {$j \in \mathcal{N}$}{
Construct the transformed outcome $\mu_j(d)$ based on Equation~(\ref{eq:to}).

Compute the probability for $j$'s treatment history to occur and the corresponding weight $w_j = \frac{\prod_{s=1}^T P\left(A_{j}^{(s)} \mid A_{j}^{(1:(s-1))}\right)}{e\left(\mathbf{A}_{j}^{(1:T)}; \mathbf{V}_{j}^{(1:T)}\right)}$, estimating $e\left(\mathbf{A}_{j}^{(1:T)}; \mathbf{V}_{j}^{(1:T)}\right)$ if unknown.
}

Estimate $\boldsymbol{\beta}(d)$ by the WLS algorithm presented in Equation~(\ref{eq:estr}).

Estimate $\hat{\boldsymbol{\beta}}(d)$'s variance, $\Sigma_N(d)$, using the heteroskedesticity and auto-correlation consistent (HAC) variance estimator in Equation~(\ref{eq:hac-gd}). 

Construct Wald-type confidence intervals using critical values from the standard normal distribution.

(Optional) Test identification assumptions by applying the estimator to a placebo outcome.}
 \caption{Implementation of the Proposed Method}
\end{algorithm}
\end{figure}

\section{Large-Sample Theory}\label{large-sample-theory}
To derive the large-sample properties of our estimator, we consider a sequence $\{\mathcal{G}_{N}\}_{N>0}$ in which $\mathcal{N}_N$ units are embedded, with $N \to \infty$ and $T$ fixed. Define the vector of transformed outcomes at proximity level $d$: $\boldsymbol{\mu}(d) = (\mu_1(d), \mu_2(d), \dots, \mu_N(d))'$, the vector of outcomes: $\mathbf{Y} = (Y_1, Y_2, \dots, Y_N)'$, and the $N\times N$ matrix: $\mathbf{D}(d) = \Big\{\frac{\mathbf{1}\{i \in \Omega_j(d)\}}{|\Omega_j(d)|}\Big\}_{N \times N}$.

From Equation~(\ref{eq:to}), we can see that $\boldsymbol{\mu}(d) = \mathbf{D}(d)\mathbf{Y}$. Next, define the matrix of regressors for the MSM as: $\mathbf{M} = \left(\mathbf{m}\left(\mathbf{A}_1^{(1:T)}\right), \mathbf{m}\left(\mathbf{A}_2^{(1:T)}\right), \dots, \mathbf{m}\left(\mathbf{A}_N^{(1:T)}\right)\right)'$, and the diagonal weight matrix: $\mathbf{W} = diag\left(w_1, w_2, \dots, w_N\right)$. Estimates from the WLS algorithm can be expressed as
\vspace{-1em}
$$
\begin{aligned}
    \hat{\boldsymbol{\beta}}(d) = & \left(\mathbf{M}'\mathbf{W}\mathbf{M}\right)^{-1}\left(\mathbf{M}'\mathbf{W}\boldsymbol{\mu}(d)\right) = \left(\mathbf{M}'\mathbf{W}\mathbf{M}\right)^{-1}\left(\mathbf{M}'\mathbf{W}\mathbf{D}(d)\mathbf{Y}\right) \\
    = &\left(\mathbf{M}'\mathbf{W}\mathbf{M}\right)^{-1}\left(\mathbf{Z}'(d)\mathbf{Y}\right) = \left(\frac{1}{N}\sum_{i=1}^N w_i\mathbf{m}\left(\mathbf{A}_i^{(1:T)}\right) \mathbf{m}'\left(\mathbf{A}_i^{(1:T)}\right)\right)^{-1}\left(\frac{1}{N}\sum_{i=1}^N \mathbf{Z}_i(d)Y_i\right),\vspace{-1em}
\end{aligned}
$$
where $\mathbf{Z}'(d) = \left(\mathbf{Z}_1(d), \mathbf{Z}_2(d), \dots, \mathbf{Z}_N(d)\right) = \mathbf{M}'\mathbf{W}\mathbf{D}(d)$. Thus, the estimates take the form of linear combinations of the outcome $Y_i$.

Since $\mathbf{A}_i^{(1:T)}$ is conditionally independent across units, we can show the convergence of the first term, $\frac{1}{N}\sum_{i=1}^N w_i\mathbf{m}\left(\mathbf{A}_i^{(1:T)}\right) \mathbf{m}'\left(\mathbf{A}_i^{(1:T)}\right)$, under standard regularity conditions. For the second term, $\mathbf{Z}_i(d) = \sum_{j=1}^N \frac{\mathbf{1}\{j \in \Omega_i(d)\}\mathbf{m}\left(\mathbf{A}_j^{(1:T)}\right)w_j}{|\Omega_j(d)|}$, which is a weighted average of the MSM regressors across $i$'s neighbors in $\Omega_i(d)$. $\mathbf{Z}_i(d)$ and $\mathbf{Z}_j(d)$ are uncorrelated if $\Omega_i(d)$ and $\Omega_j(d)$ do not overlap.\footnote{Suppose $\Omega_j(d) \coloneqq \{i \in \mathcal{N}: d_{ij} = d\}$, then this is true when $d_{ij} > 2d$.} Therefore, the large-sample performance of $\hat{\boldsymbol{\beta}}(d)$ is primarily driven by the dependence in $Y_i$, which may be caused by either interference or correlated contextual factors.

\subsection{Asymptotic Distribution}\label{asymptotic-distribution}
We restrict the dependence in $Y_i$ by assuming \textit{approximate neighborhood interference} (ANI) introduced in \citet{leung2022causal}. Let unit $i$'s \(p\)-neighborhood in $\mathcal{G}_N$ be defined as $\mathcal{N}_{\mathcal{G}_N}(i, p) = \{j \in \mathcal{N}: d_{ij} \leq p\}$, the set of units whose proximity to $i$ does not exceed $p$. We denote an independent copy of the treatment history outside \(i\)'s \(p\)-neighborhood as \(\mathbf{A}^{(1:T)}_{i,p} = \left(\mathbf{A}^{(1:T)}_{\mathcal{N}_{\mathcal{G}_N}(i, p)}, \tilde{\mathbf{A}}^{(1:T)}_{\mathcal{N}_N \setminus\mathcal{N}_{\mathcal{G}_N}(i, p)}\right)\). We then impose the following assumption:

\begin{assumption}[Approximate Neighborhood Interference]\label{assum:ani}
\vspace{-1em}
\[
  \max_{i \in \mathcal{N}_N} E\Bigg|Y_{i}\left(\mathbf{A}^{(1:T)}_{\mathcal{N}_N}\right) - Y_{i}\left(\mathbf{A}^{(1:T)}_{i,p}\right) \Bigg| \leq \theta_{N,p},
  \] where \(\sup_N \theta_{N,p} \rightarrow 0\) as \(p \rightarrow \infty\).\vspace{-0.5em}
\end{assumption}

This assumption implies that for any bounded Lipschitz function $L(\cdot)$, the covariance between $L\left(\mathbf{Z}_i(d)Y_i\right)$ and $L\left(\mathbf{Z}_j(d)Y_j\right)$ can be bounded by a transformation of $\theta_{N,p}$. This property, known as \(\psi\)-dependence \citep{doukhan1999new, kojevnikov2021limit}, ensures that the dependence between outcomes decays as the proximity between units increases, even though it may never disappear in finite samples. 

To describe the expansion rate of $\mathcal{N}_{\mathcal{G}_N}(i, p)$ as $N$ grows, we introduce the following two quantities.
\vspace{-1em}
\[
M_{N}(q, k) = \frac{1}{N}\sum_{i=1}^N \left|\mathcal{N}_{\mathcal{G}_N}(i, q)\right|^k,\vspace{-1em}
\]
and
\vspace{-1em}
\[
\mathcal{H}_{N}(p, q) = \{(i, j, i', j'): j' \in \mathcal{N}_{\mathcal{G}_N}(i, q), \, j' \in \mathcal{N}_{\mathcal{G}_N}(j, q), \, d_{\{i, i'\}, \{j, j'\}} = p\}.\footnote{Here, the distance between two sets $\mathcal{S}$ and $\mathcal{S'}$ is defined as $\min_{\{i \in \mathcal{S}, j \in \mathcal{S}'\}}d_{ij}$.}
\]
These metrics help characterize how the structure of the network or spatial proximity affects the dependence between units as $N$ increases. In particular, $M_{N}(q, k)$ describes higher-order moments of neighborhood sizes, while the set $\mathcal{H}_{N}(p, q)$ contains neighborhoods of size $q$ that are exactly $p$ away from one another.

For simplicity, we assume that the potential outcomes of $Y_i$ are always bounded:
\begin{assumption}[Bounded potential outcomes]\label{assum:bpo}
There exists a constant $\tilde y < \infty$ such that 
\vspace{-1em}
$$
\bigg|Y_{i}\left(\mathbf{a}^{(1:T)}_{\mathcal{N}_N}\right)\bigg| \leq \tilde y\vspace{-1em}
$$ 
for all units $i$ and any treatment history $\mathbf{a}^{(1:T)}$.
\end{assumption}
We discuss relaxations of this assumption in Section~\ref{extensions}. Additionally, we assume that the propensity score can be consistently estimated at the root-$N$ rate in observational studies.
\begin{assumption}[Consistent estimation of the propensity score]\label{assum:nui}
Either the propensity score is known, or there exists an estimator $\hat e\left(\mathbf{A}_{j}^{(1:T)} = \mathbf{A}^{(1:T)}; \mathbf{V}_{j}^{(1:T)}\right)$ such that
\vspace{-1em}
$$
\sqrt N \left(\hat e\left(\mathbf{A}_{j}^{(1:T)} = \mathbf{A}^{(1:T)}; \mathbf{V}_{j}^{(1:T)}\right) - e\left(\mathbf{A}_{j}^{(1:T)} = \mathbf{A}^{(1:T)}; \mathbf{V}_{j}^{(1:T)}\right)\right) = O_P\left(1\right).
$$ 
\end{assumption}

We denote the normalized variance of $\hat{\boldsymbol{\beta}}(d)$, $N \times \Var\left[\hat{\boldsymbol{\beta}}(d)\right]$, as $\Sigma_N(d)$ and define $\tilde \theta_{N,p}(d) = \theta_{N,\lfloor p /2\rfloor}\mathbf{1}\{p > 2d\} + \mathbf{1}\{p \leq 2d\}$, where $\lfloor \cdot \rfloor$ means rounding down to the nearest integer. $\tilde \theta_{N,p}(d)$ captures the dependence in both $\mathbf{Z}_i(d)$ and $Y_i$. We now present the main result on the large-sample properties of $\hat{\boldsymbol{\beta}}(d)$:
\begin{theorem}\label{thm:IPTW-asym}
Under Assumptions \ref{assum:si}-\ref{assum:nui}, if there exist $\epsilon > 0$ and a sequence $\{q_N\}$ that grows with $N$ such that
\vspace{-1em}
$$
  \frac{N^{\frac{3}{2}}}{\Sigma_{N}^{\frac{1}{2}}(d)}\tilde \theta_{N, q_N}^{1-\epsilon}(d) \to 0, \frac{1}{\sqrt N\Sigma_{N}^{\frac{3}{2}}(d)}M_{N}(q_N, 2) \to 0, \frac{1}{N^2\Sigma_{N}^2(d)}\sum_{p=0}^N |\mathcal{H}_{N}(p, q_N)|\tilde\theta_{N,p}^{1-\epsilon}(d) \to 0,\vspace{-1em}
$$
then $\left(\Sigma_N(d)\right)^{-\frac{1}{2}} \sqrt N \left(\hat{\boldsymbol{\beta}}(d) - \boldsymbol{\beta}(d)\right) \to N\left(0, \mathbf{I}\right)$ as $N \to \infty$.
\end{theorem}

The proof of Theorem~\ref{thm:IPTW-asym} is provided in Section~\ref{app:thm1} of the Supplementary Material. Intuitively, these additional conditions impose constraints on the expansion rate of $\mathcal{G}_N$, the decay in the depencence between two units, and their interactions. Consistency and asymptotic normality can be achieved when either neighborhoods in $\mathcal{G}_N$ grow slowly or the dependence between a unit and its remote neighbors declines sufficiently fast along the sequence $\{q_N\}_{N>0}$. For example, in a geographic space, it is often reasonable to assume that two units $i$ and $j$ are no longer interfering with each other if their proximity is larger than a threshold $\bar p$ that does not grow with $N$. In this case, $\theta_{N,p}(d) = 0$ if $p > \bar p$, and conditions 1 and 3 hold if $\Sigma_{N}(d) = O_P(1)$. Additionally, $\left|\mathcal{N}_{\mathcal{G}_N}(i, q_N)\right| = O\left(q_N^2\right)$, hence $M_{N}(q_N, 2) = O\left(q_N^4\right)$. Condition 2 is satisfied if $q_N = o\left(N^{\frac{1}{2}}\right)$, consistent with results in \citet{ogburn2017causal} and \citet{wang2020design}. In a network where neighborhoods expand at an exponential rate, these conditions are satisfied if $\Sigma_{N}(d) = O_P(1)$, $q_N = O_P\left(\log N\right)$, and $\theta_{N,p}(d)$ decays exponentially \citep{leung2022causal}.

\subsection{Variance Estimation}\label{variance-estimation}
Existing research has relied on HAC variance estimators for statistical inference under interference in both social network \citep{leung2022causal} and spatial settings \citep{conley99_spatial, wang2020design}. Such estimators require researchers to specify a bandwidth $\bar d_N$ that grows with $N$ and account for the covariance between the outcomes of units $i$ and $j$ only if $d_{ij} \leq \bar d_N$. We can re-weigh neighbors of unit $i$ within the bandwidth using a kernel function. In theory, $\bar d_N$ should be chosen to match the sequence $\{q_N\}_{N>0}$. In practice, it is often determined by some rules of thumb. For instance, we can set set $\bar d_N = \lfloor\max\{2\bar p, 2d\}\rfloor$ when interference disappears beyond the proximity level $\bar p$. In social networks, \citet{leung2022causal} suggests the following criterion:
\vspace{-1em}
$$
\bar d_N = \lfloor\max\{d_N^*, 2d\}\rfloor, \text{ with } d_N^* = \begin{cases} \frac{1}{2} \mathcal{L}\left(\mathcal{G}_N\right) \text{ if } \mathcal{L}\left(\mathcal{G}_N\right) < 2\frac{\log N}{\log \delta\left(\mathcal{G}_N\right)} \\ \mathcal{L}^{\frac{1}{3}}\left(\mathcal{G}_N\right) \text{ otherwise} \end{cases},\vspace{-1em}
$$
where $\delta\left(\mathcal{G}_N\right)$ and $\mathcal{L}\left(\mathcal{G}_N\right)$ denote the average degree and average path length in $\mathcal{G}_N$, respectively. Researchers should examine the robustness of the variance estimates by using various $\bar d_N$.

Directly applying these HAC variance estimators to our setting presents two challenges. First, the bandwidth is typically chosen based on the dependence in the original outcome $Y_i$, whereas our WLS estimator is constructed using the transformed outcome $\mu_j(d)$. Determining an appropriate bandwidth for $\mu_j(d)$ from $\bar d_N$ is not straightforward. Fortunately, as illustrated at the beginning of this section, the estimator $\hat{\boldsymbol{\beta}}(d)$ can be expressed as a linear combination of $Y_i$, which allows us to perform inference at the level of the original outcome. Denoting the residual from the WLS estimator for unit $i$ as $\hat \varepsilon_i(d)$, we can show that
\vspace{-1em}
$$
\begin{aligned}
    & \sum_{i=1}^N w_i\mathbf{m}\left(\mathbf{A}_i^{(1:T)}\right)\hat \varepsilon_i(d) = \sum_{i=1}^N \hat{\boldsymbol{\varepsilon}}_{Yi}(d), \\
    & \text{ where } \hat{\boldsymbol{\varepsilon}}_{Yi}(d) = \mathbf{Z}'_i(d)Y_i - \sum_{k=1}^N\frac{\mathbf{1}\{i \in \Omega_k(d)\}w_k\mathbf{m}\left(\mathbf{A}_k^{(1:T)}\right)\mathbf{m}'\left(\mathbf{A}_k^{(1:T)}\right)\hat{\boldsymbol{\beta}}(d)}{|\Omega_k(d)|}.\vspace{-1em}
\end{aligned}
$$
Let $\hat{\boldsymbol{\varepsilon}}_{Y}(d) = (\hat{\boldsymbol{\varepsilon}}_{Y1}(d), \hat{\boldsymbol{\varepsilon}}_{Y2}(d), \dots, \hat{\boldsymbol{\varepsilon}}_{YN}(d))'$, and consider the following matrix of weights based on the uniform kernel:\footnote{Both \citet{leung2022causal} and \citet{gao2023causal} find that the uniform kernel performs better than alternatives in practice.} $\mathbf{K}_{\bar d_N} = \big\{\mathbf{1}\{d_{ij} \leq \bar d_N\}\big\}_{N \times N}$. Then, the HAC variance estimator in our setting takes the form:
\vspace{-1em}
\begin{align}\label{eq:hac}
    \hat \Sigma(d) = N \left(\mathbf{M}'\mathbf{W}\mathbf{M}\right)^{-1}\left(\hat{\boldsymbol{\varepsilon}}'_{Y}(d)\mathbf{K}_{\bar d_N}\hat{\boldsymbol{\varepsilon}}_{Y}(d)\right)\left(\mathbf{M}'\mathbf{W}\mathbf{M}\right)^{-1}.
\end{align}

The second challenge is that the variance estimator may not be conservative for $\Sigma_N(d)$ due to the covariance in treatment effects when interference is present, as demonstrated in both \citet{leung2022causal} and \citet{wang2020design}. We solve this problem by adopting the proposal in \citet{gao2023causal}. Since $\mathbf{K}_{\bar d_N}$ is symmetric, it admits an eigenvalue decomposition of the form $\mathbf{Q}_N\Lambda_N\mathbf{Q}_N$, where $\mathbf{Q}_N$ is the matrix of eigenvectors and $\Lambda_N$ is a diagonal matrix of eigenvalues. \citet{gao2023causal} propose replacing $\mathbf{K}_{\bar d_N}$ in Equation~(\ref{eq:hac}) with $\mathbf{K}_{\bar d_N}^{+} = \mathbf{Q}_N\max\{\Lambda_N, 0\}\mathbf{Q}_N$, with the maximum taken element-wise. The modified HAC variance estimator can be expressed as
\vspace{-1em}
\begin{align}\label{eq:hac-gd}
    \hat \Sigma^{+}(d) = N \left(\mathbf{M}'\mathbf{W}\mathbf{M}\right)^{-1}\left(\hat{\boldsymbol{\varepsilon}}'_{Y}(d)\mathbf{K}_{\bar d_N}^{+}\hat{\boldsymbol{\varepsilon}}_{Y}(d)\right)\left(\mathbf{M}'\mathbf{W}\mathbf{M}\right)^{-1}.\vspace{-1em}
\end{align}

To establish the large-sample properties of the variance estimator, we define the following two quantities:
\vspace{-1em}
\[
\mathcal{N}_{\mathcal{G}_{N}}^{\partial}(i, p) = \{j \in \mathcal{N}_N: d_{ij} = p\},\mathcal{J}_{N}\left(p, \bar d_N\right) = \sum_{i=1}^N \sum_{j=1}^N \mathbf{1}\{d_{ij} = p\} \sum_{k=1}^N \left|K_{\bar d_N, ik}\right| \sum_{l=1}^N \left|K_{\bar d_N, jl}\right|.\vspace{-1em}
\]
Here, $\mathcal{N}_{\mathcal{G}_{N}}^{\partial}(i, p)$ represents the set of unit $i$'s neighbors whose proximity to $i$ is exactly $p$.\footnote{This set coincides with $\Omega_i(p)$ when the latter is defined as unit $i$'s $p$th-degree neighbors.} $\mathcal{J}_{N}\left(p, \bar d_N\right)$ can also be expressed as $\{(i, j, i', j'): j' \in \mathcal{N}_{\mathcal{G}_N}(i, q), \, j' \in \mathcal{N}_{\mathcal{G}_N}(j, q), \, d_{ij} = p\}$, which generalizes $\mathcal{H}_{N}(p, q)$ defined above.

We further define $\mathbf{K}_{\bar d_N}^{-} = \mathbf{Q}_N\min|\{\Lambda_N, 0\}|\mathbf{Q}_N$. It follows that $\mathbf{K}_{\bar d_N}^{+} = \mathbf{K}_{\bar d_N} + \mathbf{K}_{\bar d_N}^{-}$ and $\quad M_{N}\left(\bar d_N, k\right) = \frac{1}{N}\sum_{i=1}^N \left|\sum_{j=1}^N K_{\bar d_N, ij}\right|^k$, where $K_{\bar d_N, ij}$ denotes the $(i,j)$th entry of $\mathbf{K}_{\bar d_N}$. Analogously, we define $M_{N}^{-}\left(\bar d_N, k\right)$ and $\mathcal{J}_{N}^{-}\left(p, \bar d_N\right)$ as counterparts to $M_{N}\left(\bar d_N, k\right)$ and $\mathcal{H}_{N}\left(p, \bar d_N\right)$, respectively, using the entries of $\mathbf{K}_{\bar d_N}^{-}$. We then have the following result:
\begin{theorem}\label{thm:hac}
Under Assumptions \ref{assum:si}-\ref{assum:nui}, if 
\vspace{-1em}
$$
\begin{aligned}
    & \sum_{p = 0}^N \mathcal{N}_{\mathcal{G}_{N}}^{\partial}(i, p)\tilde \theta_{N, \bar d_N}^{1-\varepsilon}(d) = O(1), M_{N}\left(\bar d_N, 1\right) = O(N^{\frac{1}{2}}), M_{N}^{-}\left(\bar d_N, 1\right) = O(N^{\frac{1}{2}}), \\
    & M_{N}\left(\bar d_N, 2\right) = O(N), M_{N}^{-}\left(\bar d_N, 2\right) = O(N), \sum_{p = 0}^N \mathcal{J}_{N}(p, \bar d_N)\tilde \theta_{N, \bar d_N}(d) = O(N^2), \\
    & \text{and } \sum_{p = 0}^N \mathcal{J}^{-}_{N}(p, \bar d_N)\tilde \theta_{N, \bar d_N}(d) = O(N^2),\vspace{-1em}
\end{aligned}
$$
then $\lim_{N \to \infty} \hat \Sigma^{+}(d) \succeq \Sigma(d)$.
\end{theorem}
The proof of Theorem~\ref{thm:hac} is provided in Section~\ref{app:thm2} of the Supplementary Material. 

\section{Extensions}\label{extensions}
We briefly discuss several extensions of the proposed method in this section, with technical details provided in Section~\ref{app:extension} of the Supplementary Material.

\paragraph*{Placebo Tests} When certain variables in $\mathbf{L}_{i}^{(t)}$ are not expected to affect treatment, the validity of the identification assumptions can be assessed through placebo tests. For example, consider units with either of the two treatment histories, $(0, 0)$ and $(0, 1)$, in our simulated data, and suppose that $\mathbf{V}_{j}^{(t)}$ only includes variables from period $t-1$. Suppose that $\mathbf{L}_{i}^{(1)} \perp A_{j}^{(2)} \mid \mathbf{V}_{j}^{(2)} = \left(A_{j}^{(1)}, \mathbf{L}_{j}^{(2)} \right)$. Then, we should expect to find a null effect of $A_{j}^{(2)}$ on the transformed outcome constructed from $\mathbf{L}_{i}^{(1)}$ using the WLS estimator. We refer to such variables in $\mathbf{L}_{i}^{(t)}$ as placebo outcomes in Algorithm~\ref{alg}. 

\paragraph*{Relaxed Assumptions} We consider two relaxations of Assumption~\ref{assum:si}. The first allows for treatment diffusion: a unit's treatment status in period $t$ may be influenced by the histories of units whose proximity to it does not exceed $p$. When $p$ is known and fixed, researchers need only control for the histories of these nearby units when estimating the propensity scores, and our theoretical results remain unaffected. The second relaxation concerns dependence in $\mathbf{A}_{\mathcal{N}}^{(1:T)}$ across units. Assumption~\ref{assum:si} implies that $A_{i}^{(t)} \perp A_{j}^{(t)} \mid \mathbf{V}_{\mathcal{N}}^{(t)}$ for any $i$, $j$, and $t$. However, this may not hold under assignment mechanisms like complete randomization. In the Supplementary Material, we formally state the identification assumption under treatment diffusion and, following \citet{savje-etal2018-unknown-interference}, show that our theoretical results continue to hold as long as the cross-unit dependence is not too strong.

When Assumption~\ref{assum:na} fails, variables from future periods, including treatment statuses and time-varying confounders, must also be included in $\mathbf{V}_{j}^{(t)}$ for each unit $j$ in period $t$. As in the case with treatment diffusion, it suffices to control for these variables in the propensity score, and the theoretical results remain unchanged. Finally, we can replace Assumption~\ref{assum:bpo} with bounded moments for $Y_{i}\left(\mathbf{a}^{(1:T)}\right)$. Then, the asymptotic properties of our estimator still follow from the central limit theorem in \citet{kojevnikov2021limit}, although the conditions for Theorem~\ref{thm:IPTW-asym} to hold become less intuitive.

\paragraph*{A Longer Period} Our discussion so far has assumed that $T$ is fixed as $N$ increases. When $T$ also grows, the number of possible treatment histories increases exponentially, making it infeasible to summarize all AMRs with a saturated MSM. In such settings, researchers may impose structure on the MSM by using selected summary statistics of the treatment history, or focus on a shorter history, such as the most recent $s$ periods. In the Supplementary Material, we show that if the MSM is saturated with respect to the chosen statistics, the resulting estimates converge to the marginalized causal effects defined by them, even if they do not fully capture the effects of the entire treatment history. We also provide conditions under which the large-sample results continue to hold when $T$ is large.

\paragraph*{Measurement Error in the Proximity Metric} In practice, the proximity metric $d_{ij}$ may be measured with error. For example, it is common for researchers to approximate the full network using self-reported ties, which often fail to capture all relevant interactions in respondents' lives. However, such measurement error affects the interpretation of the estimand---that is, the particular form of spillover effect being studied---rather than its identification or estimation. In the extreme case where $d_{ij}$ is completely unavailable, one can still investigate the direct effect of a unit's treatment history on its own outcome, accounting for interference from others in the sample. Inference becomes more challenging in such settings, as the HAC variance estimator requires researchers to specify a cutoff value $\bar d_N$. A practical approach is to assess the robustness of variance estimates across a range of $\bar d_N$ values or to use the conservative variance estimator proposed by \citet{aronow2017estimating}.

\section{Application}\label{application}
In this section, we illustrate the application of the method using one empirical example in a spatial setting and the other in a social network setting. Simulation results on the method's performance are reported in Section~\ref{simulation} of the Supplementary Material, which show that the WLS estimator generates consistent estimates and confidence intervals with desirable coverage rates.

\subsection{Public Project and Political Support}\label{application1}
We first apply our method to the study in \citet{stokes2016electoral} on the political consequences of building wind turbines in Ontario, Canada, where the Liberal Party government passed laws since 2003 enabling individuals and corporations to build wind turbines, despite objections from local communities. The study collected election data from 6,186 precincts in Ontario over three elections: 2003, 2007, and 2011, and combined the data with the location of each proposed turbine. Precinct $i$ is considered as treated in year $t$ if a turbine project within its boundary has been proposed before the election. The data exhibit the structure of staggered adoption, where once a unit is treated, it remains treated for the remaining periods. The three treatment histories from 2007 to 2011 ($t=1$ and $t=2$) are $(0,0)$, $(0,1)$, $(1,1)$. There were 6,002 never treated precincts, 53 treated since 2007, and 131 treated since 2011.\footnote{The original analysis combines the difference-in-differences estimator with an exposure mapping, hence its validity hinges on the correct model specification.} The outcomes of interest are turnout rate and the Liberal Party's vote share in 2011.

The propensity scores are estimated by a logistic regression model, which controls for the longitude and latitude of each precinct, as well as its turnout rate and vote share for the Liberal Party in the previous election. We construct doughnuts with a radius of 2 km around the geographic center of each treated and untreated precinct, and fit the MSM in the distance range of 0 km to 16 km. The the spatial HAC variance estimator's cutoff value, $\bar d_N$, is set at $\max\{20, 2*d\}$ km for both outcomes.

\begin{figure}[htp]
 \begin{center}
 \caption{Results from Applications}
    \label{fig:app-main}
 \begin{subfigure}[t]{\textwidth}
    \caption{Political Consequences of Proposed Wind Turbines}
    \includegraphics[width=.48\linewidth, height=.48\linewidth]{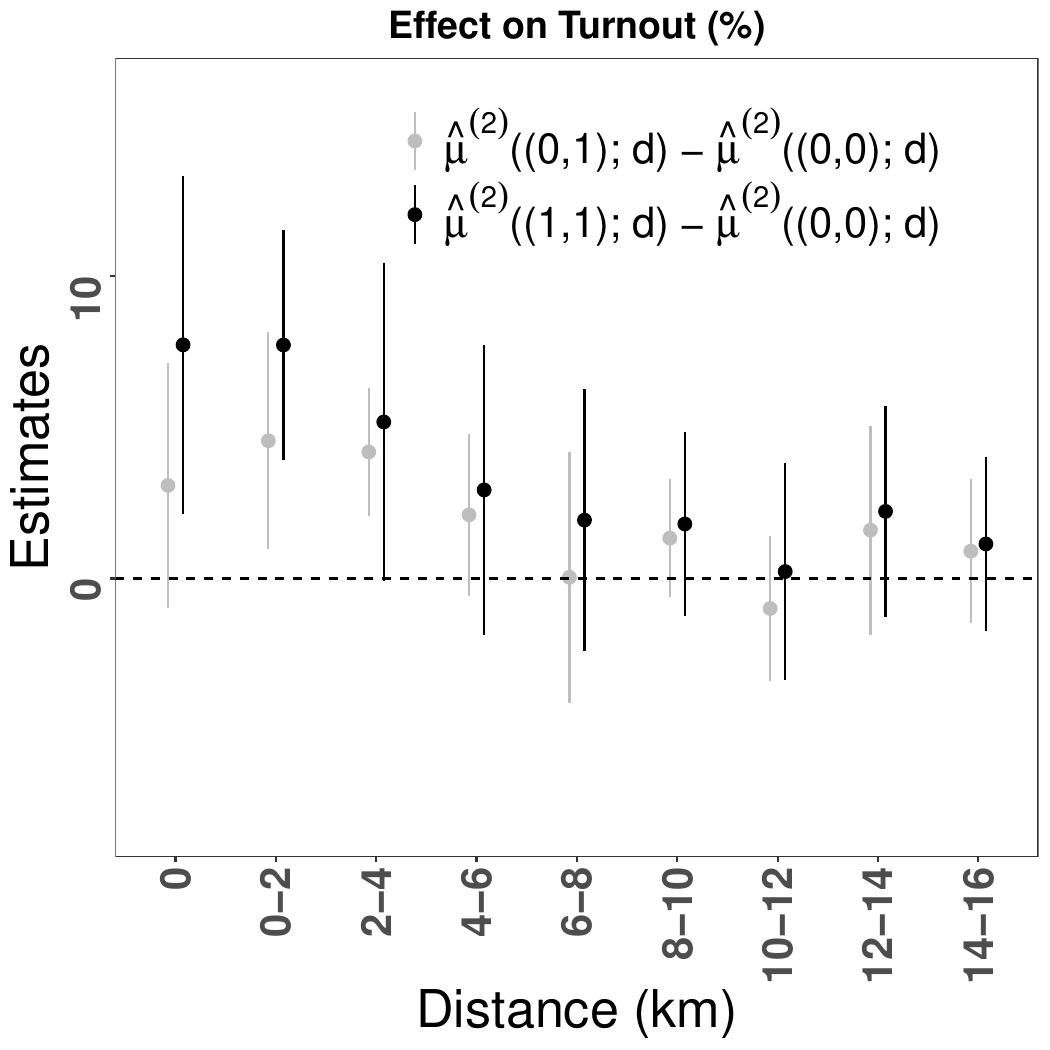}
    \includegraphics[width=.48\linewidth, height=.48\linewidth]{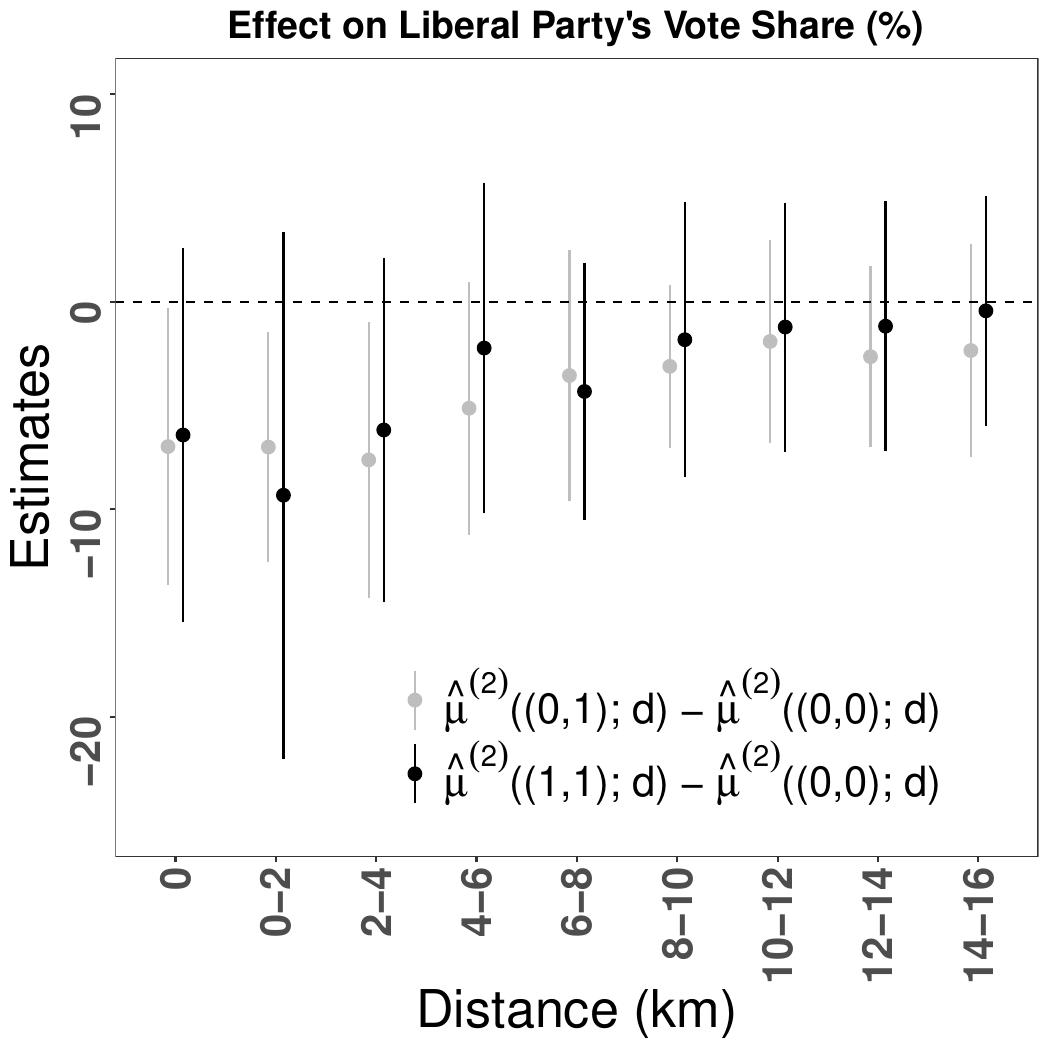}
 \end{subfigure}
 \vspace{1em} 
 \begin{subfigure}[t]{\textwidth}
    \caption{Impacts of Smoking Cessation on Health Outcomes}
    \includegraphics[width=.48\linewidth, height=.48\linewidth]{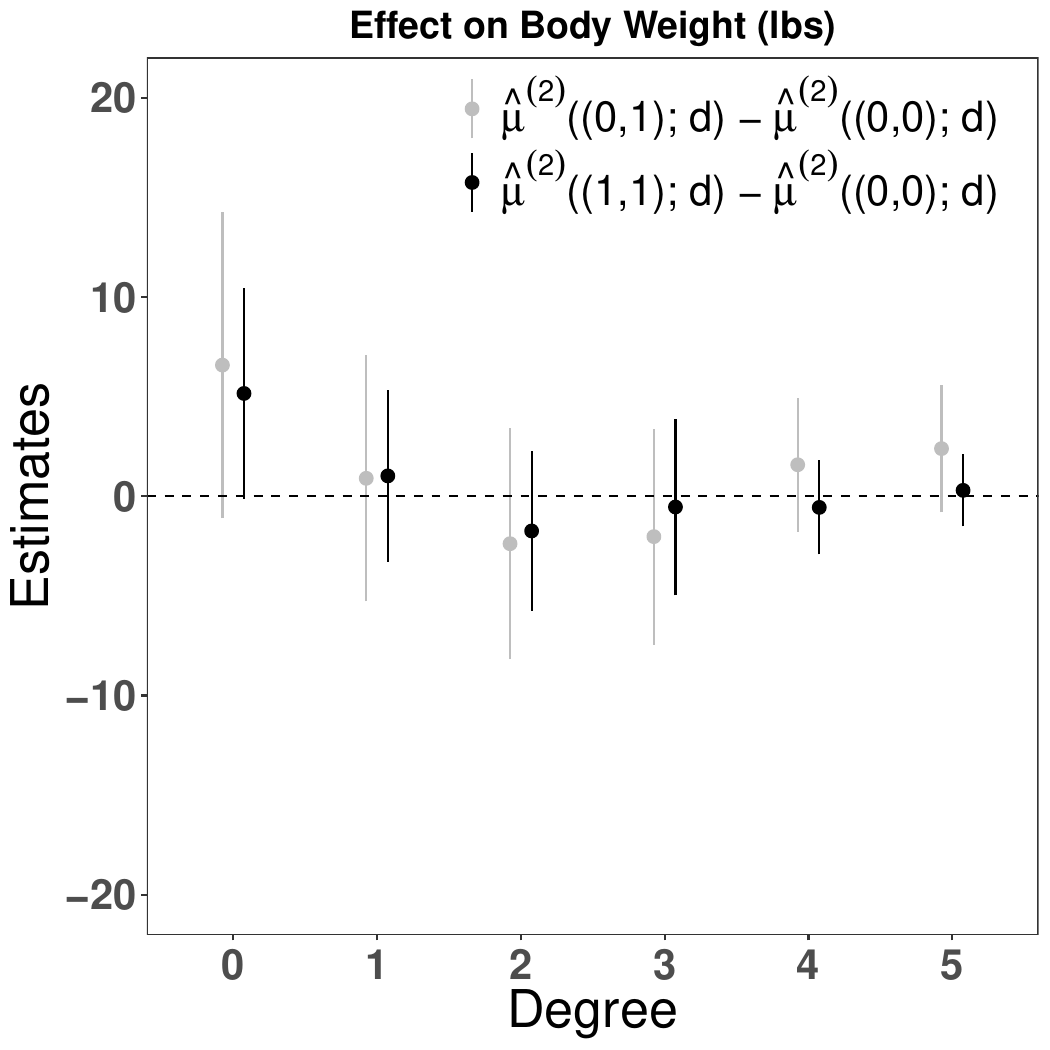}
    \includegraphics[width=.48\linewidth, height=.48\linewidth]{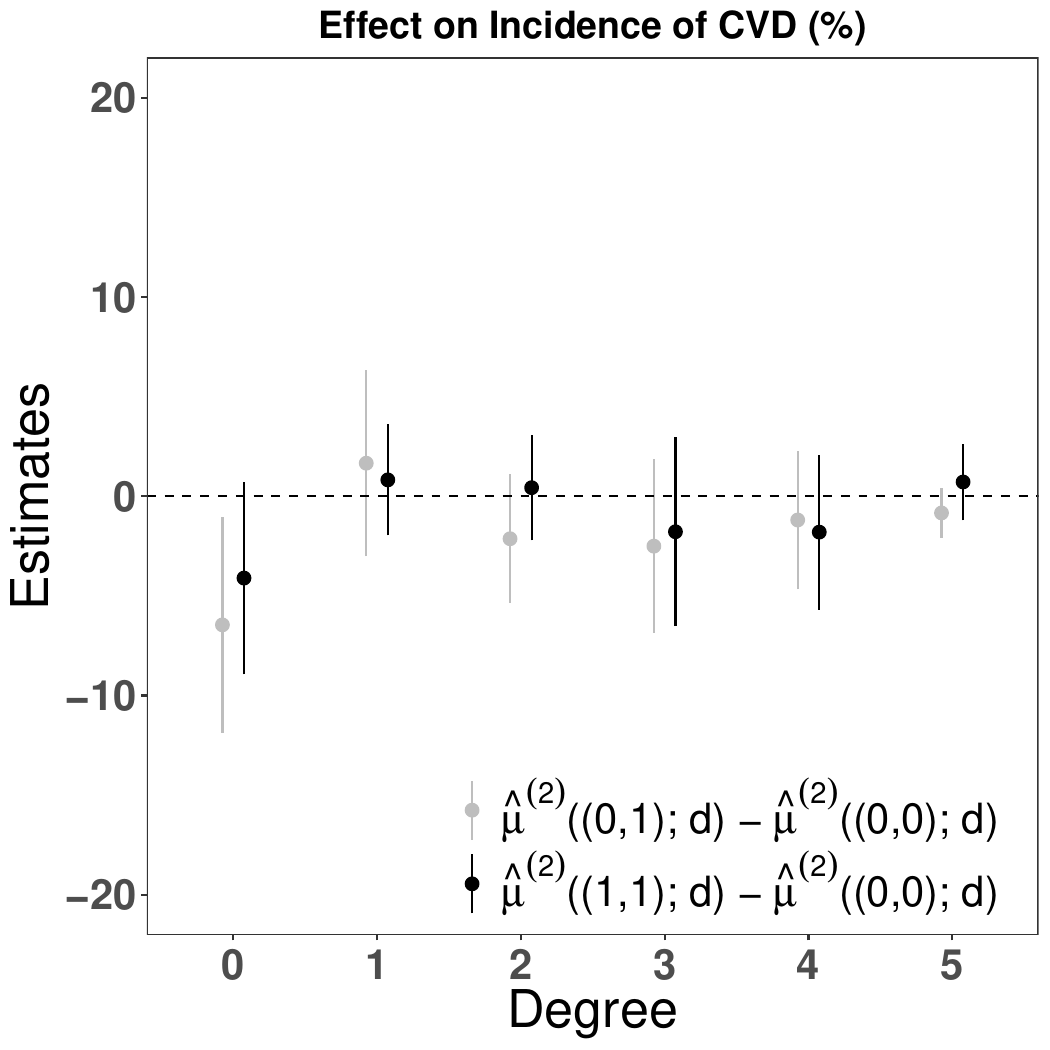}
 \end{subfigure}
 \end{center}
\textit{Notes:} Panel (a) presents results replicating \citet{stokes2016electoral}, while Panel (b) displays findings from the FHS. Black and gray dots represent point estimates for different treatment histories and proximity levels. The vertical segments show the corresponding 95\% confidence intervals, computed using the spatial/network HAC variance estimator and standard normal critical values.
\end{figure}

The results are presented in the top panel of Figure~\ref{fig:app-main}. On the left side are the estimated causal effects for the turnout rate, and on the right side are those for the Liberal Party's vote share. The black dots in the figures represent estimates for $\mu((1,1); d) - \mu((0,0); d)$, and the gray dots represent estimates for $\mu((0,1); d) - \mu((0,0); d)$. We use superscripts on $\mu(\cdot)$ to indicate the time at which the outcome is measured. The segments denote the $95\%$ confidence intervals of these estimates. The estimates provide evidence for the presence of spillover effects generated by both histories. Being treated in both $2007$ and $2011$ would increase the turnout rate by $7.7\%$ and decrease the Liberal Party's vote share by $6.4\%$ in a precinct. The effects are also observed among its neighbors that are $2$-$4$ km away, where the turnout rate would increase by $5.0\%$, and the Liberal Party's vote share would decrease by 6.0\%. In precincts treated only in $2011$, treatment assignment history caused a rise of $3.1\%$ in the turnout rate and a decline of $7.0\%$ in the Liberal Party's vote share. The spillover effects on its neighbors that are $2$-$4$ km away are of similar magnitude. The effects on the outcome variables become close to zero (and statistically insignificant) among precincts that are more than $6$ km away from the treatment, which justifies the choice of the cutoff values. 

\subsection{Impacts of Smoking Cessation on Health Outcomes}\label{application2}
Our second application examines how quitting smoking affects both an individual's health and that of their social network neighbors. Prior research has established strong links between smoking and adverse health outcomes \citep{jain2016smoking}. It is natural to conjecture that reducing exposure to second-hand smoke could yield benefits for those nearby. We use data from the Framingham Heart Study (FHS), focusing on the offspring cohort and variables from the first three exams. The average time gap is 8 years between exams 1 and 2, and 5 years between exams 2 and 3. We treat exam 1 as the initial period ($t = 0$) and define treatment history using exams 2 and 3 ($t \in \{1, 2\}$). $A_{i}^{(t)} = 1$ if the individual reports not smoking at time $t$. Outcomes include body weight, blood pressure, cholesterol level, and incidence of cardiovascular disease (CVD). The first three are measured at exam 3 ($t = 2$); CVD incidence is defined by the occurrence of symptoms (e.g., stroke) within 10 years after exam 3.\footnote{We examine whether there is any CVD incidence between exam 4 and exam 6. The attrition rate during this period is 16.4\%, therefore the influence of righ-censoring is limited.} FHS records social ties across multiple dimensions (e.g., spouse, family, friends, neighbors). We define two individuals as connected if any type of social tie exists between them and include only ties formed prior to exam 2 to avoid endogeneity in network formation.

We restrict our analysis to smokers in the initial period ($A_{i}^{(0)} = 1$) and exclude those with missing data or without any social ties, resulting in a final sample of 930 participants. Among them, 347 continued smoking through exams 2 and 3 ($\mathbf{A}_i^{(1:2)} = (0, 0)$); 413 quit smoking by exam 2 and remained non-smokers in exam 3 ($\mathbf{A}_i^{(1:2)} = (1, 1)$); 136 quit smoking only by exam 3 ($\mathbf{A}_i^{(1:2)} = (0, 1)$); and 34 quit in exam 2 but resumed smoking by exam 3 ($\mathbf{A}_i^{(1:2)} = (1, 0)$). Following the design in \citet{jain2016smoking}, we estimate propensity scores using a logistic regression model that controls for time-invariant covariates (gender, education, height, initial age), as well as treatment status and time-varying covariates (body weight, blood pressure, cholesterol level, and CVD incidence) from the previous period. We set $\mathcal{D} = \{0, 1, 2, 3, 4, 5\}$ and focus on the effects of treatment histories $(1, 1)$ and $(0, 1)$ relative to $(0, 0)$. These effects capture the impact of quitting smoking on outcomes of interest at exams 2 and 3 , respectively.

The bottom panel of Figure~\ref{fig:app-main} presents our estimates and confidence intervals for both body weight and CVD incidence. The estimand is $\mu\left((0, 1);d\right) - \mu\left((0, 0);d\right)$ on the left and $\mu\left((1, 1);d\right) - \mu\left((0, 0);d\right)$ on the right. The results suggest that smoking cessation leads to a significant increase in body weight, with a larger effect for those who quit smoking only by exam 3 (12 pounds) compared to those who quit by exam 2 (5 pounds). These findings are consistent with \citet{jain2016smoking}. Smoking cessation also leads to a significant reduction in CVD risk, with the two treatment histories yielding similar effects (6\% vs. 4\%). However, we find little evidence of spillover effects on either body weight or CVD incidence, suggesting that reductions in second-hand smoke may have limited impact on others' health.

Additional results from both applications are provided in Section~\ref{app:Stokes} of the Supplementary Material, including placebo tests supporting the identification assumptions, estimates based on a propensity score model that accounts for treatment diffusion, and estimated effects on blood pressure and cholesterol levels in the FHS.

\section{Conclusion}\label{conclusion}
This paper tackles the challenge of causal inference in longitudinal data with an unknown interference structure. We introduce a novel estimand, the AMR, to capture both direct and spillover effects generated by different treatment histories. This estimand is well-defined for any user-specified proximity metric, requiring only minimal knowledge of the underlying interference structure. It can be summarized by an MSM and identified under a slightly generalized form of sequential exchangeability. We show how to estimate any AMR and MSM parameters using a WLS estimator and establish its large-sample properties, including consistency and asymptotic normality, under mild restrictions on the dependence across units. The estimator reduces to the classic approach in longitudinal analysis when $d = 0$, suggesting that its estimates have a causal interpretation under interference. We also develop HAC variance estimators for spillover effect estimates that are ensured to be asymptotically conservative. Our approach bridges methods in longitudinal data analysis and causal inference under interference, providing researchers with a powerful tool to study a broad range of causal effects in spatial and network settings, especially when interactions among units are poorly understood. We illustrate the method's utility through simulations and two empirical applications, and discuss possible extensions.

Compared to alternative approaches based on outcome models or exposure mappings, our method avoids risks of model misspecification and the burden of computing exposure probabilities in settings with complex and unknown interference. However, it relies on researchers' substantive understanding of the prevailing treatment assignment policy so that the propensity score is either known or can be consistently estimated. Future research should investigate more flexible strategies for propensity score estimation, such as sieve estimators or highly adaptive LASSO \citep{ertefaie2020nonparametric}. Another promising direction is to combine our approach with outcome modeling to improve robustness and efficiency. Finally, extending the framework to incorporate methods that account for right-censoring and binary outcomes would further enhance its applicability across biomedical and social science research.

\bibliography{interference}

@article{jain2016smoking,
  title={Smoking cessation and long-term weight gain in the Framingham Heart Study: an application of the parametric g-formula for a continuous outcome},
  author={Jain, Priyanka and Danaei, Goodarz and Robins, James M and Manson, JoAnn E and Hern{\'a}n, Miguel A},
  journal={European journal of epidemiology},
  volume={31},
  pages={1223--1229},
  year={2016},
  publisher={Springer}
}

@article{prakasa2009conditional,
  title={Conditional independence, conditional mixing and conditional association},
  author={Prakasa Rao, Bhagavatula LS},
  journal={Annals of the Institute of Statistical Mathematics},
  volume={61},
  number={2},
  pages={441--460},
  year={2009},
  publisher={Springer}
}

@article{kennedy2019nonparametric,
  title={Nonparametric causal effects based on incremental propensity score interventions},
  author={Kennedy, Edward H},
  journal={Journal of the American Statistical Association},
  volume={114},
  number={526},
  pages={645--656},
  year={2019},
  publisher={Taylor \& Francis}
}

@article{doukhan1999new,
  title={A new weak dependence condition and applications to moment inequalities},
  author={Doukhan, Paul and Louhichi, Sana},
  journal={Stochastic processes and their applications},
  volume={84},
  number={2},
  pages={313--342},
  year={1999},
  publisher={Elsevier}
}

@book{penrose2003random,
  title={Random geometric graphs},
  author={Penrose, Mathew},
  volume={5},
  year={2003},
  publisher={OUP Oxford}
}

@article{ogburn2020causal,
  title={Causal inference, social networks and chain graphs},
  author={Ogburn, Elizabeth L and Shpitser, Ilya and Lee, Youjin},
  journal={Journal of the Royal Statistical Society Series A: Statistics in Society},
  volume={183},
  number={4},
  pages={1659--1676},
  year={2020},
  publisher={Oxford University Press}
}

@article{jetsupphasuk2025estimating,
  title={Estimating causal effects using difference-in-differences under network dependency and interference},
  author={Jetsupphasuk, Michael and Li, Didong and Hudgens, Michael G},
  journal={arXiv preprint arXiv:2502.03414},
  year={2025}
}

@article{jiang2023dynamic,
  title={Dynamic treatment regimes with interference},
  author={Jiang, Cong and Wallace, Michael P and Thompson, Mary E},
  journal={Canadian Journal of Statistics},
  volume={51},
  number={2},
  pages={469--502},
  year={2023},
  publisher={Wiley Online Library}
}

@book{hernan2010causal,
  title={Causal Inference: What If},
  author={Hernan, M.A. and Robins, J.M.},
  isbn={9781420076165},
  lccn={2022050839},
  year={2025},
  publisher={CRC Press}
}

@article{vanderweele2013social,
  title={Social networks and causal inference},
  author={VanderWeele, Tyler J and An, Weihua},
  journal={Handbook of causal analysis for social research},
  pages={353--374},
  year={2013},
  publisher={Springer}
}

@article{egami2024identification,
  title={Identification and estimation of causal peer effects using double negative controls for unmeasured network confounding},
  author={Egami, Naoki and Tchetgen Tchetgen, Eric J},
  journal={Journal of the Royal Statistical Society Series B: Statistical Methodology},
  volume={86},
  number={2},
  pages={487--511},
  year={2024},
  publisher={Oxford University Press US}
}

@article{cox1958planning, title={Planning of Experiments. By D. R. Cox. [Pp. vi+308. New York: John Wiley and Sons, Inc. London: Chapman and Hall, Ltd., 1958. 60s.]}, volume={85}, DOI={10.1017/S0020268100038063}, number={2}, journal={Journal of the Institute of Actuaries}, author={Cox, D.R.}, year={1959}, pages={317–319}}

@article{forastiere2021identification,
  title={Identification and estimation of treatment and interference effects in observational studies on networks},
  author={Forastiere, Laura and Airoldi, Edoardo M and Mealli, Fabrizia},
  journal={Journal of the American Statistical Association},
  volume={116},
  number={534},
  pages={901--918},
  year={2021},
  publisher={Taylor \& Francis}
}

@article{reich2021review,
  title={A review of spatial causal inference methods for environmental and epidemiological applications},
  author={Reich, Brian J and Yang, Shu and Guan, Yawen and Giffin, Andrew B and Miller, Matthew J and Rappold, Ana},
  journal={International Statistical Review},
  volume={89},
  number={3},
  pages={605--634},
  year={2021},
  publisher={Wiley Online Library}
}

@article{gao2023causal,
  title={Causal inference in network experiments: regression-based analysis and design-based properties},
  author={Gao, Mengsi and Ding, Peng},
  journal={arXiv preprint arXiv:2309.07476},
  year={2023}
}

@article{papadogeorgou2019adjusting,
  title={Adjusting for unmeasured spatial confounding with distance adjusted propensity score matching},
  author={Papadogeorgou, Georgia and Choirat, Christine and Zigler, Corwin M},
  journal={Biostatistics},
  volume={20},
  number={2},
  pages={256--272},
  year={2019},
  publisher={Oxford University Press}
}

@article{liu2016inverse,
  title={On inverse probability-weighted estimators in the presence of interference},
  author={Liu, Lan and Hudgens, Michael G and Becker-Dreps, Sylvia},
  journal={Biometrika},
  volume={103},
  number={4},
  pages={829--842},
  year={2016},
  publisher={Oxford University Press}
}

@article{cole2008constructing,
  title={Constructing inverse probability weights for marginal structural models},
  author={Cole, Stephen R and Hern{\'a}n, Miguel A},
  journal={American journal of epidemiology},
  volume={168},
  number={6},
  pages={656--664},
  year={2008},
  publisher={Oxford University Press}
}

@article{stokes2016electoral,
  title={Electoral backlash against climate policy: A natural experiment on retrospective voting and local resistance to public policy},
  author={Stokes, Leah C},
  journal={American Journal of Political Science},
  volume={60},
  number={4},
  pages={958--974},
  year={2016},
  publisher={Wiley Online Library}
}

@article{leung2022causal,
  title={Causal inference under approximate neighborhood interference},
  author={Leung, Michael P},
  journal={Econometrica},
  volume={90},
  number={1},
  pages={267--293},
  year={2022},
  publisher={Wiley Online Library}
}

@article{viviano2020policy,
  title={Policy design in experiments with unknown interference},
  author={Viviano, Davide},
  journal={arXiv preprint arXiv:2011.08174},
  year={2020}
}

@article{hu2022average,
  title={Average direct and indirect causal effects under interference},
  author={Hu, Yuchen and Li, Shuangning and Wager, Stefan},
  journal={Biometrika},
  year={2022}
}

@article{wang2020design,
  title={Design-based inference for spatial experiments under unknown interference},
  author={Wang, Ye and Samii, Cyrus and Chang, Haoge and Aronow, PM},
  journal={The Annals of Applied Statistics},
  volume={19},
  number={1},
  pages={744--768},
  year={2025},
  publisher={Institute of Mathematical Statistics}
}

@article{kojevnikov2021limit,
  title={Limit theorems for network dependent random variables},
  author={Kojevnikov, Denis and Marmer, Vadim and Song, Kyungchul},
  journal={Journal of Econometrics},
  volume={222},
  number={2},
  pages={882--908},
  year={2021},
  publisher={Elsevier}
}

@article{lunceford2004stratification,
  title={Stratification and weighting via the propensity score in estimation of causal treatment effects: a comparative study},
  author={Lunceford, Jared K and Davidian, Marie},
  journal={Statistics in medicine},
  volume={23},
  number={19},
  pages={2937--2960},
  year={2004},
  publisher={Wiley Online Library}
}

@article{aronow2017estimating,
  title={Estimating average causal effects under general interference, with application to a social network experiment},
  author={Aronow, Peter M and Samii, Cyrus},
  journal={The Annals of Applied Statistics},
  volume={11},
  number={4},
  pages={1912--1947},
  year={2017},
  publisher={Institute of Mathematical Statistics}
}

@article{ertefaie2020nonparametric,
  title={Nonparametric inverse probability weighted estimators based on the highly adaptive lasso},
  author={Ertefaie, Ashkan and Hejazi, Nima S and van der Laan, Mark J},
  journal={arXiv preprint arXiv:2005.11303},
  year={2020}
}

@article{papadogeorgou2020causal,
  title={Causal Inference with Spatio-temporal Data: Estimating the Effects of Airstrikes on Insurgent Violence in Iraq},
  author={Papadogeorgou, Georgia and Imai, Kosuke and Lyall, Jason and Li, Fan},
  journal={arXiv preprint arXiv:2003.13555},
  year={2020}
}

@article{ogburn2017causal,
  title={Causal inference for social network data},
  author={Ogburn, Elizabeth L and Sofrygin, Oleg and Diaz, Ivan and van der Laan, Mark J},
  journal={arXiv preprint arXiv:1705.08527},
  year={2020}
}

@article{ogburn2017vaccines,
  title={Vaccines, contagion, and social networks},
  author={Ogburn, Elizabeth L and VanderWeele, Tyler J},
  journal={The Annals of Applied Statistics},
  volume={11},
  number={2},
  pages={919--948},
  year={2017},
  publisher={Institute of Mathematical Statistics}
}

@article{savje-etal2018-unknown-interference,
  	title={Average treatment effects in the presence of unknown interference},
  	author={S{\"a}vje, Fredrik and Aronow, Peter M and Hudgens, Michael G},
  	journal={The Annals of Statistics},
  	volume={49},
  	number={2},
  	pages={673--701},
  	year={2021},
  	publisher={Institute of Mathematical Statistics}}

@article{robins2000marginal,
  title={Marginal Structural Models and Causal Inference in Epidemiology},
  author={Robins, James M and Hern{\'a}n, Miguel Angel and Brumback, Babette},
  journal={Epidemiology},
  volume={11},
  number={5},
  pages={551},
  year={2000}
}

@article{hudgens_halloran08,
	Author = {Michael Hudgens and M. Elizabeth Halloran},
	Journal = {Journal of the American Statistical Association},
	Number = {482},
	Pages = {832-842},
	Title = {Toward causal inference with interference},
	Volume = {103},
	Year = {2008}}

@article{conley99_spatial,
	Author = {Timothy G. Conley},
	Journal = {Journal of Econometrics},
	Pages = {1-45},
	Title = {{GMM} estimation with cross sectional dependence},
	Volume = {92},
	Year = {1999}}

\clearpage
\appendix
\renewcommand{\theassumption}{\thesection.\arabic{assumption}}
\renewcommand{\theproposition}{\thesection.\arabic{proposition}}
\renewcommand{\thetheorem}{\thesection.\arabic{theorem}}
\renewcommand{\thelemma}{\thesection.\arabic{lemma}}
\renewcommand{\thefigure}{\thesection.\arabic{figure}}
\renewcommand{\thetable}{\thesection.\arabic{table}}
\setcounter{figure}{0}
\setcounter{table}{0}
\setcounter{theorem}{0}
\setcounter{assumption}{0}
\setcounter{proposition}{0}
\begin{center}
{\LARGE \textit{Supplementary Material}}  
\end{center}
\renewcommand{\thesection}{A} 
\renewcommand{\thesubsection}{\thesection.\arabic{subsection}}  
\setcounter{section}{0}
\section{Additional Technical Details} \label{app:details}
This section provides additional interpretations of the estimand and identification assumptions discussed in the main text, as well as technical details for the extensions introduced in Section~\ref{extensions}.
\subsection{Structural Equations under the Identification Assumptions} \label{app:inter}
Under Assumptions~\ref{assum:si}-\ref{assum:na} in the main text, the relationships among variables in our framework can be represented by the following structural equations:
$$
\begin{aligned}\label{eq:se}
    & \mathbf{L}_{i}^{(t)} = f_{\mathbf{L}}\left(\mathbf{A}_{\mathcal{N}_{\mathcal{G}_N}(i)}^{1:(t-1)}, \mathbf{L}_{\mathcal{N}_{\mathcal{G}_N}(i)}^{1:(t-1)}, \varepsilon_{\mathbf{L}i}^{(t)}\right), \\
    & A_{i}^{(t)} = f_A\left(\mathbf{A}_i^{(1:(t-1))}, \mathbf{L}_{i}^{(1:t)}, \varepsilon_{{A}i}^{(t)}\right), \\
    & Y_{i} = f_{Y}\left(\mathbf{A}_{\mathcal{N}_{\mathcal{G}_N}(i)}^{(1:T)}, \mathbf{L}_{\mathcal{N}_{\mathcal{G}_N}(i)}^{(1:T)}, \varepsilon_{Yi}\right), \\
    & \varepsilon_{Ai}^{(t)} \perp \varepsilon_{Aj}^{(t)}, \text{ and } \left(\varepsilon_{Yi}, \varepsilon_{\mathbf{L}i}^{(t+1)}, \mathbf{A}_i^{(1:T)}, \mathbf{L}_{i}^{(1:T)}\right) \perp \varepsilon_{Aj}^{(t)} \mid \left(\mathbf{A}_j^{(1:(t-1))}, \mathbf{L}_{j}^{(1:t)}\right) \text{ for any } i,j,t.
\end{aligned}
$$
Here, we consider $\mathcal{G}_N$ as fixed. $\varepsilon_{\mathbf{L}i}^{(t)}$, $\varepsilon_{Ai}^{(t)}$, and $\varepsilon_{Yi}$ represent other factors that may influence $\mathbf{L}_{i}^{(t)}$, $A_{i}^{(t)}$, and $Y_{i}$, respectively. They may include a unit's own attributes, characteristics of its neighbors, and measurement errors. We do not assume that these error terms are identically distributed, nor do we impose independence on either $\varepsilon_{Yi}$ or $\varepsilon_{\mathbf{L}i}^{(t)}$. 

$\mathcal{N}_{\mathcal{G}_N}(i)$ in the subscripts represents a unit-specific and unknown neighborhood for unit $i$ in $\mathcal{G}_N$. The equations are compatible with the DAG in Figure~\ref{fig:dag} of the main text and closely resemble those in \citet{ogburn2017causal}. However, our framework allows for arbitrary heterogeneity in how units respond to treatment histories and does not impose a predefined exposure mapping that summarizes the effects of $\mathbf{A}_{\mathcal{N}}^{(1:T)}$. These equations formally define what unknown interference means and reflect a more realistic setting for spatial or network data, where both $Y_{i}$ and $\mathbf{L}_{i}^{(t)}$ may be shaped by unobserved and idiosyncratic factors, such as unit-specific neighborhood structures. This flexibility introduces the additional challenges for inference, as discussed in Section~\ref{large-sample-theory}.

Let's denote the collection $\left(\varepsilon_{\mathbf{L},\mathcal{i}}^{(1:T)}, \varepsilon_{Y,\mathcal{i}}\right)$ as $\mathbf{F}_{i}^{(1:T)}$ and expectation over $\mathbf{F}_{i}^{(1:T)}$ as $\mathbb{E}[\cdot]$. Analyses in the main text are conditional on $\mathbf{F}_{i}^{(1:T)}$, with the expectation $\E[\cdot] = \E\left[\cdot \mid \mathbf{F}_{i}^{(1:T)}\right]$. It is worth noting that $\E\left[Y_{i;j}\left(\mathbf{a}^{(1:T)}\right) \mid \mathbf{F}_{i}^{(1:T)}, \mathbf{F}_{j}^{(1:T)}\right] = \E\left[Y_{i;j}\left(\mathbf{a}^{(1:T)}\right) \mid \mathbf{F}_{i}^{(1:T)}\right]$ by definition. We can further marginalize $Y_{i;j}\left(\mathbf{a}^{(1:T)}\right)$ each AMR over $\mathbf{F}_{i}^{(1:T)}$ and obtain
\vspace{-1em}
\begin{align}\label{eq:PAMR}
& \mu^{*}\left(\mathbf{a}^{(1:T)};d\right) \coloneqq \frac{1}{N}\sum_{j=1}^N \mu_j \left(\Big\{\mathbb{E}\left[Y_{i;j}\left(\mathbf{a}^{(1:T)}\right)\right]\Big\}_{i \in \mathcal{N}}; d\right), \\
& \mu^{*}\left(\mathbf{a}^{(1:T)};d\right) = \mathbf{m}\left(\mathbf{a}^{(1:T)}\right)'\boldsymbol{\beta}^{*}(d).\vspace{-1em}
\end{align}
Below, we show that the main results hold for this estimand under additional restrictions.

\subsection{Welfare Implication of the AMR} \label{app:welfare}
Consider the utilitarian welfare function: $W = \frac{1}{N} \sum_{i = 1}^N \E\left[Y_{i}\left(\mathbf{A}_{\mathcal{N}}^{(1:T)}\right)\right]$,
and assume that the treatment assignment policy in period $t$ is governed by a parameter $\gamma_t$: $P\left(A_{j}^{(t)} = 1 \mid \mathbf{V}_{j}^{(t)}\right) = g\left(\mathbf{V}_{j}^{(t)};\gamma_t\right)$. Then, we have
$$
\begin{aligned}
    & \E\left[Y_{i}\left(\mathbf{A}_{\mathcal{N}}^{(1:T)}\right)\right] = \E\left[\E\left[Y_{i}\left(\mathbf{A}_{\mathcal{N}}^{(1:T)}\right) \mid \mathbf{V}_{j}^{(t)}\right]\right] \\
    = & \E\left[\sum_{a_{j}^{(t)} \in \{0,1\}} \E\left[Y_{i}\left(a_{j}^{(t)}, \mathbf{A}_{j}^{(1:(t-1), (t+1):T)}, \mathbf{A}_{\mathcal{N}\setminus \{j\}}^{(1:T)}\right) \mid \mathbf{V}_{j}^{(t)}, A_{j}^{(t)} = a_{j}^{(t)}\right]P\left(A_{j}^{(t)} = a_{j}^{(t)} \mid \mathbf{V}_{j}^{(t)}\right)\right].
\end{aligned}
$$
Using the fact that $P\left(A_{j}^{(t)} = 0 \mid \mathbf{V}_{j}^{(t)}\right) + P\left(A_{j}^{(t)} = 1 \mid \mathbf{V}_{j}^{(t)}\right) = 1$, we know that
$$
\begin{aligned}
   & \frac{\partial \E\left[Y_{i}\left(\mathbf{A}_{\mathcal{N}}^{(1:T)}\right)\right]}{\partial P\left(A_{j}^{(t)} = 1 \mid \mathbf{V}_{j}^{(t)}\right)} = \E\left[Y_{i}\left(1, \mathbf{A}_{j}^{(1:(t-1), (t+1):T)}, \mathbf{A}_{\mathcal{N}\setminus \{j\}}^{(1:T)}\right)\right] - \E\left[Y_{i}\left(0, \mathbf{A}_{j}^{(1:(t-1), (t+1):T)}, \mathbf{A}_{\mathcal{N}\setminus \{j\}}^{(1:T)}\right)\right].
\end{aligned}
$$
Then, for any two period $t' < t$, we can similarly obtain
$$
\begin{aligned}
   & \frac{\partial^T \E\left[Y_{i}\left(\mathbf{A}_{\mathcal{N}}^{(1:T)}\right)\right]}{\partial P\left(\mathbf{A}_{j}^{(t')} = 1 \mid \mathbf{V}_{j}^{(t')}\right)\dots\partial P\left(\mathbf{A}_{j}^{(t)} = 1 \mid \mathbf{V}_{j}^{(t)}\right)} \\
   = & \sum_{\mathbf{a}^{(t':t)} \in \mathcal{A}^{(s:t)}} (-1)^{t - \sum_{s=t'}^t a^{(s)}} \E\left[Y_{i}\left(\mathbf{a}^{(t':t)}, \mathbf{A}_{j}^{(1:t',t:T)}, \mathbf{A}_{\mathcal{N} \setminus \{j\}}^{(1:T)}\right)\right].
\end{aligned}
$$
Averaging units in the sample, we can see that
$$
\begin{aligned}
   & \sum_{j=1}^N \frac{\partial^T W}{\partial P\left(\mathbf{A}_{j}^{(t')} = 1 \mid \mathbf{V}_{j}^{(t')}\right)\dots\partial P\left(\mathbf{A}_{j}^{(t)} = 1 \mid \mathbf{V}_{j}^{(t)}\right)} \\
   = & \frac{1}{N} \sum_{j=1}^N \sum_{i = 1}^N \sum_{\mathbf{a}^{(t':t)} \in \mathcal{A}^{(t':t)}} (-1)^{t - \sum_{s=t'}^t a^{(s)}} \E\left[Y_{i}\left(\mathbf{a}^{(t':t)}, \mathbf{A}_{j}^{(1:t',t:T)}, \mathbf{A}_{\mathcal{N} \setminus \{j\}}^{(1:T)}\right)\right] \\
   = & \sum_{\mathbf{a}^{(t':t)} \in \mathcal{A}^{(t':t)}} (-1)^{t - \sum_{s=t'}^t a^{(s)}} \sum_{d \in \mathcal{D}}\frac{1}{N} \sum_{j=1}^N  \sum_{i = 1}^N   \mathbf{1}\{d_{ij} \in \Omega_j(d)\}\E\left[Y_{i}\left(\mathbf{a}^{(t':t)}, \mathbf{A}_{j}^{(1:t',t:T)}, \mathbf{A}_{\mathcal{N} \setminus \{j\}}^{(1:T)}\right)\right] \\
   = & \sum_{\mathbf{a}^{(t':t)} \in \mathcal{A}^{(t':t)}} (-1)^{t - \sum_{s=t'}^t a^{(s)}} \sum_{d \in \mathcal{D}} \frac{1}{N} \sum_{j=1}^N \mu_j \left(\Big\{Y_{i;j}\left(\mathbf{a}^{(t':t)}\right)\Big\}_{i \in \mathcal{N}}; d\right)|\Omega_j(d)| \\ 
   \to & \sum_{\mathbf{a}^{(t':t)} \in \mathcal{A}^{(t':t)}} (-1)^{t - \sum_{s=t'}^t a^{(s)}} \sum_{d \in \mathcal{D}} \mu\left(\mathbf{a}^{(t':t)};d\right)| \Omega(d) |.
\end{aligned}
$$
Finally, consider the policy parameters $(\gamma_1, \dots, \gamma_T)$, we have
$$
\begin{aligned}
   & \sum_{j=1}^N \frac{\partial^T W}{\partial \gamma_{t'}\dots\partial \gamma_t} = \sum_{j=1}^N \frac{\partial^T W}{\partial P\left(\mathbf{A}_{j}^{(t')} = 1 \mid \mathbf{V}_{j}^{(t)}\right)\dots\partial P\left(\mathbf{A}_{j}^{(t)} = 1 \mid \mathbf{V}_{j}^{(t)}\right)} \prod_{s=t'}^t \frac{\partial P\left(\mathbf{A}_{j}^{(s)} = 1 \mid \mathbf{V}_{j}^{(s)}\right)}{\partial \gamma_t} \\
   = & \sum_{\mathbf{a}^{(t':t)} \in \mathcal{A}^{(t':t)}} (-1)^{t - \sum_{s=t'}^t a^{(s)}} \sum_{d \in \mathcal{D}} \tilde \mu\left(\mathbf{a}^{(t':t)};d\right),
\end{aligned}
$$
where $\tilde \mu\left(\mathbf{a}^{(t':t)};d\right) = \frac{1}{N} \sum_{j=1}^N \mu_j \left(\Big\{Y_{i;j}\left(\mathbf{a}^{(t':t)}\right)\Big\}_{i \in \mathcal{N}}; d\right)| \Omega_j(d) |\prod_{s=t'}^t \frac{\partial P\left(\mathbf{A}_{j}^{(s)} = 1 \mid \mathbf{V}_{j}^{(s)}\right)}{\partial \gamma_s}$ is a re-weighted AMR conditional on the treatment history from period $t'$ to period $t$. In particular, when $g\left(\mathbf{V}_{j}^{(t)};\gamma_t\right) = \gamma_t$ and $|\Omega_j(d)| = |\Omega(d)| + o(1)$, the expression above can be approximated by $\sum_{\mathbf{a}^{(t':t)} \in \mathcal{A}^{(t':t)}} (-1)^{t - \sum_{s=t'}^t a^{(s)}} \sum_{d \in \mathcal{D}} \mu\left(\mathbf{a}^{(t':t)};d\right)| \Omega(d) |$. This result suggests that any partial derivative of the welfare function with regards to the policy parameters can be approximated by a linear combination of the (reweighted) sums of AMRs across proximity levels. Therefore, even though the AMRs do not identify optimal policies, they are informative about the directions of policy changes that improve welfare \citep{hu2022average, viviano2020policy}.

\subsection{Technical Details of the Extensions} \label{app:extension}
\paragraph*{Placebo Tests} We demonstrate the logic of conducting placebo tests using the DAG in Figure~\ref{fig:dag-p}. For simplicity, we only keep paths that are relevant for causal identification. In this toy example, two units, $\{i, j\}$, are observed over three periods, $\{0, 1, 2\}$. We assume that sequential exchangeability holds conditional on each unit's treatment status in the previous period and the current value of time-varying confounders. Under this assumption, there is no direct path from $\mathbf{L}_{i}^{(1)}$ to $A_{i}^{(2)}$.

Now consider a time-varying unobserved variable, $U_i^{(t)}$, that influences both $A_i^{(t)}$ and $\mathbf{L}_{i}^{(t+1)}$, as shown by dashed arrows in the DAG. In this case, sequential exchangeability is violated. A backdoor path arises between $A_i^{(2)}$ and $\mathbf{L}_{i}^{(1)}$, even when conditioning on $A_{i}^{(1)}$ and $\mathbf{L}_{i}^{(2)}$. As a result, we would expect to find a spurious effect generated by $A_i^{(2)}$ on $\mathbf{L}_{i}^{(1)}$ using the WLS estimator, which can serve as evidence against the identification assumptions.

\begin{figure}[htp]
 \begin{center}
 \caption{A DAG illustration for placebo tests}
 \label{fig:dag-p}
\begin{tikzpicture}[
scale=0.85,dot/.style={fill,draw,circle,minimum width=1pt},
arrow style/.style={->,line width=1pt, shorten <=2pt,shorten >=2pt, lightgray},
arrow1 style/.style={->,line width=1.2pt, shorten <=2pt,shorten >=2pt},
arrow2 style/.style={->,line width=1.2pt, shorten <=2pt,shorten >=2pt, red},
arrow3 style/.style={->,line width=1.2pt, shorten <=2pt,shorten >=2pt, blue},
arrow4 style/.style={->,line width=1.2pt, shorten <=2pt,shorten >=2pt, purple},
arrow5 style/.style={->,line width=1pt, dashed, shorten <=2pt,shorten >=2pt },]

\node [fill=black, dot, label=below right: $Y_{i}$] (y2) at (8,5) {};
\node [fill=black, dot, label=above right: $Y_{j}$] (y4) at (8,1) {};
\node [fill=gray, dot, label=below right: $\mathbf{L}_{i}^{(2)}$] (y1) at (2,5) {} edge [arrow style, bend left=20] (y2) edge [arrow style, bend left=20] (y4);
\node [fill=gray, dot, label=above right: $\mathbf{L}_{j}^{(2)}$] (y3) at (2,1) {} edge [arrow style, bend right=20] (y4) edge [arrow style, bend right=20] (y2);
\node [fill=gray, dot, label=below right: $\mathbf{L}_{i}^{(1)}$] (y5) at (-4,5) {} edge [arrow style, bend left=20] (y1) edge [arrow style, bend left=20] (y3);
\node [fill=gray, dot, label=above right: $\mathbf{L}_{j}^{(1)}$] (y6) at (-4,1) {} edge [arrow style, bend right=20] (y3) edge [arrow style, bend right=20] (y1);

\node [fill=white, dot, label=below left: $A_{i}^{(2)}$] (a2) at (6,5) {};
\node [fill=white, dot, label=below left: $A_{i}^{(1)}$] (a1) at (0,5) {} edge [arrow1 style, bend left=20] (a2);
\node [fill=white, dot, label=above left: $A_{j}^{(2)}$] (a4) at (6,1) {};
\node [fill=white, dot, label=above left: $A_{j}^{(1)}$] (a3) at (0,1) {} edge [arrow1 style, bend right=20] (a4);

\draw[arrow1 style] (y1) -- (a2);
\draw[arrow1 style] (y3) -- (a4);

\draw[arrow1 style] (y5) -- (a1);
\draw[arrow1 style] (y6) -- (a3);

\node [fill=gray, dot, label=above: $U_{i}^{(2)}$] (u2) at (7,7) {} edge [arrow5 style] (y2) edge [arrow5 style] (a2);
\node [fill=gray, dot, label=above: $U_{i}^{(1)}$] (u1) at (1,7) {} edge [arrow5 style] (y1) edge [arrow5 style] (a1);
\node [fill=gray, dot, label=above: $U_{i}^{(0)}$] (u5) at (-5,7) {} edge [arrow5 style] (y5);
\node [fill=gray, dot, label=below: $U_{j}^{(2)}$] (u4) at (7,-1) {} edge [arrow5 style] (y4) edge [arrow5 style] (a4);
\node [fill=gray, dot, label=below: $U_{j}^{(1)}$] (u3) at (1,-1) {} edge [arrow5 style] (y3) edge [arrow5 style] (a3);
\node [fill=gray, dot, label=below: $U_{j}^{(0)}$] (u6) at (-5,-1) {} edge [arrow5 style] (y6);

\draw[arrow1 style] (u1) -- (u2);
\draw[arrow1 style] (u3) -- (u4);
\draw[arrow1 style] (u5) -- (u1);
\draw[arrow1 style] (u6) -- (u3);

\end{tikzpicture}
 \end{center}
 \textit{Notes:} The DAG depicts two units, $\{i, j\}$, observed over three periods, $\{0, 1, 2\}$. Variables are represented by circles and causal paths by arrows. White circles denote treatment, gray circles represent time-varying confounders, and black circles correspond to outcomes. Black arrows indicate relationships influencing treatment assignment, gray arrows represent other potential dependencies between variables, and dashed arrows show potential influences from an unobservable variable.
\end{figure}

\paragraph*{Diffusion and Dependence in Treatment Assignment} To allow for diffusion in treatment, we can modify Assumption~\ref{assum:si} as follows:
\begin{assumption}[Sequential exchangeability with treatment diffusion]\label{assum:si1}
\vspace{-1em}
\[
  \Big\{Y_{i}\left(\mathbf{a}^{(1:T)}, \mathbf{A}_{\mathcal{N}\setminus \{j\}}^{(1:T)}\right), \mathbf{L}_{i}^{(t+1)}\left(\mathbf{a}^{(1:T)}, \mathbf{A}_{\mathcal{N}\setminus \{j\}}^{(1:T)}\right)\Big\} \perp A_{j}^{(t)} \mid \left(\mathbf{A}_{\{j\} \cup \mathcal{N}_{\mathcal{G}_N}(j, p)}^{(1:(t-1))}, \mathbf{L}_{\{j\} \cup \mathcal{N}_{\mathcal{G}_N}(j, p)}^{(1:(t-1))}\right).\vspace{-1em}
\]
for any $i$, $j$, $t$, $\mathbf{a}^{(1:T)}$, and known $p$.
\end{assumption}
Here, $\mathcal{N}_{\mathcal{G}_N}(j, p)$ denotes the set of units whose proximity to $j$ does not exceed $p$, as defined in Section~\ref{asymptotic-distribution}. Under Assumption~\ref{assum:si1}, the variables $\mathbf{A}_{\mathcal{N}_{\mathcal{G}_N}(j, p)}^{(1:(t-1))}$ and $\mathbf{L}_{\mathcal{N}_{\mathcal{G}_N}(j, p)}^{(1:(t-1))}$ must be included in $\mathbf{V}_{j}^{(t)}$ and controlled for when estimating unit $j$'s propensity score at time $t$. This adjustment is straightforward when $p$ is known and fixed, and it does not affect our theoretical results. We leave the case where $p$ is unknown or grows with $N$ to future work. 

To characterize the degree of dependence in treatment status between two units in period $t$, we follow \citet{savje-etal2018-unknown-interference} and define the conditional alpha-mixing coefficient of two random variables, \(X_1\) and \(X_2\), given covariates $\mathbf{V}$, as 
$$
\alpha(X_1, X_2; \mathbf{V}) = \sup_{\substack{x_1 \in \sigma(X_1 | \mathbf{V}), \\ x_2 \in \sigma(X_2 | \mathbf{V})}}| P(x_1 \cap x_2 | \mathbf{V}) - P(x_1 | \mathbf{V})P(x_2 | \mathbf{V})|.
$$ 
Let \(\check{\mathbf{A}}_{\mathcal{N}\setminus \{j\}}^{(t)}(d)\) the subset of treatment assignments in period $t$, excluding \(A_{j}^{(t)}\), that affect $\mu_j(d)$. We use an indicator $k_{ij}$ to represent whether $\mu_i(d)$ and $\mu_j(d)$ are dependent. The conditional internal average mixing coefficient in period \(t\) is then defined as $\alpha_{INT, t} = \sum_{j=1}^N \alpha\left(A_{j}^{(t)}, \check{\mathbf{A}}_{\mathcal{N}\setminus \{j\}}^{(t)}(d) \mid \mathbf{V}_{\mathcal{N}}^{(t)}\right)$, and the conditional external average mixing coefficient in period \(t\) is defined as $\alpha_{EXT, t} = \frac{1}{N} \sum_{i=1}^N\sum_{j=1}^N (1 - g_{ij})\alpha\left(\check{\mathbf{A}}_{\mathcal{N}\setminus \{i\}}^{(t)}(d), \check{\mathbf{A}}_{\mathcal{N}\setminus \{j\}}^{(t)}(d) \mid \mathbf{V}_{\mathcal{N}}^{(t)}\right)$. Theorem 1 holds holds under the following assumption:

\begin{assumption}[Design mixing and design separation]\label{assum:des}
$$
\alpha_{INT, t} = o(N), \alpha_{EXT, t} = o(N), 
$$ 
for any period $t$.
\end{assumption}

This assumption is satisfied by many common treatment assignment mechanisms, such as complete randomization. Formal proofs can be found in \citet{savje-etal2018-unknown-interference}, where the authors also show that Assumption~\ref{assum:des} holds for common designs such as complete randomization.

\paragraph*{Marginal Structural Models with a Large $T$}
When $T$ is large, we can consider summarizing the effects of any treatment history using a statistic $\mathcal{T}\left(\mathbf{a}^{(1:T)}\right)$. A common choice is the number of periods under treatment: $\mathcal{T}\left(\mathbf{a}^{(1:T)}\right) = \sum_{t=1}^T a^{(t)}$. We then define AMRs for each $d$ and each possible value $\mathcal{T}$ of $\mathcal{T}\left(\mathbf{a}^{(1:T)}\right)$:
$$
\begin{aligned}
    & Y_{i;j}\left(\mathcal{T}\right) \coloneqq \E\left[Y_{i}\left(\mathcal{T}\left(\mathbf{A}_j^{(1:T)}\right)=\mathcal{T}; \mathbf{A}_{\mathcal{N}\setminus \{j\}}^{(1:T)}\right)\right], \\
    & \mu \left(\big\{Y_{i;j}\left(\mathcal{T}\right)\big\}_{i \in \mathcal{N}}; \Omega_j(d)\right) \coloneqq \frac{\sum_{i=1}^N \mathbf{1}\{i \in \Omega_j(d)\}Y_{i;j}\left(\mathcal{T}\right)}{|\Omega_j(d)|}, \\
    & \mu\left(\mathcal{T};d\right) \coloneqq \frac{1}{N}\sum_{j=1}^N \mu_j \left(\big\{Y_{i;j}\left(\mathcal{T}\right)\big\}_{i \in \mathcal{N}}; d\right).
\end{aligned}
$$
As $\mu\left(\mathcal{T};d\right)$ only depends on $\mathcal{T}$, we can summarize all its possible values using a saturated MSM:
$$
\begin{aligned}
\mu\left(\mathcal{T};d\right) = \mathbf{m}\left(\mathcal{T}\right)'\beta(d).
\end{aligned}
$$
By the same logic, each $\mu\left(\mathcal{T};d\right)$ can be identified under Assumption~\ref{assum:si}, and we can estimate $\beta(d)$ using the WLS estimator.

When $T$ is large, Assumption~\ref{assum:ani} may be questionable if effects from remote neighbors accumulate over time and eventually become non-negligible. To avoid this possibility, we denote an independent copy of the treatment history beyond the recent $s$ periods as \(\mathbf{A}^{(1:T),s}_{\mathcal{N}_N} = \left(\mathbf{A}^{((T-s+1):T)}_{\mathcal{N}_N}, \tilde{\mathbf{A}}^{(1:(T-s))}_{\mathcal{N}_N}\right)\). Then, Assumption~\ref{assum:ani} can be replaced by the following version:
\begin{assumption}[Approximate Neighborhood Interference in Two Dimensions]\label{assum:ani1}
\vspace{-1em}
\[
\begin{aligned}
  & \max_{i \in \mathcal{N}_N} E\Bigg|Y_{i}\left(\mathbf{A}^{(1:T)}_{\mathcal{N}_N}\right) - Y_{i}\left(\mathbf{A}^{(1:T),s}_{\mathcal{N}_N}\right) \Bigg| \leq \theta_{T,s}, \\
  & \max_{i \in \mathcal{N}_N} E\Bigg|Y_{i}\left(\mathbf{A}^{((T-s+1):T)}_{\mathcal{N}_N}\right) - Y_{i}\left(\mathbf{A}^{((T-s+1):T)}_{i,p}\right) \Bigg| \leq \theta_{N,p},
\end{aligned}
  \] where \(\sup_T \theta_{T,s} \rightarrow 0\) as \(s \rightarrow \infty\) and \(\sup_N \theta_{N,p} \rightarrow 0\) as \(p \rightarrow \infty\).\vspace{-1em}
\end{assumption}
This assumption explicitly stipulates that the influence of remote observations, whether in the temporal or cross-unit dimension, will eventually be negligible in large samples. It is straightforward to verify that our main theorems hold if Assumption~\ref{assum:ani} is replaced by Assumption~\ref{assum:ani1}.

\newpage 
\renewcommand{\thesection}{B} 
\renewcommand{\thesubsection}{\thesection.\arabic{subsection}}  
\setcounter{section}{0}
\section{Proofs} \label{app:proofs}
This section provides the proofs of the theorems stated in the main text.
\subsection{Proof for Proposition~\ref{thm:id}} \label{app:id}
For any units $i$ and $j$, we know that
$$
\begin{aligned}
& \E\left[\frac{\mathbf{1}\{\mathbf{A}_j^{(1:T)} = \mathbf{a}^{(1:T)}\}Y_{i}}{e\left(\mathbf{A}_j^{(1:T)} = \mathbf{a}^{(1:T)}; \mathbf{V}_{j}^{(1:T)}\right)}\right] = \E\left[\frac{\mathbf{1}\{\mathbf{A}_j^{(1:T)} = \mathbf{a}^{(1:T)}\}Y_{i}\left(\mathbf{A}_{\mathcal{N}}^{(1:T)}\right)}{e\left(\mathbf{A}_j^{(1:T)} = \mathbf{a}^{(1:T)}; \mathbf{V}_{j}^{(1:T)}\right)}\right] \\
= & \E\left[\E\left[\frac{\mathbf{1}\{\mathbf{A}_j^{(1:T)} = \mathbf{a}^{(1:T)}\}Y_{i}\left(\mathbf{A}_{\mathcal{N}}^{(1:T)}\right)}{e\left(\mathbf{A}_j^{(1:T)} = \mathbf{a}^{(1:T)}; \mathbf{V}_{j}^{(1:T)}\right)} \mid \mathbf{V}_{j}^{(T)}\right]\right] \\
= & \E\left[\frac{\mathbf{1}\{\mathbf{A}_j^{(1:(T-1))} = \mathbf{a}^{(1:(T-1))}\}}{e\left(\mathbf{A}_j^{(1:(T-1))} = \mathbf{a}^{(1:(T-1))}; \mathbf{V}_{j}^{(1:(T-1))}\right)}\E\left[\frac{\mathbf{1}\{A_{j}^{(T)} = a^{(T)}\}Y_{i}\left(\mathbf{A}_{\mathcal{N}}^{(1:T)}\right)}{P\left(A_{j}^{(T)} = a^{(T)} \mid \mathbf{V}_{j}^{(T)}\right)} \mid \mathbf{V}_{j}^{(T)}\right]\right] \\
= & \E\left[\frac{\mathbf{1}\{\mathbf{A}_j^{(1:(T-1))} = \mathbf{a}^{(1:(T-1))}\}}{e\left(\mathbf{A}_j^{(1:(T-1))} = \mathbf{a}^{(1:(T-1))}; \mathbf{V}_{j}^{(1:(T-1))}\right)}\E\left[Y_i\left(a^{(T)}, \mathbf{A}_j^{(1:(T-1))}, \mathbf{A}_{\mathcal{N}\setminus \{j\}}^{(1:T)}\right) \mid \mathbf{V}_{j}^{(T)}, A_{j}^{(T)} = a^{(T)}\right]\right] \\
= & \E\left[\frac{\mathbf{1}\{\mathbf{A}_j^{(1:(T-1))} = \mathbf{a}^{(1:(T-1))}\}}{e\left(\mathbf{A}_j^{(1:(T-1))} = \mathbf{a}^{(1:(T-1))}; \mathbf{V}_{j}^{(1:(T-1))}\right)}\E\left[Y_i\left(a^{(T)}, \mathbf{A}_j^{(1:(T-1))}, \mathbf{A}_{\mathcal{N}\setminus \{j\}}^{(1:T)}\right) \mid \mathbf{V}_{j}^{(T)}\right]\right] \\
= & \E\left[\frac{\mathbf{1}\{\mathbf{A}_j^{(1:(T-1))} = \mathbf{a}^{(1:(T-1))}\}Y_i\left(a^{(T)}, \mathbf{A}_j^{(1:(T-1))}, \mathbf{A}_{\mathcal{N}\setminus \{j\}}^{(1:T)}\right)}{e\left(\mathbf{A}_j^{(1:(T-1))} = \mathbf{a}^{(1:(T-1))}; \mathbf{V}_{j}^{(1:(T-1))}\right)}\right] = \dots \\
= & \E\left[Y_i\left(\mathbf{a}^{(1:T)}, \mathbf{A}_{\mathcal{N}\setminus \{j\}}^{(1:T)}\right)\right] = Y_{i;j}\left(\mathbf{a}^{(1:T)}\right).
\end{aligned} 
$$
The $\dots$ part iterates the same step from period $T-1$ to period $1$. The first equality uses the definition of the potential outcome, while the fourth equality uses the law of total expectation. The fifth equality holds because of sequential exchaneagibility. Therefore, for any $d \in \mathcal{D}$,
$$
\begin{aligned}
& \frac{1}{N} \sum_{j=1}^N\E\left[\frac{\mathbf{1}\{\mathbf{A}_j^{(1:T)} = \mathbf{a}^{(1:T)}\}\mu_j(d)}{e\left(\mathbf{A}_j^{(1:T)} = \mathbf{a}^{(1:T)}; \mathbf{V}_{j}^{(1:T)}\right)}\right] = \frac{1}{N} \sum_{j=1}^N\E\left[\frac{\mathbf{1}\{\mathbf{A}_j^{(1:T)} = \mathbf{a}^{(1:T)}\}\mu\left(\big\{Y_{i}\big\}_{i \in \mathcal{N}};d\right)}{e\left(\mathbf{A}_j^{(1:T)} = \mathbf{a}^{(1:T)}; \mathbf{V}_{j}^{(1:T)}\right)}\right] \\
= & \frac{1}{N} \sum_{j=1}^N\frac{1}{|\Omega_j(d)|}\sum_{i=1}^N \mathbf{1}\{i \in \Omega_j(d)\}\E\left[\frac{\mathbf{1}\{\mathbf{A}_j^{(1:T)} = \mathbf{a}^{(1:T)}\}Y_{i}}{e\left(\mathbf{A}_j^{(1:T)} = \mathbf{a}^{(1:T)}; \mathbf{V}_{j}^{(1:T)}\right)}\right] \\
= & \frac{1}{N} \sum_{j=1}^N\frac{\sum_{i=1}^N \mathbf{1}\{i \in \Omega_j(d)\}Y_{i;j}\left(\mathbf{a}^{(1:T)}\right)}{|\Omega_j(d)|} = \mu\left(\mathbf{a}^{(1:T)};d\right).
\end{aligned} 
$$
In a saturated MSM, each of the parameters is a linear combination of the AMRs thus can also be identified from data. We can similarly prove that
$$
\begin{aligned}
& \frac{1}{N} \sum_{j=1}^N\mathbb{E}\left[\E\left[\frac{\mathbf{1}\{\mathbf{A}_j^{(1:T)} = \mathbf{a}^{(1:T)}\}\mu_j(d)}{e\left(\mathbf{A}_j^{(1:T)} = \mathbf{a}^{(1:T)}; \mathbf{V}_{j}^{(1:T)}\right)}\right]\right] = \mu^{*}\left(\mathbf{a}^{(1:T)};d\right).
\end{aligned} 
$$

\subsection{Proof for Theorem~\ref{thm:IPTW-asym}} \label{app:thm1}
We start from the case where $e\left(\mathbf{A}_{i}^{(1:T)} = \mathbf{A}^{(1:T)}; \mathbf{V}_{j}^{(1:T)}\right)$ is known. Consider the first term in the expression of $\hat \beta(d)$, namely, $\frac{1}{N}\sum_{i=1}^N w_i\mathbf{m}\left(\mathbf{A}_i^{(1:T)}\right) \mathbf{m}'\left(\mathbf{A}_i^{(1:T)}\right)$. We can establish the following lemma:

\begin{lemma}\label{lemma:ind-A}
Under Assumption 1, 
$$
\begin{aligned}
& \E\left[\frac{f\left(\mathbf{A}_i^{(1:T)}\right) f\left(\mathbf{A}_j^{(1:T)}\right)}{e\left(\mathbf{A}_{i}^{(1:T)} = \mathbf{A}^{(1:T)}; \mathbf{V}_{j}^{(1:T)}\right)e\left(\mathbf{A}_{j}^{(1:T)} = \mathbf{A}^{(1:T)}; \mathbf{V}_{j}^{(1:T)}\right)} \right] \\
= & \E\left[\frac{f\left(\mathbf{A}_i^{(1:T)}\right) }{e\left(\mathbf{A}_{i}^{(1:T)} = \mathbf{A}^{(1:T)}; \mathbf{V}_{j}^{(1:T)}\right)} \right]\E\left[\frac{f\left(\mathbf{A}_j^{(1:T)}\right) }{e\left(\mathbf{A}_{j}^{(1:T)} = \mathbf{A}^{(1:T)}; \mathbf{V}_{j}^{(1:T)}\right)} \right]
\end{aligned}
$$
for any function $f(\cdot)$ and units $i, j \in \mathcal{N}_N$ and $i \neq j$.
\end{lemma}
\textit{Proof:} Let $\mathcal{A}^{(1:T)}$ denote all possible values that $\mathbf{A}_i^{(1:T)}$ can take, then
$$
\begin{aligned}
& \E\left[\frac{f\left(\mathbf{A}_i^{(1:T)}\right) }{e\left(\mathbf{A}_{i}^{(1:T)} = \mathbf{A}^{(1:T)}; \mathbf{V}_{j}^{(1:T)}\right)}\right] = \E\left[\E\left[\frac{f\left(\mathbf{A}_i^{(1:T)}\right) }{e\left(\mathbf{A}_{i}^{(1:T)} = \mathbf{A}^{(1:T)}; \mathbf{V}_{j}^{(1:T)}\right)} \mid \mathbf{V}_{i}^{(T)}\right]\right] \\
= & \E\left[\frac{\E\left[\frac{f\left(\mathbf{A}_i^{(1:T)}\right) }{P\left(A_{i}^{(T)} = A^{(T)} | \mathbf{V}_{i}^{(T)}\right)} \mid \mathbf{V}_{i}^{(T)}, \right]}{\prod_{s=1}^{T-1} P\left(A_{i}^{(s)} = A^{(s)} | \mathbf{V}_{j}^{(s)}\right)}\right] \\
= & \E\left[\frac{\sum_{a_{i}^{(T)} = 0}^1\E\left[\frac{f\left(a_{i}^{(T)}, \mathbf{A}_i^{(1:(T - 1))}\right)}{P\left(A_{i}^{(T)} = a_{i}^{(T)} | \mathbf{V}_{i}^{(T)}\right)}  \mid \mathbf{V}_{i}^{(T)}, A_{i}^{(T)} = a_{i}^{(T)}\right]P\left(A_{i}^{(T)} = a_{i}^{(T)} \mid \mathbf{V}_{i}^{(T)}\right)}{\prod_{s=1}^{T-1} P\left(A_{i}^{(s)} | \mathbf{V}_{j}^{(s)}\right)}\right] \\
= & \sum_{a_{i}^{(T)} = 0}^1\E\left[\frac{f\left(a_{i}^{(T)}, \mathbf{A}_i^{1:(T - 1)}\right)}{\prod_{s=1}^{T-1} P\left(A_{i}^{(s)} | \mathbf{V}_{j}^{(s)}\right)}\right] = \dots = \sum_{\mathbf{a}_i^{(1:T)} \in \mathcal{A}^{(1:T)}} f\left(\mathbf{a}_i^{(1:T)}\right).
\end{aligned} 
$$

Assumption~\ref{assum:si} implies that $A_{i}^{(T)} \perp A_{j}^{(T)} \mid \left(\mathbf{V}_{i}^{(T)}, \mathbf{V}_{j}^{(T)}\right)$ for any $t$ and $s$ by the rule of weak union. Therefore, for any $a_{i}^{(T)}, a_{j}^{(T)} \in \{0, 1\}$, $P\left(A_{i}^{(T)} = a_{i}^{(T)}, A_{j}^{(T)} = a_{j}^{(T)} \mid \mathbf{V}_{i}^{(T)}, \mathbf{V}_{j}^{(T)}\right) = P(A_{i}^{(T)} = a_{i}^{(T)} | \mathbf{V}_{i}^{(T)})P(A_{j}^{(T)} = a_{j}^{(T)} | \mathbf{V}_{j}^{(T)})$, and
$$
\begin{aligned}
& \E\left[\frac{f\left(\mathbf{A}_i^{(1:T)}\right) f\left(\mathbf{A}_j^{(1:T)}\right)}{e\left(\mathbf{A}_{i}^{(1:T)} = \mathbf{A}^{(1:T)}; \mathbf{V}_{j}^{(1:T)}\right)e\left(\mathbf{A}_{j}^{(1:T)} = \mathbf{A}^{(1:T)}; \mathbf{V}_{j}^{(1:T)}\right)} \right] \\
= & \E\left[\E\left[\frac{f\left(\mathbf{A}_i^{(1:T)}\right) f\left(\mathbf{A}_j^{(1:T)}\right)}{e\left(\mathbf{A}_{i}^{(1:T)} = \mathbf{A}^{(1:T)}; \mathbf{V}_{j}^{(1:T)}\right)e\left(\mathbf{A}_{j}^{(1:T)} = \mathbf{A}^{(1:T)}; \mathbf{V}_{j}^{(1:T)}\right)} \mid \mathbf{V}_{i}^{(T)}, \mathbf{V}_{j}^{(T)}\right]\right] \\
= & \E\left[\frac{\E\left[\frac{1}{P\left(A_{i}^{(T)} | \mathbf{V}_{i}^{(T)}\right)P\left(A_{j}^{(T)} | \mathbf{V}_{j}^{(T)}\right)}f\left(\mathbf{A}_i^{(1:T)}\right) f\left(\mathbf{A}_j^{(1:T)}\right)  \mid \mathbf{V}_{i}^{(T)}, \mathbf{V}_{j}^{(T)}\right]}{\prod_{s=1}^{T-1} P\left(A_{i}^{(s)} | \mathbf{V}_{j}^{(s)}\right)\prod_{s=1}^{T-1} P\left(A_{j}^{(s)} | \mathbf{V}_{j}^{(s)}\right)}\right] \\
= & \E\left[\frac{\sum_{a_{i}^{(T)} = 0}^1\sum_{a_{j}^{(T)} = 0}^1\E\left[\frac{f\left(a_{i}^{(T)}, \mathbf{A}_i^{(1:(T - 1))}\right) f\left(a_{j}^{(T)}, \mathbf{A}_j^{(1:(T - 1))}\right)}{P\left(A_{i}^{(T)} = a_{i}^{(T)} | \mathbf{V}_{i}^{(T)}\right)P\left(A_{j}^{(T)} = a_{j}^{(T)} | \mathbf{V}_{j}^{(T)}\right)}  \mid \mathbf{V}_{i}^{(T)}, \mathbf{V}_{j}^{(T)}, A_{i}^{(T)} = a_{i}^{(T)}, A_{j}^{(T)} = a_{j}^{(T)}\right]}{\frac{1}{P\left(A_{i}^{(T)} = a_{i}^{(T)}, A_{j}^{(T)} = a_{j}^{(T)} \mid \mathbf{V}_{i}^{(T)}, \mathbf{V}_{j}^{(T)}\right)}\prod_{s=1}^{T-1} P\left(A_{i}^{(s)} | \mathbf{V}_{j}^{(s)}\right)\prod_{s=1}^{T-1} P\left(A_{j}^{(s)} | \mathbf{V}_{j}^{(s)}\right)}\right] \\
= & \sum_{a_{i}^{(T)} = 0}^1\sum_{a_{j}^{(T)} = 0}^1\E\left[\frac{f\left(a_{i}^{(T)}, \mathbf{A}_i^{(1:(T - 1))}\right) f\left(a_{j}^{(T)}, \mathbf{A}_j^{(1:(T - 1))}\right)}{\prod_{s=1}^{T-1} P\left(A_{i}^{(s)} | \mathbf{V}_{j}^{(s)}\right)\prod_{s=1}^{T-1} P\left(A_{j}^{(s)} | \mathbf{V}_{j}^{(s)}\right)}\right] = \dots \\
= & \sum_{\mathbf{a}_i^{(1:T)} \in \mathcal{A}^{(1:T)}}\sum_{\mathbf{a}_j^{(1:T)} \in \mathcal{A}^{(1:T)}}f\left(\mathbf{a}_i^{(1:T)}\right) f\left(\mathbf{a}_j^{(1:T)}\right) \\
= & \E\left[\frac{f\left(\mathbf{A}_i^{(1:T)}\right) }{e\left(\mathbf{A}_{i}^{(1:T)} = \mathbf{A}^{(1:T)}; \mathbf{V}_{j}^{(1:T)}\right)} \right]\E\left[\frac{f\left(\mathbf{A}_j^{(1:T)}\right) }{e\left(\mathbf{A}_{j}^{(1:T)} = \mathbf{A}^{(1:T)}; \mathbf{V}_{j}^{(1:T)}\right)} \right].
\end{aligned} 
$$

In our framework, $\mathbf{L}_{i}^{(T)}$ and $\mathbf{L}_{j}^{(T)}$ are dependent because both are functions of $\mathbf{A}_{\mathcal{N}}^{(1:T)}$. As a result, for any function $f(\cdot)$, $f\left(\mathbf{A}_i^{(1:T)}\right)$ and $f\left(\mathbf{A}_j^{(1:T)}\right)$ are generally dependent as well. Nevertheless, Lemma~\ref{lemma:ind-A} suggests that once normalized by the weights, $\frac{1}{e\left(\mathbf{A}_{i}^{(1:T)} = \mathbf{A}^{(1:T)}; \mathbf{V}_{j}^{(1:T)}\right)}$ and $\frac{1}{e\left(\mathbf{A}_{j}^{(1:T)} = \mathbf{A}^{(1:T)}; \mathbf{V}_{j}^{(1:T)}\right)}$, they become uncorrelated. Due to positivity and the binary nature of $A_{i}^{(T)}$, $w_i\mathbf{m}\left(\mathbf{A}_i^{(1:T)}\right) \mathbf{m}'\left(\mathbf{A}_i^{(1:T)}\right)$ is uniformly bounded. Then, using Lemma~\ref{lemma:ind-A}, we can see that
$$
\Var\left[\frac{1}{N}\sum_{i=1}^N w_i\mathbf{m}\left(\mathbf{A}_i^{(1:T)}\right) \mathbf{m}'\left(\mathbf{A}_i^{(1:T)}\right)\right] = \frac{1}{N^2}\sum_{i=1}^N \Var\left[w_i\mathbf{m}\left(\mathbf{A}_i^{(1:T)}\right) \mathbf{m}'\left(\mathbf{A}_i^{(1:T)}\right)\right] \to 0,
$$
as $N \to \infty$. Therefore, the term $\frac{1}{N}\sum_{i=1}^N w_i\mathbf{m}\left(\mathbf{A}_i^{(1:T)}\right) \mathbf{m}'\left(\mathbf{A}_i^{(1:T)}\right)$ is consistent for its expectation, $\frac{1}{N}\sum_{i=1}^N \E\left[w_i\mathbf{m}\left(\mathbf{A}_i^{(1:T)}\right) \mathbf{m}'\left(\mathbf{A}_i^{(1:T)}\right)\right]$. We use $\Sigma_A$ to represent the normalized variance $\frac{1}{N}\sum_{i=1}^N \Var\left[w_i\mathbf{m}\left(\mathbf{A}_i^{(1:T)}\right) \mathbf{m}'\left(\mathbf{A}_i^{(1:T)}\right)\right]$.

Next, we turn to the second term, $\frac{1}{N}\sum_{i=1}^N \mathbf{Z}_i(d)Y_i$. Let's define $U_i(d) = \kappa' \mathbf{Z}_i(d)Y_i$, where $\kappa$ is a vector of weights that sum up to $1$. We first show that the sequence $\{U_i(d)\}_{i=1}^N$ satisfies a property termed as $\psi$-dependence by \citet{kojevnikov2021limit}. For any subset $\mathcal{H} \subset \mathcal{N}_N$ with $|\mathcal{H}| = h$, we denote the collection of bounded Lipchitz functions on it with the associated Lipchitz constant $L_h$ as $\mathcal{L}_h$. Consider an independent copy of the entire treatment history, $\left(\mathbf{A}^{(1:T)}_{\mathcal{N}_N}\right)^{\dagger}$. Since both $\kappa' \mathbf{Z}_i(d)$ and $Y_i$ are functions of the entire treatment history, we can define $U_i^{(p,\dagger)}(d)$ as the value of $U_i(d)$ under $\left(\mathbf{A}^{(1:T)}_{\mathcal{N}_{\mathcal{G}_N}(i, p)}, \left(\mathbf{A}^{(1:T)}_{\mathcal{N}_N \setminus \mathcal{N}_{\mathcal{G}_N}(i, p)}\right)^{\dagger}\right)$. $U_i^{(p,\ddagger)}(d)$ is similarly defined under a different copy, $\left(\mathbf{A}^{(1:T)}_{\mathcal{N}_N}\right)^{\ddagger}$. 

We further define $\mathbf{U}_{\mathcal{H}}(d) = (\dots, U_i(d), \dots)'_{i \in \mathcal{H}}$ and $\mathbf{U}_{\mathcal{H}}^{(p,\dagger)}(d)$ similarly. For any subset $\tilde{\mathcal{H}} \subset \mathcal{N}_N$ such that $\min_{i \in \mathcal{H}, j \in \tilde{\mathcal{H}}} d_{ij} \geq p$ and $\mathbf{U}_{\tilde{\mathcal{H}}}^{(p,\ddagger)}(d)$ defined on it, it is straightforward to see that $\Cov\left[f_h\left(\mathbf{U}_{\mathcal{H}}^{(\lfloor\frac{p}{2}\rfloor,\dagger)}(d)\right), f_{\tilde h}\left(\mathbf{U}_{\tilde{\mathcal{H}}}^{(\lfloor\frac{p}{2}\rfloor,\ddagger)}(d)\right)\right] = 0$ if $p > 2*d$. Then, for any $f_h \in \mathcal{L}_h$ and $f_{\tilde h} \in \mathcal{L}_{\tilde h}$, we know that $\Bigg|\Cov\left[f_h\left(\mathbf{U}_{\mathcal{H}}(d)\right), f_{\tilde h}\left(\mathbf{U}_{\tilde{\mathcal{H}}}(d)\right)\right]\Bigg| \leq 2||f_{h}||_{\infty}||f_{\tilde h}||_{\infty}$ for $p \leq 2*d$, and for $p > 2*d$,
$$
\begin{aligned}
& \Bigg|\Cov\left[f_h\left(\mathbf{U}_{\mathcal{H}}(d)\right), f_{\tilde h}\left(\mathbf{U}_{\tilde{\mathcal{H}}}(d)\right)\right]\Bigg| \\
= & \Bigg|\Cov\left[f_h\left(\mathbf{U}_{\mathcal{H}}(d)\right), f_{\tilde h}\left(\mathbf{U}_{\tilde{\mathcal{H}}}(d)\right)\right] - \Cov\left[f_h\left(\mathbf{U}_{\mathcal{H}}^{(\lfloor\frac{p}{2}\rfloor,\dagger)}(d)\right), f_{\tilde h}\left(\mathbf{U}_{\tilde{\mathcal{H}}}^{(\lfloor\frac{p}{2}\rfloor,\ddagger)}(d)\right)\right]\Bigg| \\
= & \Bigg|\Cov\left[f_h\left(\mathbf{U}_{\mathcal{H}}(d)\right), f_{\tilde h}\left(\mathbf{U}_{\tilde{\mathcal{H}}}(d)\right)\right] - \Cov\left[f_h\left(\mathbf{U}_{\mathcal{H}}^{(\lfloor\frac{p}{2}\rfloor,\dagger)}(d)\right), f_{\tilde h}\left(\mathbf{U}_{\tilde{\mathcal{H}}}(d)\right)\right] \\
& + \Cov\left[f_h\left(\mathbf{U}_{\mathcal{H}}^{(\lfloor\frac{p}{2}\rfloor,\dagger)}(d)\right), f_{\tilde h}\left(\mathbf{U}_{\tilde{\mathcal{H}}}(d)\right)\right] - \Cov\left[f_h\left(\mathbf{U}_{\mathcal{H}}^{(\lfloor\frac{p}{2}\rfloor,\dagger)}(d)\right), f_{\tilde h}\left(\mathbf{U}_{\tilde{\mathcal{H}}}^{(\lfloor\frac{p}{2}\rfloor,\ddagger)}\right)(d)\right]\Bigg| \\
\leq & \Bigg|\Cov\left[f_{h}\left(\mathbf{U}_{\mathcal{H}}(d)\right)-f_h\left(\mathbf{U}_{\mathcal{H}}^{(\lfloor\frac{p}{2}\rfloor,\dagger)}(d)\right), f_{\tilde h}\left(\mathbf{U}_{\tilde{\mathcal{H}}}(d)\right)\right]\Bigg| \\
& + \Bigg|\Cov\left[ f_h\left(\mathbf{U}_{\mathcal{H}}^{(\lfloor\frac{p}{2}\rfloor,\dagger)}(d)\right), f_{\tilde h}\left(\mathbf{U}_{\tilde{\mathcal{H}}}(d)\right) - f_{\tilde h}\left(\mathbf{U}_{\tilde{\mathcal{H}}}^{(\lfloor\frac{p}{2}\rfloor,\ddagger)}(d)\right)\right]\Bigg| \\
\leq & 2\Big| ||f_{\tilde h}||_{\infty}\E\left[f_{h}\left(\mathbf{U}_{\mathcal{H}}(d)\right)-f_h\left(\mathbf{U}_{\mathcal{H}}^{(\lfloor\frac{p}{2}\rfloor,\dagger)}(d)\right)\right]\Big| + 2\Big| ||f_{ h}||_{\infty}\E\left[f_{\tilde h}\left(\mathbf{U}_{\tilde{\mathcal{H}}}(d)\right) - f_{\tilde h}\left(\mathbf{U}_{\tilde{\mathcal{H}}}^{(\lfloor\frac{p}{2}\rfloor, \ddagger)}(d)\right)\right]\Big| \\
\leq & 2||f_{\tilde h}||_{\infty}hL_h \theta_{N,\lfloor\frac{p}{2}\rfloor}(d) + 2||f_{h}||_{\infty}\tilde hL_{\tilde h} \theta_{N,\lfloor\frac{p}{2}\rfloor}(d).
\end{aligned} 
$$

As $Y_i$ is uniformly bounded, so is $U_i$. Then, Assumption 3.3 in \citet{kojevnikov2021limit} is satisfied. It is straightforward to verify that conditions in Theorem 1 imply Assumption 3.4 in \citet{kojevnikov2021limit}. The asymptotic normality of $\frac{1}{N}\sum_{i=1}^N U_i$ follows from Theorem 3.2 in \citet{kojevnikov2021limit}. Using the Cramer-Wold device, we know that $\frac{1}{N}\sum_{i=1}^N \mathbf{Z}_i(d)Y_i$ is also asymptotically normal:
$$
\frac{\left(\Sigma_Y(d)\right)^{-\frac{1}{2}}}{\sqrt N}\sum_{i=1}^N \left[\mathbf{Z}_i(d)Y_i - \E\left[\mathbf{Z}_i(d)Y_i\right]\right] \to \mathcal{N}(0, \mathbf{I}),
$$
Theorem~\ref{thm:IPTW-asym} can then be derived from Slutsky's Theorem, with $\Sigma_N(d)= \left(\Sigma_A\right)^{-1}\Sigma_Y(d)\left(\Sigma_A\right)^{-1}$.

Consider our simulated example in the main text with $t \in \{0, 1, 2\}$ and $\mathbf{m}\left(\mathbf{A}^{(1:T)}\right) = \left(1, A_{1}, A_{2}, A_{1}A_{2}\right)'$. Then,
$$
\left(\frac{1}{N}\sum_{i=1}^N w_i\mathbf{m}\left(\mathbf{A}_i^{(1:T)}\right) \mathbf{m}'\left(\mathbf{A}_i^{(1:T)}\right)\right)^{-1} \to \begin{pmatrix}
1 & -1 & -1 & 1 \\ -1 & 2 & 1 & -2 \\ -1 & 1 & 2 & -2 \\ 1 & -2 & -2 & 4
\end{pmatrix},
$$
while 
$$
\frac{1}{N}\sum_{i=1}^N \mathbf{Z}_i(d)Y_i \to \begin{pmatrix}
\mu\left((0, 0); d\right) + \mu\left((0, 1); d\right) + \mu\left((1, 0); d\right) + \mu\left((1, 1); d\right) \\
\mu\left((1, 0); d\right) + \mu\left((1, 1); d\right) \\
\mu\left((0, 1); d\right) + \mu\left((1, 1); d\right) \\
\mu\left((1, 1); d\right)
\end{pmatrix}.
$$
Therefore, $\hat \beta(d) \to \beta(d) = \left(\frac{1}{N}\sum_{i=1}^N \E\left[w_i\mathbf{m}\left(\mathbf{A}_i^{(1:T)}\right) \mathbf{m}'\left(\mathbf{A}_i^{(1:T)}\right)\right]\right)^{-1}\frac{1}{N}\sum_{i=1}^N \E\left[\mathbf{Z}_i(d)Y_i\right]$, where
$$
\beta(d) = \begin{pmatrix}
\mu\left((0, 0); d\right) \\
\mu\left((0, 1); d\right) - \mu\left((0, 0); d\right)\\
\mu\left((1, 0); d\right) - \mu\left((0, 0); d\right)\\
\mu\left((1, 1); d\right) - \mu\left((0, 1); d\right) - \mu\left((1, 0); d\right) + \mu\left((0, 0); d\right)
\end{pmatrix}.
$$

When the propensity score needs to be estimated from data, there is an additional part in the variance, driven by the uncertainty from estimating $e\left(\mathbf{A}_{j}^{(1:T)} = \mathbf{A}^{(1:T)}; \mathbf{V}_{j}^{(1:T)}\right)$. Its expression can be derived using the standard M-estimation theory, as in \citet{lunceford2004stratification}. We present the expression of the additional part when the propensity score is estimated via a logistic regression:
$$
P\left(A_{j}^{(t)} = 1 \mid \mathbf{V}_{j}^{(t)}\right) = \frac{\exp\left(\mathbf{V}_{j}^{(t)'}\gamma_t\right)}{1 + \exp\left(\mathbf{V}_{j}^{(t)'}\gamma_t\right)} = g\left(\mathbf{V}_{j}^{(t)};\gamma_t\right).
$$

We know that $\frac{}{}$ $\Sigma_{\gamma_t} = \Var\left[\hat \gamma_t\right] = \sum_{j = 1}^N \mathbf{V}_{j}^{(t)'}g\left(\mathbf{V}_{j}^{(t)};\gamma_t\right)\left(1 - g\left(\mathbf{V}_{j}^{(t)};\gamma_t\right)\right)\mathbf{V}_{j}^{(t)}$. Remember that the regression residual for unit $j$ from the WLS is denoted as $\hat \varepsilon_j$, and define $w_j^{SW} = \prod_{s=1}^T P\left(A_{j}^{(s)} = A^{(s)} \mid \mathbf{A}_{j}^{(1:(s-1))} = \mathbf{A}^{(1:(s-1))}\right)$. Further define $\mathbf{H}_t(d)$ as
$$
 \sum_{j = 1}^N \frac{w_j^{SW}}{\prod_{s\neq t} g\left(\mathbf{V}_{j}^{(s)};\gamma_s\right)}\left(\frac{\left(1 - g\left(\mathbf{V}_{j}^{(t)};\gamma_t\right)\right)A_{j}^{(t)}}{g\left(\mathbf{V}_{j}^{(t)};\gamma_t\right)} + \frac{g\left(\mathbf{V}_{j}^{(t)};\gamma_t\right)\left(1 - A_{j}^{(t)}\right)}{1 - g\left(\mathbf{V}_{j}^{(t)};\gamma_t\right)}\right)\hat \varepsilon_j\mathbf{V}_{j}^{(t)'}\mathbf{m}\left(\mathbf{A}_j^{(1:T)}\right).
$$
Then, the additional part in the variance equals
$$
\begin{aligned}
    \Sigma_{adj}(d) = - \sum_{t = 1}^T \mathbf{H}_t'(d) \Sigma_{\gamma_t}\mathbf{H}_t(d).
\end{aligned}
$$
The final variance equals $\Sigma_N(d) + \Sigma_{adj}(d)$.

To establish the convergence of $\hat{\boldsymbol{\beta}}(d)$ to $\boldsymbol{\beta}^{*}(d)$ defined in Section~\ref{app:inter}, we impose additional restrictions to bound the dependence in $Y_i$ induced by the dependence in $\mathbf{F}_{i}^{(1:T)}$. For any two $\sigma$-fields $\sigma$ and $\tilde \sigma$, define 
$$
\alpha\left(\sigma, \tilde \sigma\right) = \sup_{\mathcal{A} \in \sigma, \mathcal{B} \in \tilde \sigma}\bigg|\mathbbm{Cov}\left(\mathbf{1}\{\mathcal{A}\}, \mathbf{1}\{\mathcal{B}\}\right) \bigg|,
$$
where $\mathbbm{Cov}(\cdot)$ is defined using $\mathbb{E}[\cdot]$. For any two sets $\mathcal{H}, \tilde{\mathcal{H}} \in \mathcal{N}_N$ such that $d\left(\mathcal{H}, \tilde{\mathcal{H}}\right) > p$, we further define the strong mixing coefficients as
$$
\begin{aligned}
    \alpha_{N, p} = \sup_{\mathcal{H}, \tilde{\mathcal{H}}} \alpha\left(\sigma\left(\Big\{\mathbf{F}_{i}^{(1:T)}\Big\}_{i \in \mathcal{H}}\right), \sigma\left(\Big\{\mathbf{F}_{j}^{(1:T)}\Big\}_{j \in \tilde{\mathcal{H}}}\right)\right).
\end{aligned}
$$
We then impose the following assumption:
\begin{assumption}[Strong mixing of stochastic factors]\label{assum:wd}
\vspace{-1em}
$$
\sup_N \alpha_{N, p} \rightarrow 0,
$$
as \(p \rightarrow \infty\).\vspace{-0.5em}
\end{assumption}
Assumption~\ref{assum:wd} indicates that any dependence in the stochastic factors declines to zero as the proximity between two units increases. 

For any set $\mathcal{H}$, $\E\left[f_h\left(\mathbf{U}_{\mathcal{H}}^{(\lfloor\frac{p}{2}\rfloor,\dagger)}(d)\right)\mid \mathbf{F}_{\mathcal{H}}^{(1:T)}\right]$ is a random variable defined on $\sigma\left(\Big\{\mathbf{F}_{i}^{(1:T)}\Big\}_{i \in \mathcal{H}}\right)$, which is bounded by $\bigg|\bigg|\E\left[f_h\left(\mathbf{U}_{\mathcal{H}}^{(\lfloor\frac{p}{2}\rfloor,\dagger)}(d)\right)\mid \mathbf{F}_{\mathcal{H}}^{(1:T)}\right]\bigg|\bigg|_{\infty}$. This quantity is finite since $Y_i$ is uniformly bounded. Then, under Assumption~\ref{assum:wd}, when $p > 2*d$, we have
$$
\begin{aligned}
& \mathbbm{Cov}\left[f_h\left(\mathbf{U}_{\mathcal{H}}^{(\lfloor\frac{p}{2}\rfloor,\dagger)}(d)\right), f_{\tilde h}\left(\mathbf{U}_{\tilde{\mathcal{H}}}^{(\lfloor\frac{p}{2}\rfloor,\ddagger)}(d)\right)\right] = \mathbb{E}\left[\Cov\left[f_h\left(\mathbf{U}_{\mathcal{H}}^{(\lfloor\frac{p}{2}\rfloor,\dagger)}(d)\right), f_{\tilde h}\left(\mathbf{U}_{\tilde{\mathcal{H}}}^{(\lfloor\frac{p}{2}\rfloor,\ddagger)}(d)\right)\right]\right] \\
& + \mathbbm{Cov}\left[\E\left[f_h\left(\mathbf{U}_{\mathcal{H}}^{(\lfloor\frac{p}{2}\rfloor,\dagger)}(d)\right)\mid \mathbf{F}_{\mathcal{H}}^{(1:T)}\right], \E\left[f_{\tilde h}\left(\mathbf{U}_{\tilde{\mathcal{H}}}^{(\lfloor\frac{p}{2}\rfloor,\ddagger)}(d)\right) \mid \mathbf{F}_{\tilde{\mathcal{H}}}^{(1:T)}\right]\right] \\
& \leq 4\bigg|\bigg|\E\left[f_h\left(\mathbf{U}_{\mathcal{H}}^{(\lfloor\frac{p}{2}\rfloor,\dagger)}(d)\right)\mid \mathbf{F}_{\mathcal{H}}^{(1:T)}\right]\bigg|\bigg|_{\infty}\bigg|\bigg|\E\left[f_{\tilde h}\left(\mathbf{U}_{\tilde{\mathcal{H}}}^{(\lfloor\frac{p}{2}\rfloor,\ddagger)}(d)\right)\mid \mathbf{F}_{\tilde{\mathcal{H}}}^{(1:T)}\right]\bigg|\bigg|_{\infty} \alpha_{N, p} = \tilde \alpha_{N, p} \to 0.
\end{aligned}
$$
The last inequality uses Theorem 9 in \citet{prakasa2009conditional}. Consequently, we can similarly show that
$$
\begin{aligned}
& \Bigg|\mathbbm{Cov}\left[f_h\left(\mathbf{U}_{\mathcal{H}}(d)\right), f_{\tilde h}\left(\mathbf{U}_{\tilde{\mathcal{H}}}(d)\right)\right]\Bigg| \\
\leq & 2||f_{\tilde h}||_{\infty}hL_h \theta_{N,\lfloor\frac{p}{2}\rfloor}(d) + 2||f_{h}||_{\infty}\tilde hL_{\tilde h} \theta_{N,\lfloor\frac{p}{2}\rfloor}(d) + \tilde \alpha_{N, p}.
\end{aligned} 
$$
Assumption 3.3 in \citet{kojevnikov2021limit} is thus satisfied, and consistency and asymptotic normality follow.

\subsection{Proof for Theorem~\ref{thm:hac}} \label{app:thm2}
First, note that $\frac{\mathbf{M}'\mathbf{W}\mathbf{M}}{N} \to \Sigma_A$ as $N \to \infty$ by Lemma~\ref{lemma:ind-A} and standard theory. Define $\mathbf{Q}'_i(d) = \sum_{k=1}^N\frac{\mathbf{1}\{i \in \Omega_k(d)\}w_k\mathbf{m}\left(\mathbf{A}_k^{(1:T)}\right)\mathbf{m}'\left(\mathbf{A}_k^{(1:T)}\right)}{|\Omega_k(d)|}$. Then, $\hat{\boldsymbol{\varepsilon}}_{Yi}(d) = \mathbf{Z}_i(d)Y_i - \mathbf{Q}'_i(d)\hat{\boldsymbol{\beta}}(d)$. It is straightforward to verify that $\frac{1}{N}\sum_{i=1}^N\mathbf{Q}'_i(d) = \frac{1}{N}\sum_{i=1}^N w_i\mathbf{m}\left(\mathbf{A}_i^{(1:T)}\right) \mathbf{m}'\left(\mathbf{A}_i^{(1:T)}\right)$, which implies that $\boldsymbol{\beta}(d) = \left(\frac{1}{N}\sum_{i=1}^N\E\left[\mathbf{Q}'_i(d)\right]\right)^{-1}\left(\frac{1}{N}\sum_{i=1}^N\E\left[\mathbf{Z}_i(d)Y_i\right]\right)$ and $\hat{\boldsymbol{\beta}}(d) = \left(\frac{1}{N}\sum_{i=1}^N\mathbf{Q}'_i(d)\right)^{-1}\left(\frac{1}{N}\sum_{i=1}^N\mathbf{Z}_i(d)Y_i\right)$. Next, we prove the following lemma concerning the terms $\{\mathbf{Z}_i(d)Y_i\}_{i \in \mathcal{N}_N}$ and $\{\mathbf{Q}_i(d)\}_{i \in \mathcal{N}_N}$:
\begin{lemma}\label{lemma:cov}
Under Assumptions~\ref{assum:si}-\ref{assum:nui}, 
$$
\begin{aligned}
    & \frac{1}{N}\sum_{i=1}^N\sum_{j:\{d_{ij} \leq \bar d_N\}} \kappa_i \left[\mathbf{Q}_i(d) - \E\left[\mathbf{Q}_i(d)\right]\right] \to 0, \frac{1}{N}\sum_{i=1}^N\sum_{j:\{d_{ij} \leq \bar d_N\}} \kappa_i \left[\mathbf{Z}_i(d)Y_i - \E\left[\mathbf{Z}_i(d)Y_i\right]\right] \to 0,
\end{aligned}
$$
for any uniformly bounded weights $\{\kappa_i\}_{i \in \mathcal{N}_N}$ and units $i, j \in \mathcal{N}_N$.
\end{lemma}

\textit{Proof:} See the proof of Theorem 4 in \citet{leung2022causal}. Note that $\Cov\left[\mathbf{Q}_i(d), \mathbf{Q}'_j(d)\right] = 0$ if $d_{ij} \geq 2*d$.

Using the same derivation as in Section 4 and standard theory on estimating equations, we have
$$
\begin{aligned}
    \Sigma_Y(d) = & \frac{1}{N} \sum_{i=1}^N \sum_{j:\{d_{ij} \leq \bar d_N\}} \E\left[\left[\boldsymbol{\varepsilon}_{Yi}(d) - \E\left[\boldsymbol{\varepsilon}_{Yi}(d)\right]\right] \left[\boldsymbol{\varepsilon}_{Yj}(d) - \E\left[\boldsymbol{\varepsilon}_{Yj}(d)\right]\right]\right]' ,
\end{aligned}
$$
where $\boldsymbol{\varepsilon}_{Yi}(d) = \mathbf{Z}_i(d)Y_i - \mathbf{Q}'_i(d)\boldsymbol{\beta}(d)$. Note that $\E\left[\boldsymbol{\varepsilon}_{Yi}(d)\right] \neq 0$ as the outcomes are not identically distributed. Let's define
$$
\begin{aligned}
    & \widehat \Sigma_Y(d) = \frac{1}{N} \hat{\boldsymbol{\varepsilon}}'_{Y}(d)\mathbf{K}_{\bar d_N}\hat{\boldsymbol{\varepsilon}}_{Y}(d) = \frac{1}{N} \sum_{i=1}^N \sum_{j:\{d_{ij} \leq \bar d_N\}} \hat{\boldsymbol{\varepsilon}}_{Yi}(d) \hat{\boldsymbol{\varepsilon}}'_{Yj}(d),  \\
    & \tilde \Sigma_Y(d) = \frac{1}{N} \sum_{i=1}^N \sum_{j:\{d_{ij} \leq \bar d_N\}} \left[\boldsymbol{\varepsilon}_{Yi}(d) - \E\left[\boldsymbol{\varepsilon}_{Yi}(d)\right]\right] \left[\boldsymbol{\varepsilon}_{Yj}(d) - \E\left[\boldsymbol{\varepsilon}_{Yj}(d)\right]\right]', \text{ and }\\
    & \check \Sigma_Y(d) = \frac{1}{N} \sum_{i=1}^N \sum_{j:\{d_{ij} \leq \bar d_N\}} \boldsymbol{\varepsilon}_{Yi}(d)\boldsymbol{\varepsilon}'_{Yj}(d).
\end{aligned}
$$
\citet{kojevnikov2021limit} show that $\tilde \Sigma_Y(d) \to \Sigma_Y(d)$ as $N \to \infty$. Let $\mathbf{R}(d) = \frac{1}{N} \sum_{i=1}^N \sum_{j:\{d_{ij} \leq \bar d_N\}}\E\left[\boldsymbol{\varepsilon}_{Yi}(d)\right]\E\left[\boldsymbol{\varepsilon}_{Yj}'(d)\right]$. Then, by Lemma~\ref{lemma:cov},
$$
\begin{aligned}
    & \check \Sigma_Y(d) - \tilde \Sigma_Y(d) \\
    = & \frac{1}{N} \sum_{i=1}^N \sum_{j:\{d_{ij} \leq \bar d_N\}} \left(\boldsymbol{\varepsilon}_{Yi}(d) \E\left[\boldsymbol{\varepsilon}_{Yj}'(d)\right] + \E\left[\boldsymbol{\varepsilon}_{Yi}(d)\right]\boldsymbol{\varepsilon}'_{Yj}(d) - \E\left[\boldsymbol{\varepsilon}_{Yi}(d)\right]\E\left[\boldsymbol{\varepsilon}_{Yj}'(d)\right]\right) \\
    = & \frac{1}{N} \sum_{i=1}^N \sum_{j:\{d_{ij} \leq \bar d_N\}} \left(\boldsymbol{\varepsilon}_{Yi}(d) - \E\left[\boldsymbol{\varepsilon}_{Yi}(d)\right]\right)\E\left[\boldsymbol{\varepsilon}_{Yj}'(d)\right] \\
    & + \frac{1}{N} \sum_{i=1}^N \sum_{j:\{d_{ij} \leq \bar d_N\}} \E\left[\boldsymbol{\varepsilon}_{Yi}(d)\right]\left(\boldsymbol{\varepsilon}_{Yj}(d) - \E\left[\boldsymbol{\varepsilon}_{Yj}(d)\right]\right)' + \frac{1}{N} \sum_{i=1}^N \sum_{j:\{d_{ij} \leq \bar d_N\}}\E\left[\boldsymbol{\varepsilon}_{Yi}(d)\right]\E\left[\boldsymbol{\varepsilon}_{Yj}'(d)\right] \\
    = & \frac{1}{N} \sum_{i=1}^N \left[\mathbf{Z}_i(d)Y_i - \E\left[\mathbf{Z}_i(d)Y_i\right]\right]\left(\sum_{j = 1}^N \mathbf{1}\{d_{ij} \leq \bar d_N\}\right) \E\left[\boldsymbol{\varepsilon}_{Yj}'(d)\right] \\
    & + \frac{1}{N} \sum_{i=1}^N \left(\sum_{j = 1}^N \mathbf{1}\{d_{ij} \leq \bar d_N\}\right) \E\left[\boldsymbol{\varepsilon}_{Yi}(d)\right]\left[\mathbf{Z}_j(d)Y_j - \E\left[\mathbf{Z}_j(d)Y_j\right]\right]' \\
    & + \frac{1}{N} \sum_{i=1}^N \left[\mathbf{Q}_i(d) - \E\left[\mathbf{Q}_i(d)\right]\right]\left(\sum_{j = 1}^N \mathbf{1}\{d_{ij} \leq \bar d_N\}\right) \E\left[\boldsymbol{\varepsilon}_{Yj}'(d)\right] \\
    & + \frac{1}{N} \sum_{i=1}^N \left(\sum_{j = 1}^N \mathbf{1}\{d_{ij} \leq \bar d_N\}\right) \E\left[\boldsymbol{\varepsilon}_{Yi}(d)\right]\left[\mathbf{Q}_j(d) - \E\left[\mathbf{Q}_j(d)\right]\right]' + \mathbf{R}(d) \to \mathbf{R}(d) \\
\end{aligned}
$$
Moreover,
$$
\begin{aligned}
    & \mid\widehat \Sigma_Y(d) - \check \Sigma_Y(d)\mid \\
    = & \Bigg| \frac{1}{N} \sum_{i=1}^N \sum_{j:\{d_{ij} \leq \bar d_N\}} \left(\mathbf{Q}'_i(d)\left(\boldsymbol{\beta}(d) + \hat{\boldsymbol{\beta}}(d)\right)/2 - \mathbf{Z}_i(d)Y_i \right)\left(\hat{\boldsymbol{\beta}}(d) - \boldsymbol{\beta}(d)\right)'\mathbf{Q}_j(d) \\
    & + \frac{1}{N} \sum_{i=1}^N \sum_{j:\{d_{ij} \leq \bar d_N\}} \mathbf{Q}'_i(d)\left(\hat{\boldsymbol{\beta}}(d) - \boldsymbol{\beta}(d)\right)\left(\left(\boldsymbol{\beta}(d) + \hat{\boldsymbol{\beta}}(d)\right)'\mathbf{Q}_j(d)/2 - \mathbf{Z}_i(d)Y_i \right)\Bigg| \\
    \leq & \frac{1}{N} \sum_{i=1}^N \sum_{j:\{d_{ij} \leq \bar d_N\}} \frac{C}{\sqrt N} = \frac{CM_{N}\left(\bar d_N, 1\right)}{\sqrt N} \to 0.
\end{aligned}
$$
Combining these results, we conclude that $\widehat \Sigma_Y(d) \to \Sigma_Y(d) + \mathbf{R}(d)$.

Finally, note that $\widehat \Sigma^{+}_Y(d) = \widehat \Sigma_Y(d) + \widehat \Sigma^{-}_Y(d)$. We can similarly show that
$$
\begin{aligned}
    & \widehat \Sigma^{-}_Y(d) \to \Sigma^{-}_Y(d) + \mathbf{R}^{-}(d).
\end{aligned}
$$
Therefore,
$$
\begin{aligned}
    \widehat \Sigma^{+}_Y(d) \to & \Sigma_Y(d) + \mathbf{R}(d) + \Sigma^{-}_Y(d) + \mathbf{R}^{-}(d) \\
    = & \Sigma_Y(d) + \mathbf{R}^{+}(d) + \Sigma^{-}_Y(d) \succeq \Sigma_Y(d).
\end{aligned}
$$

\newpage
\renewcommand{\thesection}{C} 
\renewcommand{\thesubsection}{\thesection.\arabic{subsection}}  
\setcounter{section}{0}
\section{Extra Evidence from Simulation and Applications} \label{app:evi}
In this section, we present the results from our simulation study along with additional findings from the two empirical applications.
\subsection{Examples of $d_{ij}$ and $\Omega_j(d)$} \label{app:exp}
\begin{table}[H]
\begin{center}
    \caption{Examples of $d_{ij}$ and $\Omega_j(d)$} \label{tab:proximity}
\begin{tabular}{|c|c|}
  \hline
  $d_{ij}$ & $\Omega_j(d)$ \\
  \hline
  Euclidean or geodesic distance & Donut: $\{i \in \mathcal{N}: d - \kappa < d_{ij} \leq d\}$  \\
  \hline
  $d_{ij} =  \begin{cases} 
    1, & \text{same street/block} \\ 
    2, & \text{same district/town} \\ 
    3, & \text{otherwise} 
  \end{cases}$ & $\{i \in \mathcal{N}: i \text{ belongs to } j\text{'s block or cluster} \}$ \\
  \hline
  Travel time or accessibility & Disk: $\{i \in \mathcal{N}: d_{ij} \leq d\}$ \\
  \hline
  Shortest path length in a graph & $d$th-degree neighbors: $\{i \in \mathcal{N}: d_{ij} = d\}$ \\
  \hline
  Cultural similarity or social distance & Conditional set: $\{i \in \mathcal{N}: d_{ij} = d, \mathbf{X}_i = \mathbf{x}\}$ \\
  \hline
\end{tabular}
\end{center}
\vspace{0.5em}
\textit{Notes:} This table lists common choices for the proximity metric $d_{ij}$ (left column) and the set $\Omega_j(d)$ (right column). The two columns are not aligned row-by-row: each entry illustrates a standalone option and may be combined with others in practice. $\mathbf{X}_i$ in the final row denotes covariates of unit $i$, which allows researchers to define neighborhoods conditioned on observable characteristics and estimate conditional causal effects.
\end{table}

\subsection{Simulation Evidence}\label{simulation}
In this section, we test the performance of the proposed method using the simulated dataset examined in the main text, where $N = 400$ and $T = 2$. We focus on the setting in which the units are embedded in a social network (top-left plot of Figure~\ref{fig:space}), generated by a random geometric graph (RGG) model \citep{penrose2003random, leung2022causal}. Let $G_{ij}$ indicate the presence of an edge between units $i$ and $j$. Then $G_{ij} = \mathbf{1}\Big\{||\alpha_i - \alpha_j|| \leq \sqrt{\frac{8}{\pi N}}\Big\}$, where $\alpha_i = (\alpha_{1i}, \alpha_{2i})'$ and each component is independently drawn from a uniform distribution on $[0, 1]$. This specification yields a network $\mathcal{G}_N$ with an expected degree of 8.

We assume that there is one time-varying confounder in each $t \in \{1, 2\}$ that takes the form of the outcome's value in the previous period: $L_{i}^{(t)} = Y_{i}^{(t-1)}$, and the final outcome $Y_i = Y_{i}^{(2)}$. In the absence of any treatment, the potential values of $Y_{i}^{(t)}$ are generated from the following process:
$$
\begin{aligned}
    & Y_{i}^{(t)}\left(\mathbf{0}_{\mathcal{N}}^{(1:2)}\right) = 0.3 + \xi^{(t)} + \varepsilon_{i}^{(t)}, 
\end{aligned}
$$
where $\xi^{(t)}$ is the period-specific intercept and $\varepsilon_{i}^{(t)}$ is the idiosyncratic error term for unit $i$ in period $t$. Both $\xi^{(t)}$ and $\varepsilon_{i}^{(t)}$ are independently drawn from the standard normal distribution. In each period $t$, unit $j$'s treatment generates a contemporaneous effect on unit $i$'s outcome with the magnitude of
$$
\begin{aligned}
    & g\left(d_{ij}, A_{i}^{(t)}, A_{j}^{(t)}\right) = A_{j}^{(t)} \rho_{1i} \exp\left(-\frac{d_{ij}^2}{3}\right)  + A_{i}^{(t)}A_{j}^{(t)}\rho_{2i} \exp\left(-\frac{d_{ij}^2}{3}\right),
\end{aligned}
$$
where $\rho_{1i} = \frac{\alpha_{1i}}{2} + 1$ and $\rho_{2i} = \frac{\alpha_{2i}}{2}$ are unit-specific constants governing treatment effect heterogeneity. The effect's magnitude depends on the proximity between two units and their treatment statues. It decreases exponentially as $d$ increases. We can verify that conditions required for Theorems~\ref{thm:IPTW-asym} and~\ref{thm:hac} are satisfied in the data-generating process.

The effect received by each unit $i$ in period $t$ consists of four components: the average of contemporaneous effects from all the units in the sample, carryover effects from the previous period, influence of its own outcome (time-varying confounder) in the previous period, and the direct impacts of its first-degree neighbors' outcome in the previous period (contagion):
$$
\begin{aligned}
\tau_{i}^{(t)}\left(\mathbf{A}_{\mathcal{N}}^{(1:2)}\right) = & \underbrace{\sum_{d = 0}^{d_{\max}}\frac{\sum_{j=1}^{N} \mathbf{1}\{d_{ij} = d\} g\left(d_{ij}, A_{i}^{(t)}, A_{j}^{(t)}\right)}{\sum_{j=1}^{N} \mathbf{1}\{d_{ij} = d\}}}_{\text{Contemporaneous effects}} + \underbrace{0.5 * \tau_{i}^{(t-1)}\left(\mathbf{A}_{\mathcal{N}}^{(1:2)}\right)}_{\text{Carryover effects}} \\
& + 0.4*Y_{i}^{(t-1)} + \underbrace{0.1*\frac{\sum_{j=1}^{N} \mathbf{1}\{d_{ij} = 1\} Y_{j}^{(t-1)}}{\sum_{j=1}^{N} \mathbf{1}\{d_{ij} = 1\}}}_{\text{Congation}}.
\end{aligned}
$$
As mentioned in the main text, the carryover effect equals $50\%$ of the average contemporaneous effect in the previous period. Then, the observed value of unit $i$'s outcome in period $t$ equals 
$$
\begin{aligned}
    & Y_{i}^{(t)} = Y_{i}^{(t)}\left(\mathbf{A}_{\mathcal{N}}^{(1:2)}\right) = Y_{i}^{(t)}\left(\mathbf{0}_{\mathcal{N}}^{(1:2)}\right) + \tau_{i}^{(t)}\left(\mathbf{A}_{\mathcal{N}}^{(1:2)}\right).
\end{aligned}
$$
Structural equations in Section~\ref{app:inter} can be expressed as $\mathbf{L}_{i}^{(t)} = f_{\mathbf{L}}\left(\mathbf{A}_{\mathcal{N}_{\mathcal{G}_N}(i)}^{1:(t-1)}, \mathbf{L}_{\mathcal{N}_{\mathcal{G}_N}(i)}^{1:(t-1)}, \varepsilon_{\mathbf{L}i}^{(t)}\right) = Y_{i}^{(t-1)}\left(\mathbf{0}_{\mathcal{N}}^{(1:2)}\right) + \tau_{i}^{(t-1)}\left(\mathbf{A}_{\mathcal{N}}^{(1:2)}\right)$ and $Y_{i} = f_{Y}\left(\mathbf{A}_{\mathcal{N}_{\mathcal{G}_N}(i)}^{(1:T)}, \mathbf{L}_{\mathcal{N}_{\mathcal{G}_N}(i)}, \varepsilon_{Yi}\right) = Y_{i}^{(2)}\left(\mathbf{0}_{\mathcal{N}}^{(1:2)}\right) + \tau_{i}^{(2)}\left(\mathbf{A}_{\mathcal{N}}^{(1:2)}\right)$. ${\mathcal{N}_{\mathcal{G}_N}(i)}$'s form depends on the effect's magnitude for each unit $i$, which is in turn decided by the idiosyncratic $\rho_i$.

In each period $t \in \{1, 2\}$, unit $i$'s propensity score is generated by a logistical regression function of its own outcome and treatment status in the previous period:
$$
P(A_{i}^{(t)} = 1) = Logit\left(-0.1 + 0.2*Y_{i}^{(t-1)} - 0.6*A_{i}^{(t-1)} + \nu_{i}^{(t)}\right),
$$
where $\nu_{it}$ is drawn from a normal distribution with mean zero and standard deviation of $0.5$. Sequential exchangeability is clearly satisfied in this setting. As mentioned in the main text, we repeat the assignment process $1,000$ times and approximate each AMR using the average over the simulations. On average, the number of units under treatment histories $(0, 0)$, $(1, 0)$, $(0, 1)$ and $(1, 1)$ is $61$, $91$, $114$, and $134$, respectively.

\begin{figure}[htp]
 \begin{center}
 \caption{Bias of the Proposed Method}
 \label{fig:IPTW-bias}
\includegraphics[width=\linewidth, height=\linewidth]{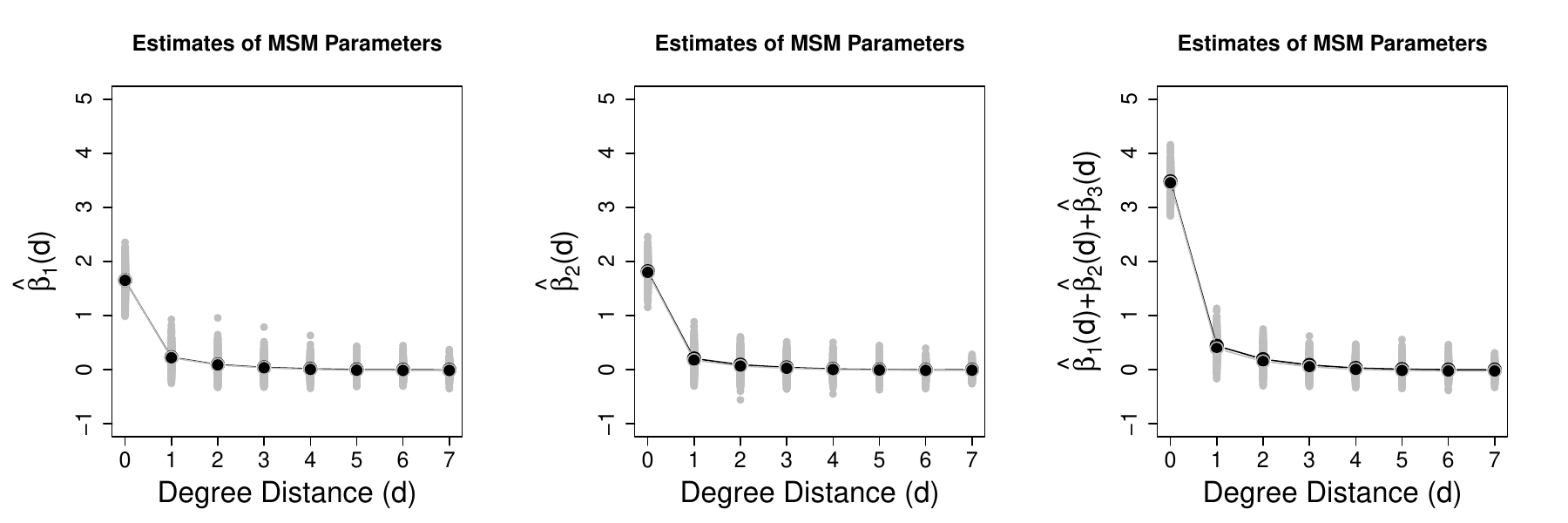}
 \end{center}
\textit{Notes:} These plots show the bias of the WLS estimator for three combinations of MSM parameters. Black dots indicate the simulated true values of the causal effects at each proximity level, while gray dots represent estimates from each of the $1,000$ treatment assignments.
\end{figure}

For each unit $i$, its transformed outcome at each $d \in \{0,1,\dots,7\}$ is constructed as the average outcome across its $d$th-degree neighbors in the network. The propensity scores are estimated via logistic regressions fitted for each period. In each of the $1,000$ assignments, we rely on WLS to obtain $\hat{\boldsymbol{\beta}}(d)$ for each $d$ and plot them against their simulated true values in Figure \ref{fig:IPTW-bias}. Despite the moderate sample size ($400$ units), the bias is negligible, confirming the consistency of the proposed estimator.

\begin{figure}[htp]
 \begin{center}
 \caption{Asymptotic Performance of the Proposed Method}
 \label{fig:wls-mse-coverage}
\includegraphics[width=\linewidth, height=\linewidth]{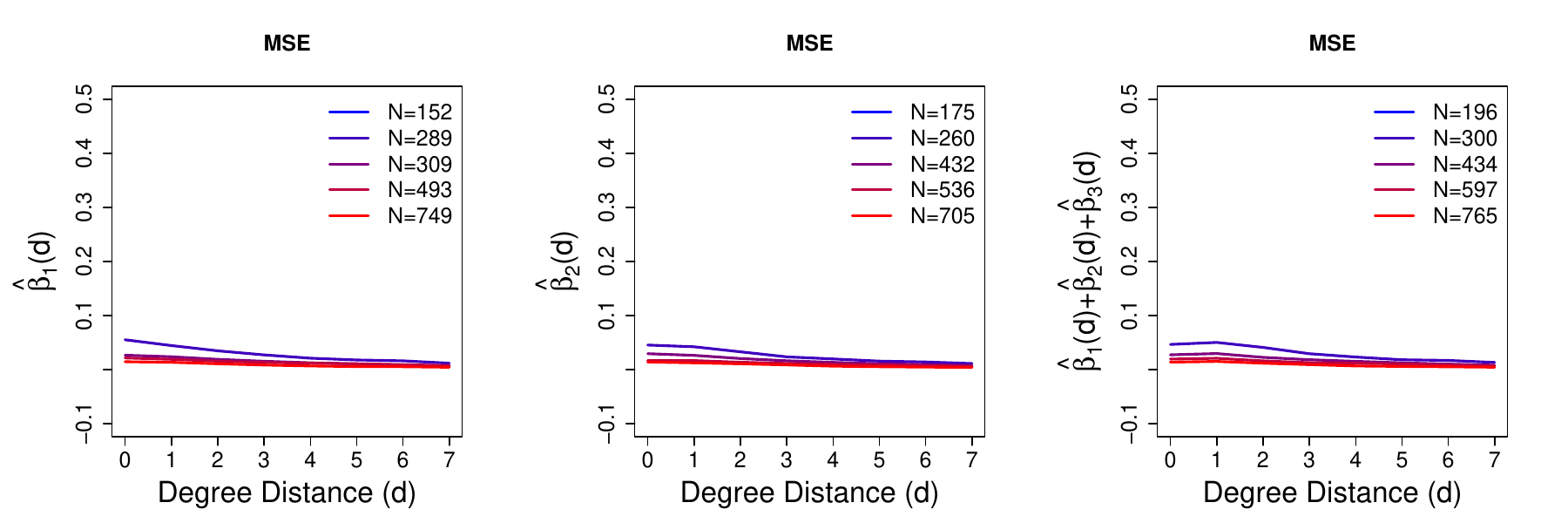}
\includegraphics[width=\linewidth, height=\linewidth]{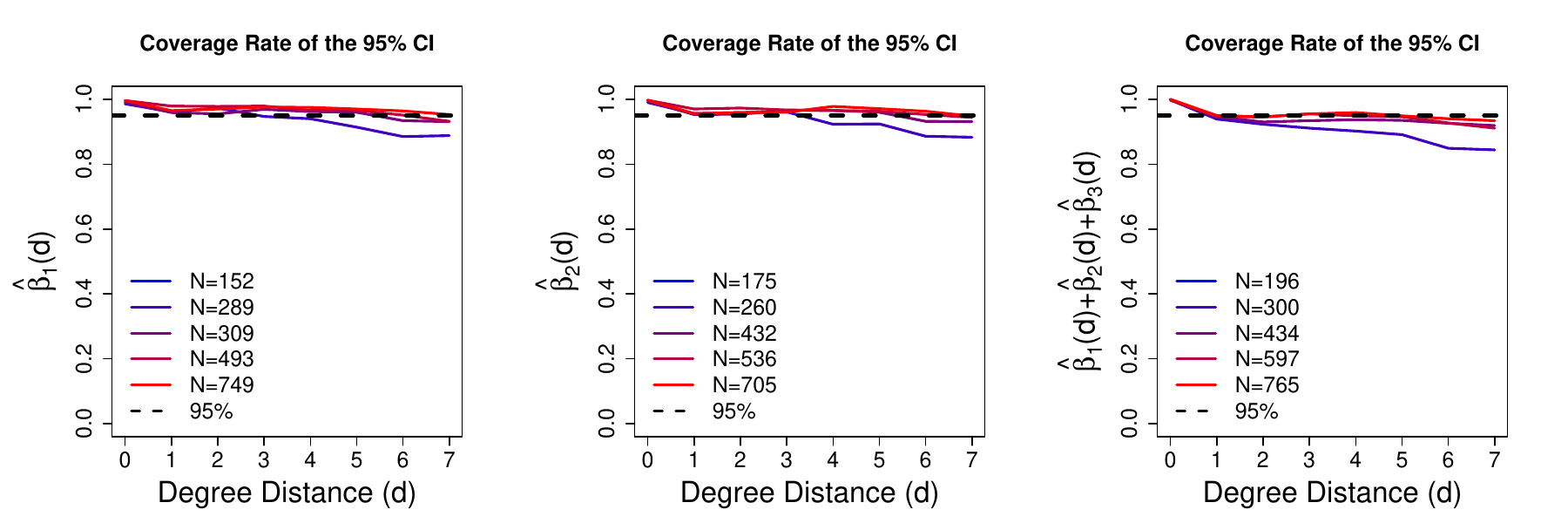}
 \end{center}
\textit{Notes:} The top panels display the average mean squared error (MSE) of the WLS estimator across $1,000$ treatment assignments, while the bottom panels show the coverage rates of the proposed 95\% confidence intervals. Different colors correspond to different numbers of units used in estimation.
\end{figure}

We further investigate how the performance of the method changes with an increasing sample size: $N \in \{400, 625, 900, 1225, 1600\}$. The top row of Figure~\ref{fig:wls-mse-coverage} shows the average mean squared error (MSE) of the estimates for each $d$ and $N$, while the bottom row displays the coverage rates. As the plots indicate, MSEs decline across all proximity levels as $N$ grows, and the coverage rates approach or exceed the nominal 95\% level when the sample size is sufficiently large.

\subsection{Extra Results from the Applications} \label{app:Stokes}
\begin{figure}[htp]
 \begin{center}
 \caption{Placebo Tests}
    \label{fig:app-p}
 \begin{subfigure}[t]{\textwidth}
    \caption{Placebo Tests in the Replication of \citet{stokes2016electoral}}
    \includegraphics[width=.48\linewidth, height=.48\linewidth]{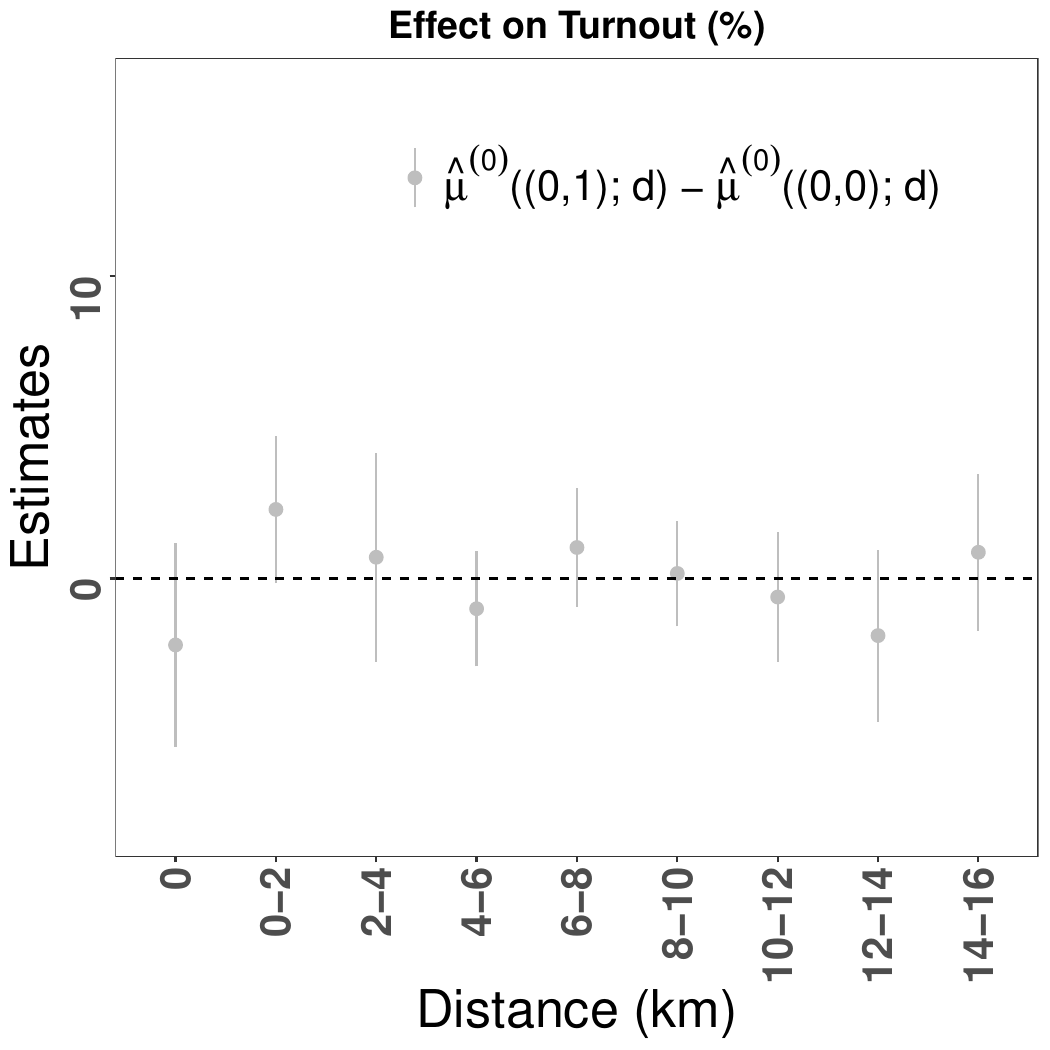}
    \includegraphics[width=.48\linewidth, height=.48\linewidth]{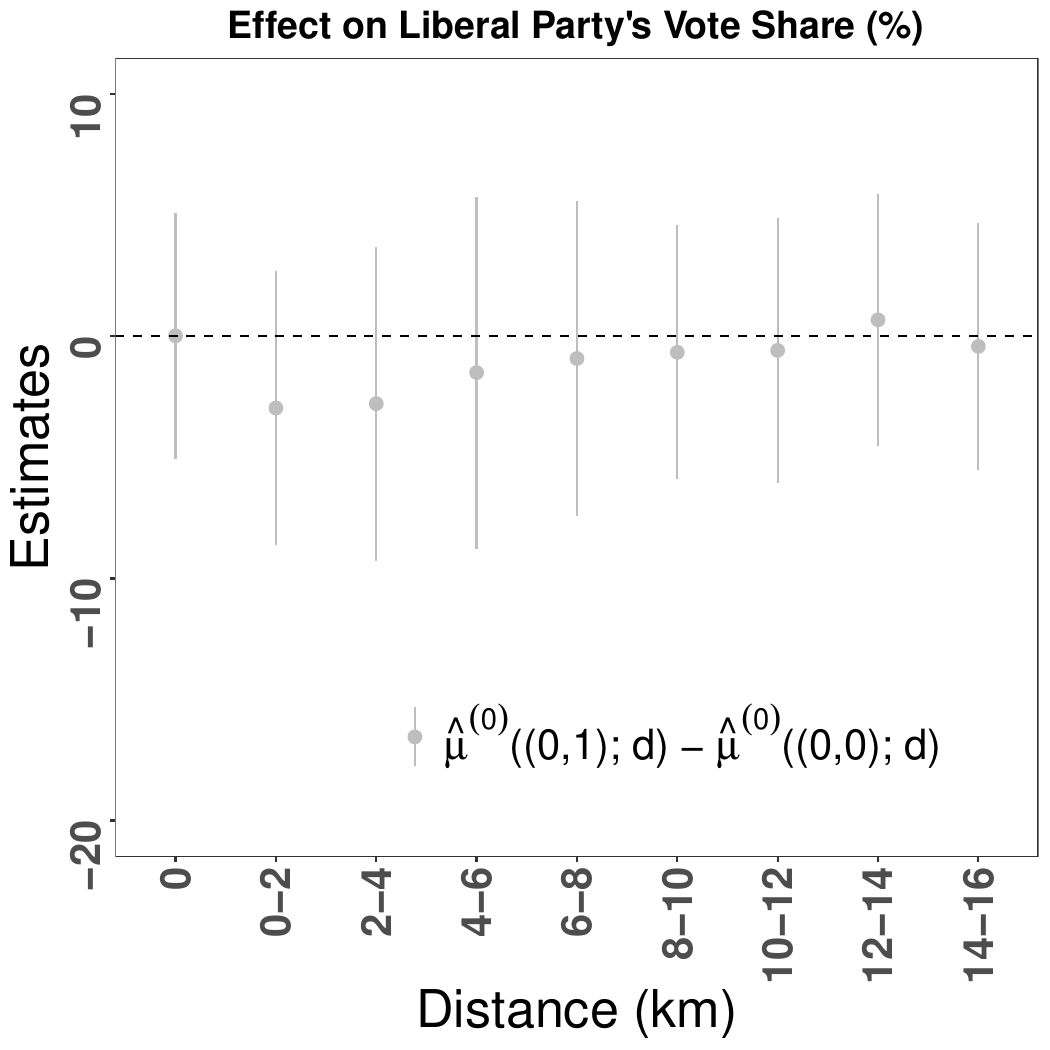}
 \end{subfigure}
 \vspace{1em} 
 \begin{subfigure}[t]{\textwidth}
    \caption{Placebo Tests in the FHS Analysis}
    \includegraphics[width=.48\linewidth, height=.48\linewidth]{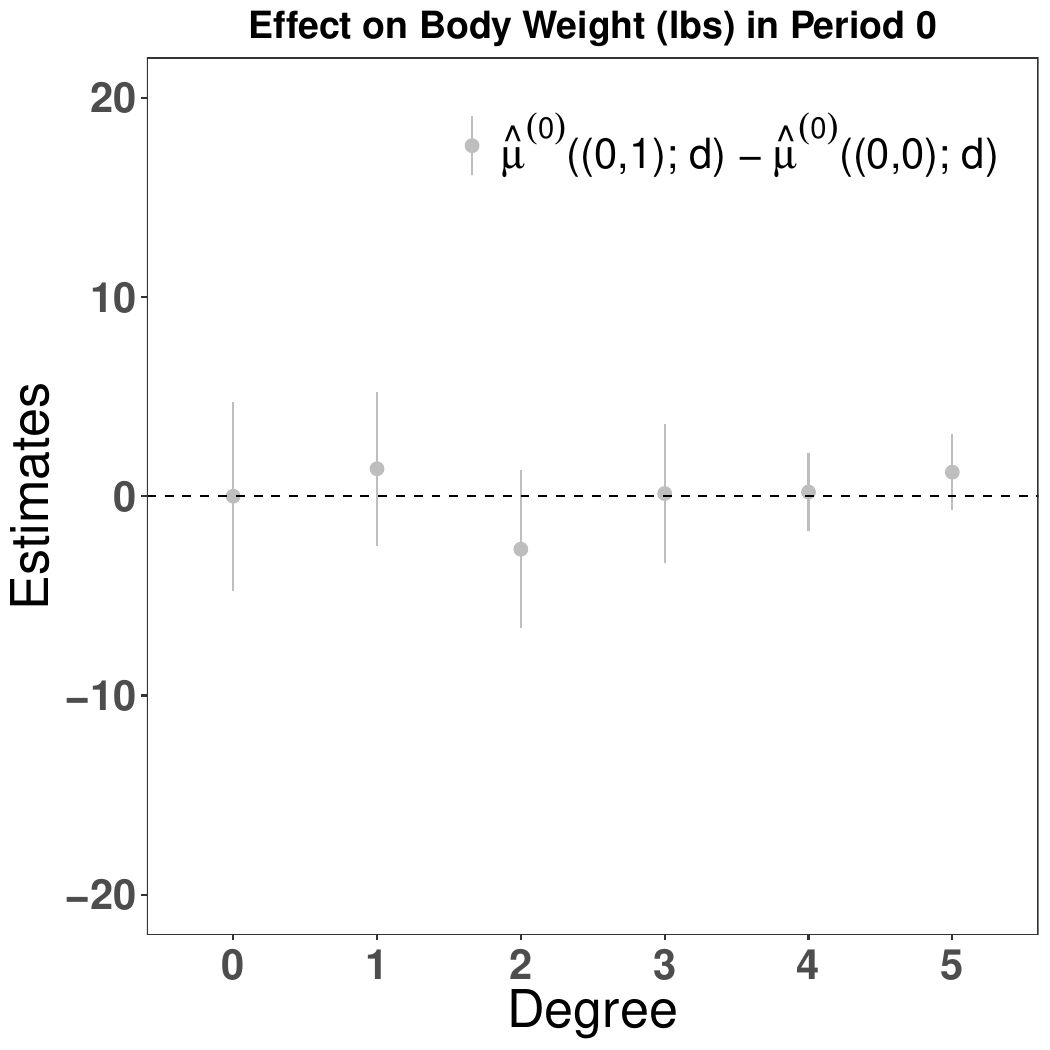}
    \includegraphics[width=.48\linewidth, height=.48\linewidth]{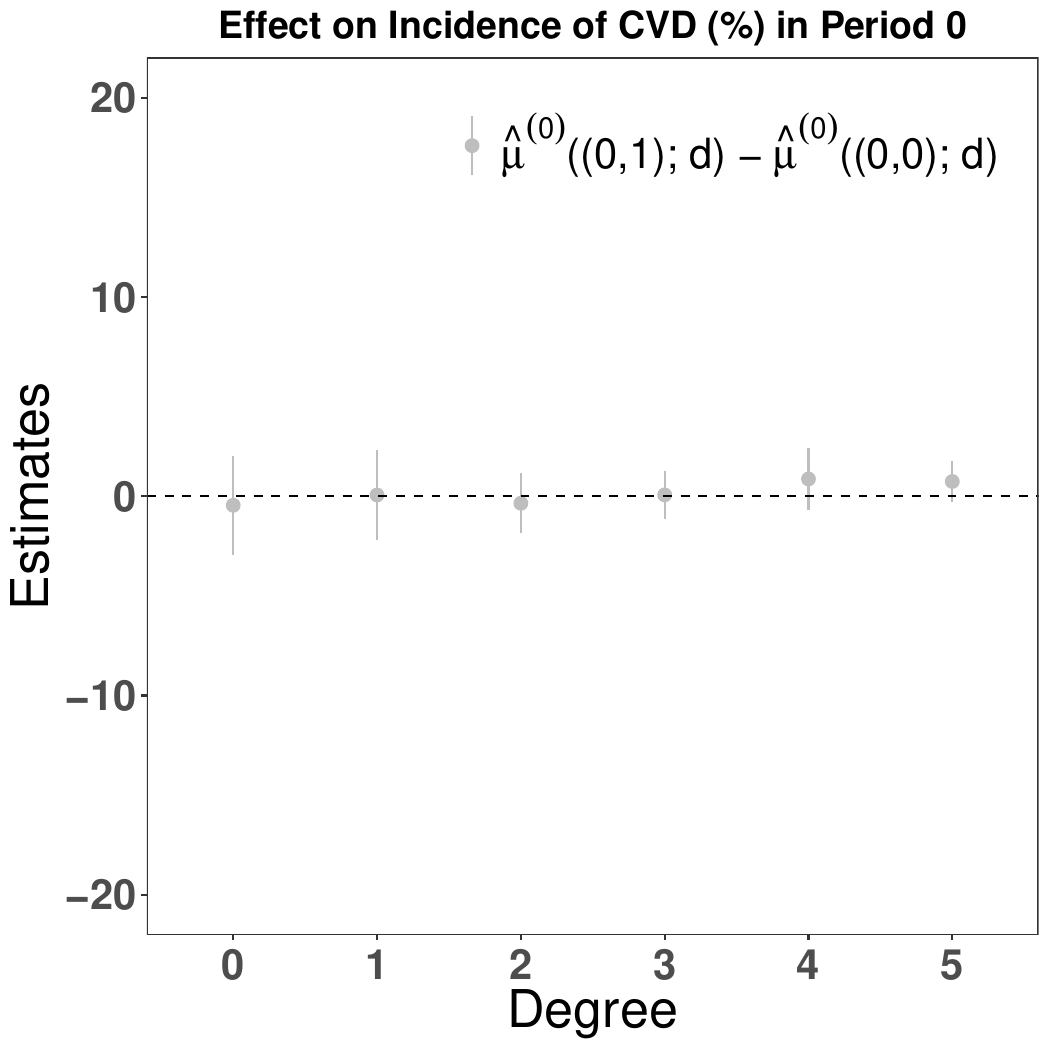}
 \end{subfigure}
 \end{center}
\textit{Notes:} Panel (a) presents results from conducting placebo tests based on \citet{stokes2016electoral}, while Panel (b) displays the results from the FHS. Black and gray dots represent point estimates for different treatment histories and proximity levels. The vertical segments show the corresponding 95\% confidence intervals, computed using the spatial/network HAC variance estimator and standard normal critical values.
\end{figure}

To assess the validity of our identification assumptions in the two applications, we conduct placebo tests as described in Section~\ref{app:extension}, estimating the effect of treatment history $(0, 1)$ on outcomes in period $0$. The main analysis assumes sequential exchangeability conditional on each unit's treatment and outcomes from the previous period, implying that outcomes in period $0$ should not be influenced by the treatment status in period $2$. The results are presented in Figure~\ref{fig:app-p}. Consistent with our expectation, there is no significant effect on any of the four outcomes studied in the main text, thereby supporting the identification assumptions.

\begin{figure}[htp]
 \begin{center}
 \caption{Results with Diffusion in Treatment}
    \label{fig:app-d}
 \begin{subfigure}[t]{\textwidth}
    \caption{Political Consequences of Proposed Wind Turbines}
    \includegraphics[width=.48\linewidth, height=.48\linewidth]{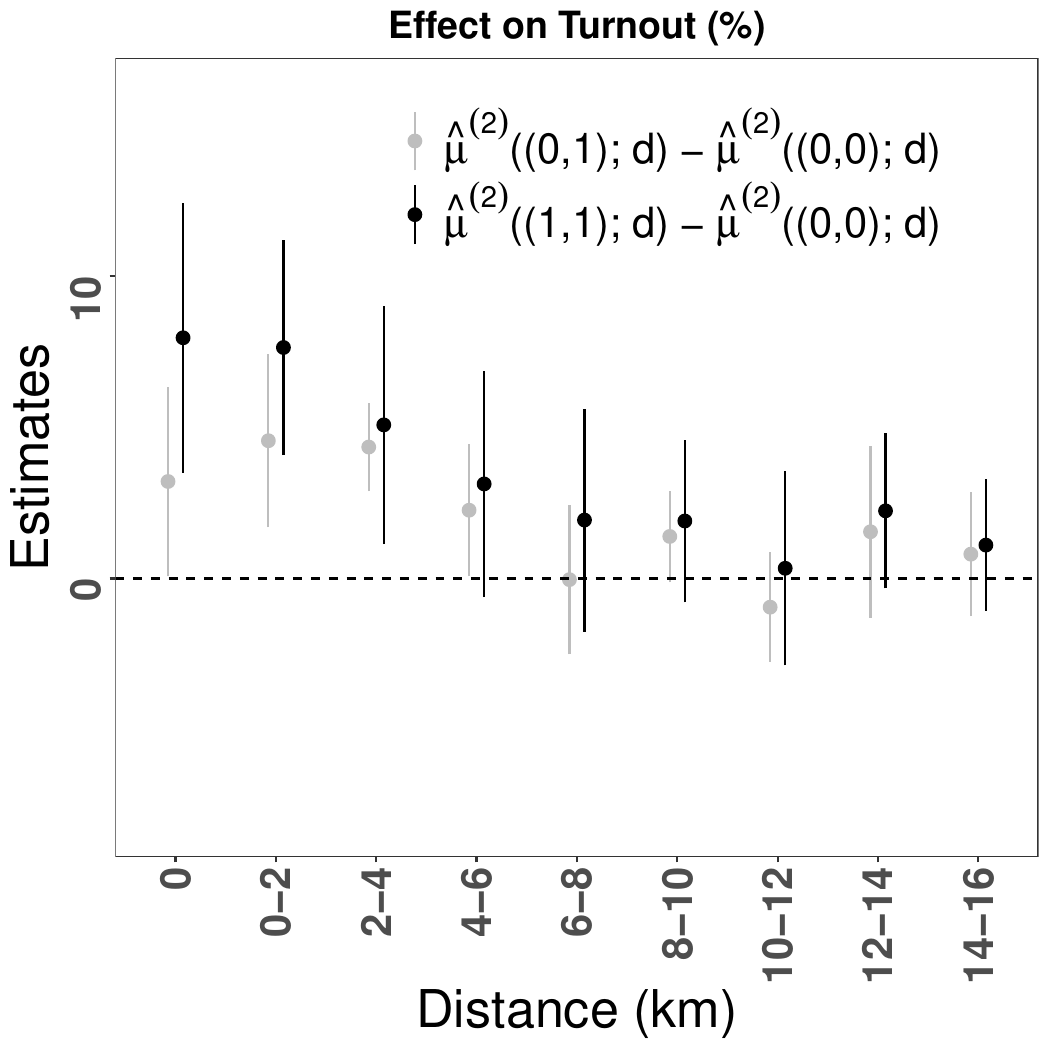}
    \includegraphics[width=.48\linewidth, height=.48\linewidth]{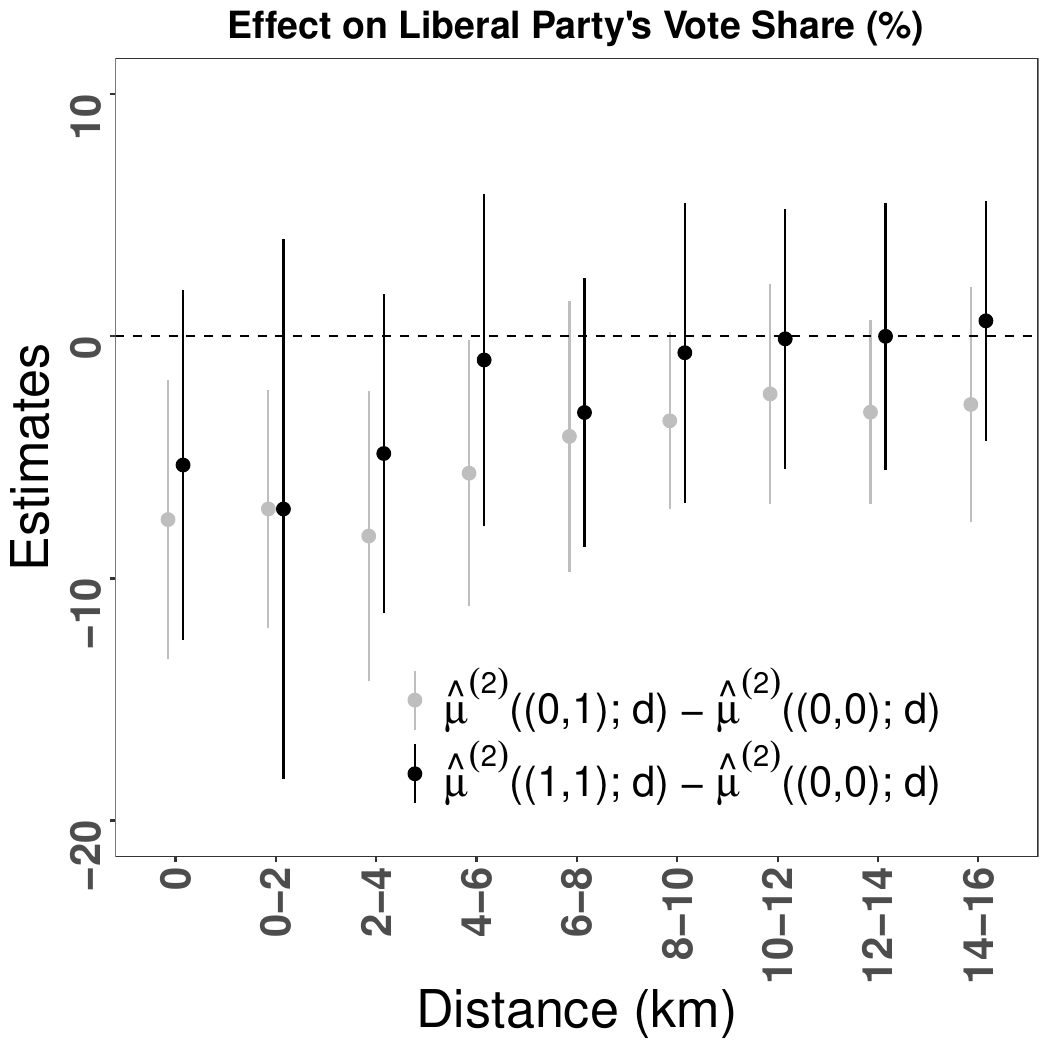}
 \end{subfigure}
 \vspace{1em} 
 \begin{subfigure}[t]{\textwidth}
    \caption{Impacts of Smoking Cessation on Health Outcomes}
    \includegraphics[width=.48\linewidth, height=.48\linewidth]{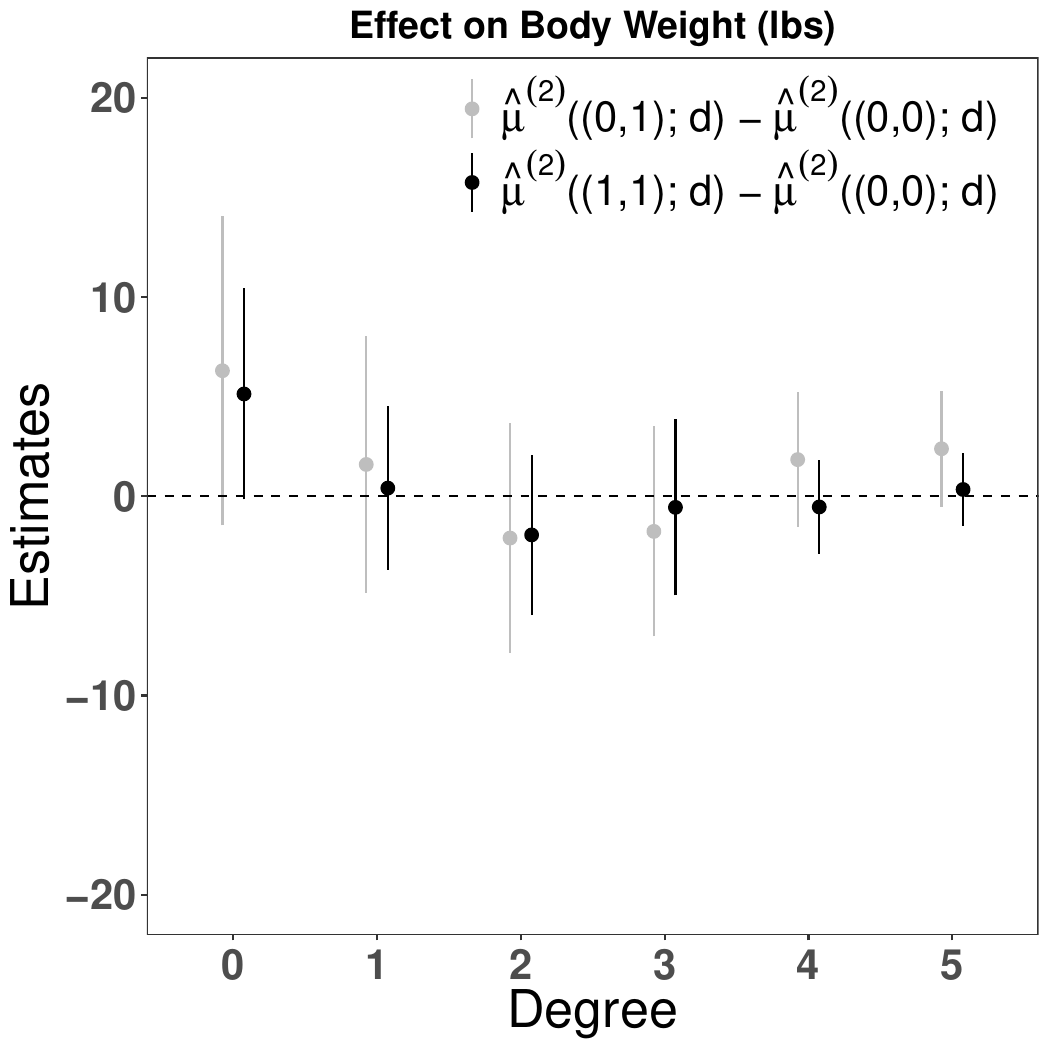}
    \includegraphics[width=.48\linewidth, height=.48\linewidth]{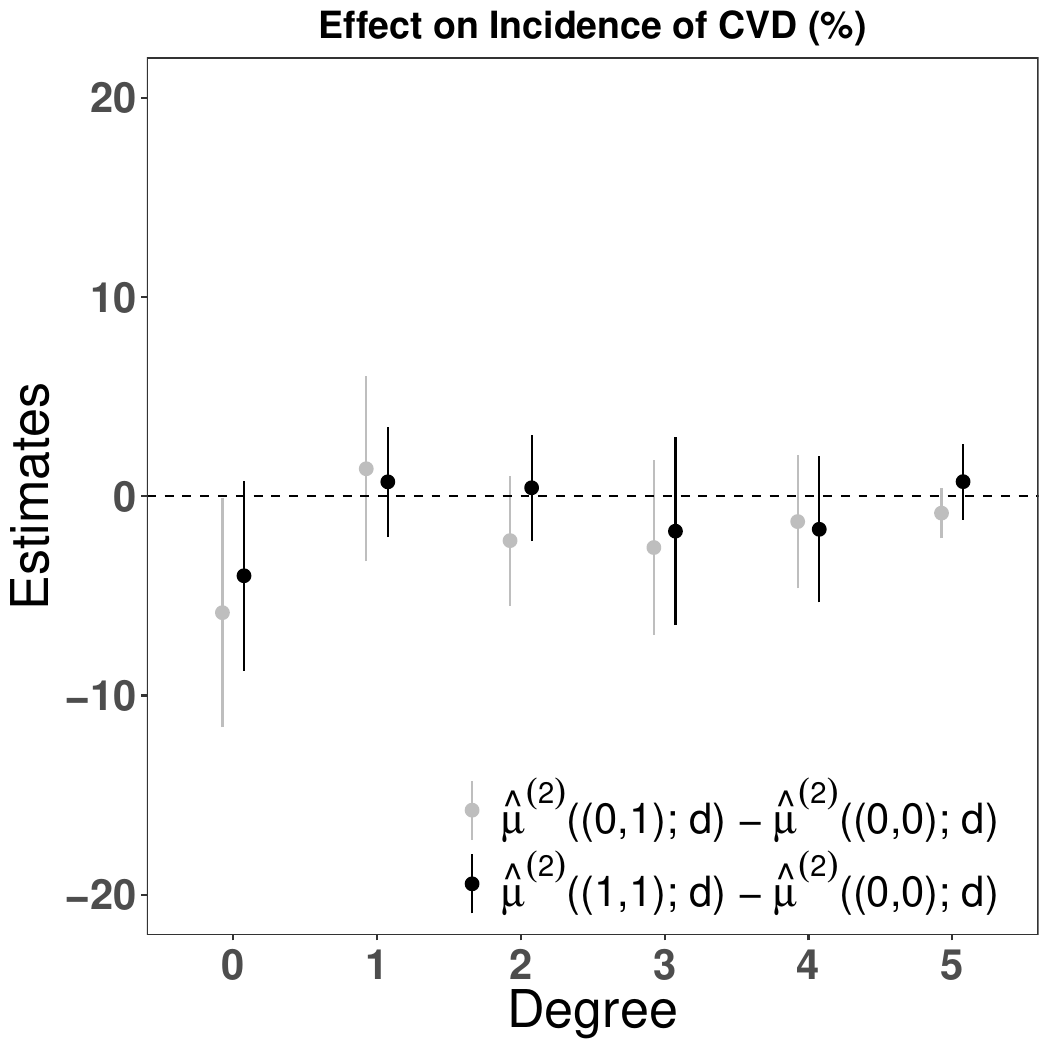}
 \end{subfigure}
 \end{center}
\textit{Notes:} Panel (a) presents results accounting for diffusion in treatment based on \citet{stokes2016electoral}, while Panel (b) displays the results from the FHS. Black and gray dots represent point estimates for different treatment histories and proximity levels. The vertical segments show the corresponding 95\% confidence intervals, computed using the spatial/network HAC variance estimator and standard normal critical values.
\end{figure}

Next, we examine the robustness of our findings to potential diffusion in treatment by modifying the propensity score model to incorporate the treatment statuses and time-varying confounders of nearby neighbors. In the replication of \citet{stokes2016electoral}, we include constituencies within a 5 km radius. For the FHS analysis, we consider up to three degrees of neighbors. Variables from each neighbor $j$ are weighted by the inverse of $d_{ij}$ when estimating the propensity score for unit $i$. Results are presented in Figure~\ref{fig:app-d}. Most point estimates and their confidence intervals remain nearly identical to those obtained without adjusting for treatment diffusion.

\begin{figure}[htp]
 \begin{center}
 \caption{mpacts of Smoking Cessation on Additional Health Outcomes}
 \label{fig:fhs-app1}
\includegraphics[width=.48\linewidth, height=.48\linewidth]{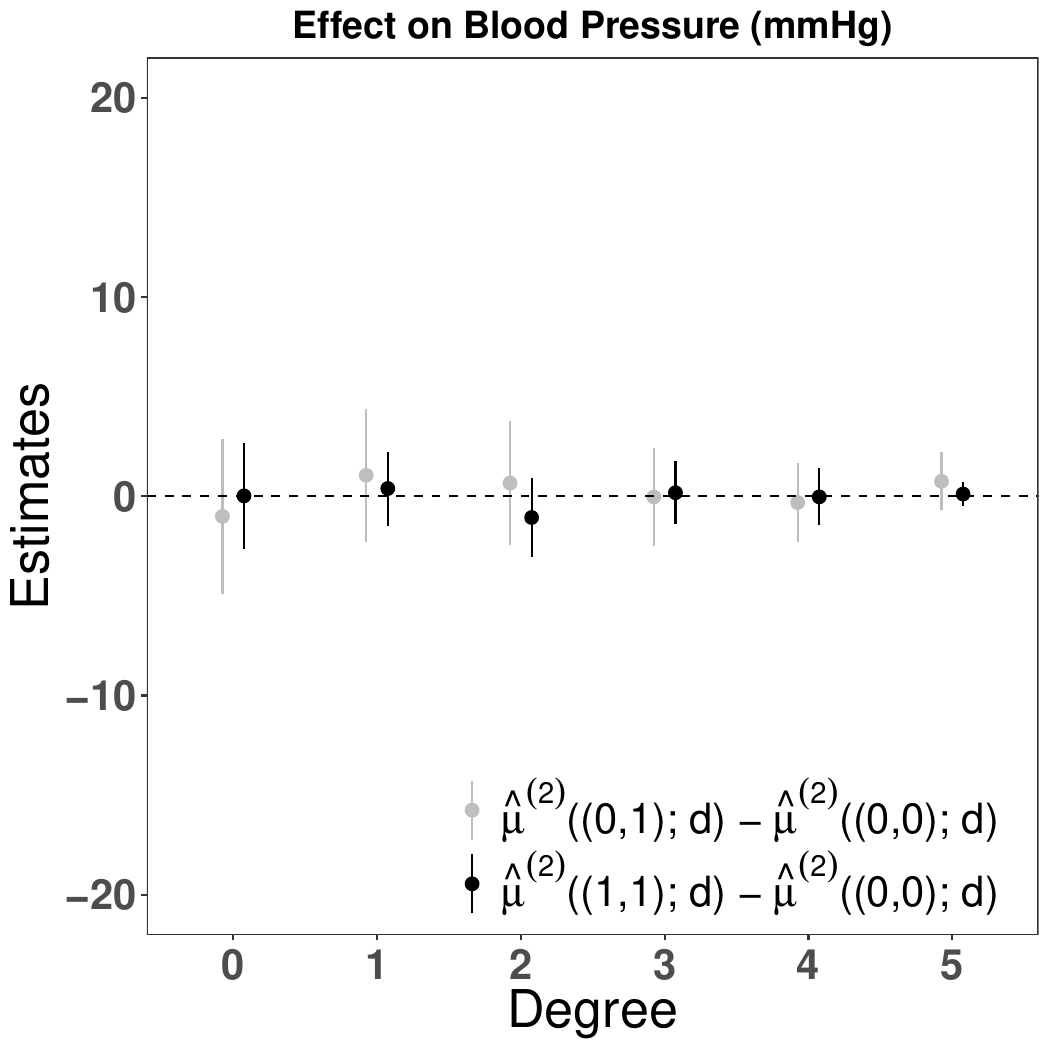}
\includegraphics[width=.48\linewidth, height=.48\linewidth]{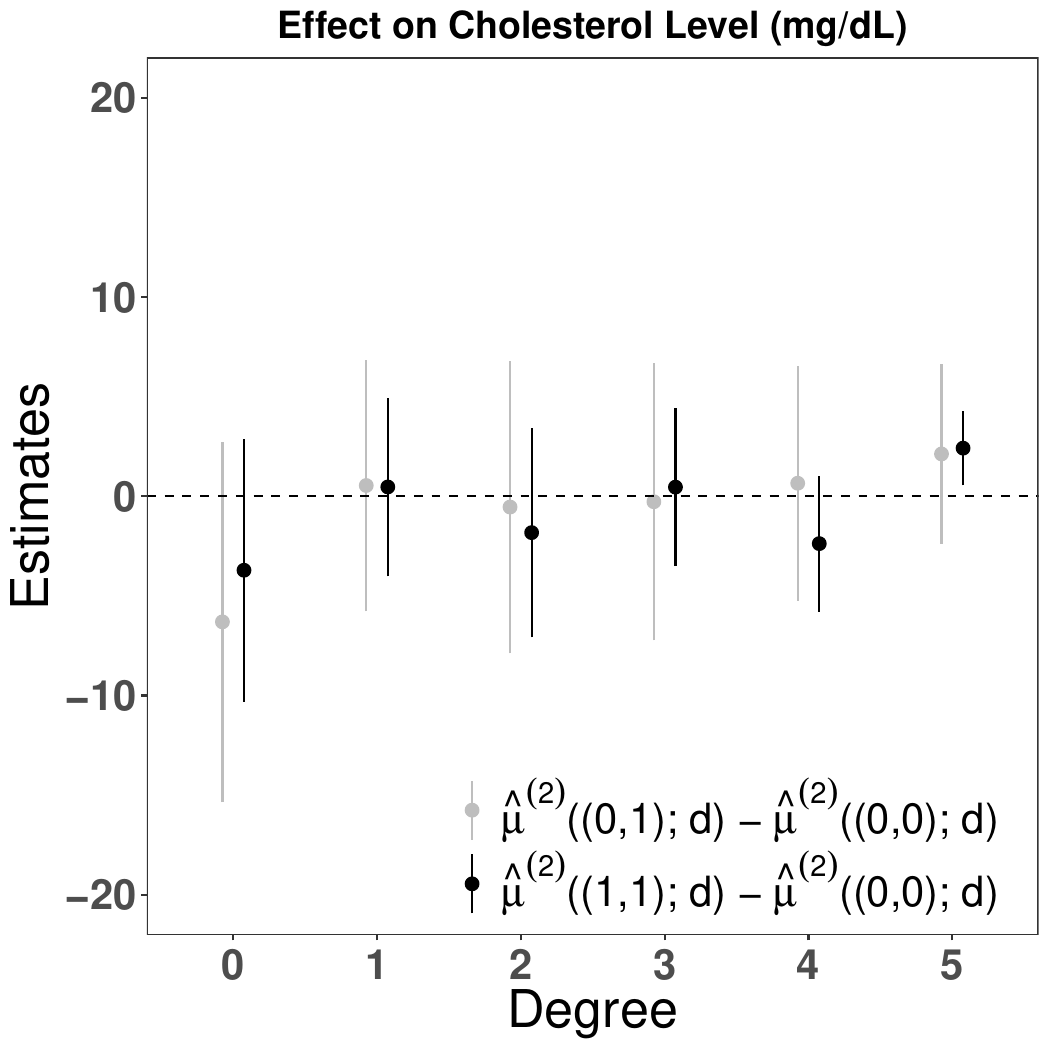} \\
\includegraphics[width=.48\linewidth, height=.48\linewidth]{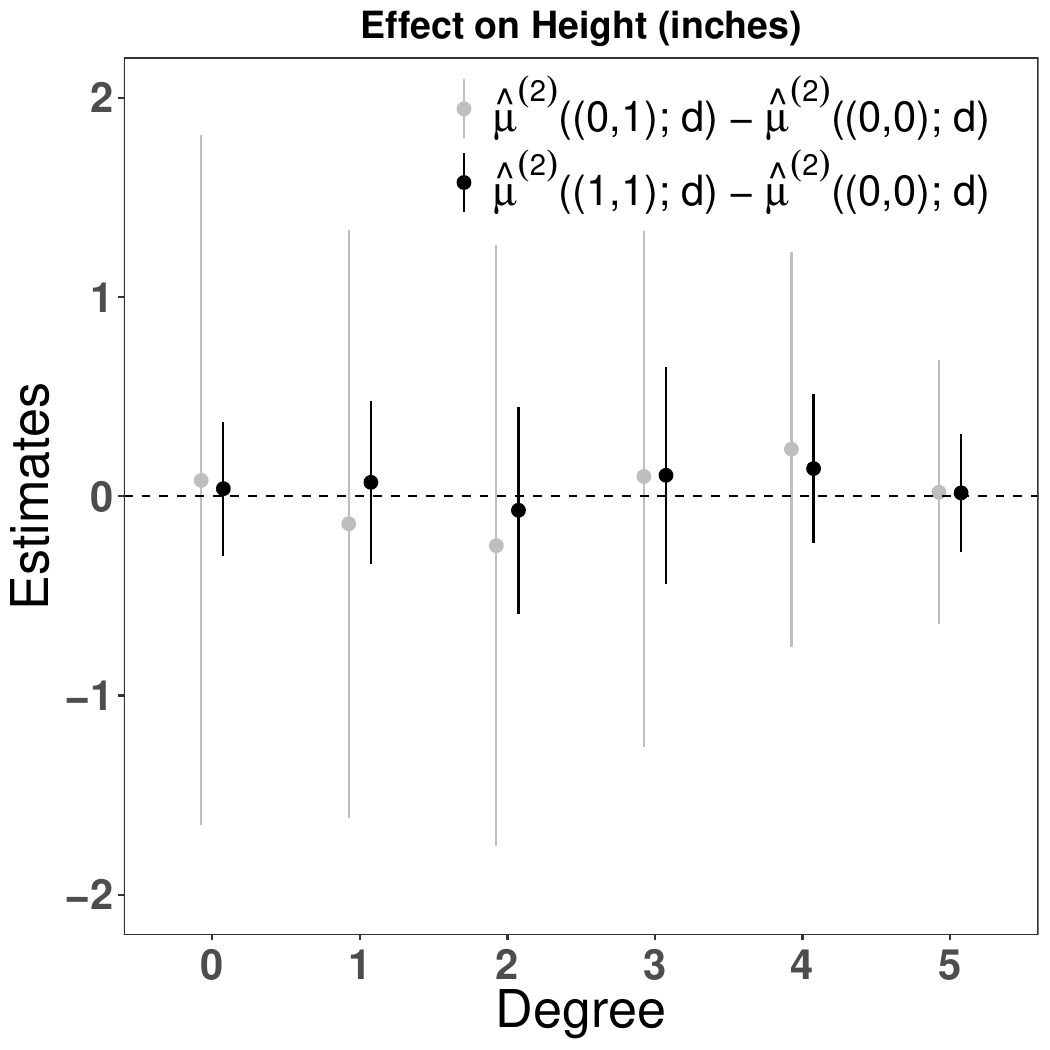}
 \end{center}
\textit{Notes:} These plots present effect estimates for additional outcomes in the FHS. Black and gray dots represent point estimates for different treatment histories and proximity levels. The vertical segments show the corresponding 95\% confidence intervals, computed using the network HAC variance estimator and standard normal critical values.
\end{figure}

Finally, we report effect estimates for three additional outcomes in the FHS: blood pressure, cholesterol level, and height. Among them, height serves as a placebo outcome, as it is unlikely to be influenced by smoking cessation. The results, presented in Figure~\ref{fig:fhs-app1}, suggest that smoking cessation leads to a moderate reduction in an individual’s cholesterol level but has little effect on their social network neighbors. In contrast, neither blood pressure nor height appears to be significantly affected by the treatment.

\end{document}